\documentclass[a4paper,fleqn]{cas-sc}

\usepackage[authoryear]{natbib}
\usepackage[mathlines,displaymath]{lineno} 
\usepackage{color}
\usepackage{etoolbox}
\BeforeBeginEnvironment{equation}{\begin{linenomath}}
\AfterEndEnvironment{equation}{\end{linenomath}}

\def\tsc#1{\csdef{#1}{\textsc{\lowercase{#1}}\xspace}}
\tsc{WGM}
\tsc{QE}
\tsc{EP}
\tsc{PMS}
\tsc{BEC}
\tsc{DE}

\begin{document}
\let\WriteBookmarks\relax
\def\floatpagepagefraction{1}
\def\textpagefraction{.001}

\shorttitle{Constraining Ganymede's Interior via Dynamo}

\shortauthors{J. Kimura et~al.}

\title [mode = title]{Constraining Ganymede's Interior Structure and Bulk Composition from Its Present-Day Magnetic Field Strength}

\tnotemark[1] 


%

\author[1]{Jun Kimura}[type=editor,
                        auid=000,bioid=1,
                        prefix=Dr,
                        role=Researcher,
                        orcid=0000-0002-5825-0454]
\cormark[1]


\ead{junkim@ess.sci.osaka-u.ac.jp}

\ead[url]{}

\credit{Conceptualization, Methodology, Calculation, Visualization, Investigation, Writing - Original draft}

\affiliation[1]{organization={Department of Earth and Space Science, The University of Osaka},
            addressline={1-1 Machikaneyama-cho}, 
            city={Toyonaka},
            postcode={560-0043}, 
            state={Osaka},
            country={Japan}}

\author[1]{Ryohei Kawakami}


\ead{kawakami120604@ess.sci.osaka-u.ac.jp}

\ead[url]{}

\credit{Conceptualization, Methodology, Calculation, Investigation}

\author[1]{Taro Matsuo}[type=editor,
                        orcid=0009-0003-7304-7512]
\ead{matsuo@ess.sci.osaka-u.ac.jp}
\credit{Investigation, Writing - review and editing}

\author[1]{Sho Sasaki}[type=editor,
                        orcid=0000-0002-7646-5248]
\ead{sasakisho2@gmail.com}
\credit{Investigation, Writing - review and editing}

\cortext[1]{Investigation, Writing - review and editing}


%

\begin{abstract}
Ganymede is the only Solar System satellite with a global magnetic field driven by a core dynamo. 
Although its small moment-of-inertia (MoI) factor indicates a differentiated structure of a metallic core, rocky mantle, and outer hydrosphere, its exact bulk composition and layer thicknesses remain highly uncertain. 
We developed a comprehensive, long-term thermochemical evolution model that couples thermal history with magnetic field generation. 
Building upon previous theoretical frameworks, this study systematically evaluates how diverse chondritic precursor compositions constrain Ganymede's internal thermochemical evolution and core solidification processes.
Based on interior structures initially permitted by a wide range of chondritic compositions and geodetic constraints, we applied the observed equatorial surface magnetic field strength ($\sim$719~nT) as a dynamic filter. 
Our evolutionary simulations demonstrate that while magnetic dynamos can operate under entire melting, Fe-snow, and FeS-layer regimes, the requirement to sustain the observed field strength strongly restricts the viable interior models. 
Satisfying this constraint preferentially requires a relatively large core radius and a specific, non-eutectic bulk core sulfur concentration. 
Consequently, this coupled thermochemical approach highlights the main novelty of this study: characterizing the hydrosphere's density serves as an independent constraint to break the structural degeneracy of Ganymede's bulk composition. 
Our model indicates that L and CI chondrites yield parameter spaces capable of sustaining comparable field strengths under a low-density hydrosphere scenario (1,200~$\mathrm{kg\,m^{-3}}$), whereas the H chondrite composition can satisfy this criterion under a high-density scenario (1,300~$\mathrm{kg\,m^{-3}}$).
These predictions provide a critical framework to be tested by the JUICE mission arriving in 2031.
\end{abstract}


\begin{highlights}
\item Ganymede's core evolution is linked to its chondritic parent-body composition.
\item A thinner rocky mantle enhances deep core cooling, prolonging the dynamo lifetime.
\item Broad compositional space search maps Ganymede's origin to its hydrosphere density.
\item Fe-snow sustains dynamos in low-sulfur cores under specific structural conditions.
\item Synergistic JUICE data can identify core regimes and planetary bulk origins.
\end{highlights}

\begin{keywords}
Ganymede \sep Interiors \sep Thermal histories \sep Magnetic fields
\end{keywords}

\maketitle

\section{Introduction}\label{}
The Galileo spacecraft's discovery of an intrinsic dipole magnetic field at Jupiter's moon Ganymede fundamentally shifted our understanding of icy satellites \citep{Kivelson1996, Kivelson2002}. 
Generating an equatorial surface field of approximately 719~nT, this magnetic field strongly implies the existence of an actively convecting liquid metallic core \citep{Kivelson1996}. 
Furthermore, while early gravity data from the Galileo mission established that Ganymede possesses a highly differentiated interior \citep{Anderson1996, Schubert2004}, recent high-precision measurements refined by the Juno mission have tightly constrained its normalized moment of inertia (MoI) factor to $0.3156_{-0.0133}^{+0.0117}$ \citep{GomezCasajus2022}.
This value unambiguously confirms that Ganymede possesses a highly differentiated interior comprising an outer hydrosphere (H$_2$O layer), a rocky mantle, and a central metallic core.

However, the MoI factor represents a highly degenerate constraint. 
Because the specific densities and thicknesses of these internal layers are poorly constrained by gravity data alone, the inferred radius of the metallic core spans a massive parameter space \citep{Anderson1996,Sohl2002,GomezCasajus2022}.
For instance, structural models satisfying the MoI indicate that depending on whether the core is assumed to be composed of pure Fe or a lighter FeS alloy, estimates for the core radius vary drastically, ranging from approximately 600~km up to 1150~km assuming the density of 1,300~kg\,m$^{-3}$ for the hydrosphere \citep{kimura2009}.
Such structural ambiguity of over 500~km directly reflects a profound uncertainty regarding the core's bulk chemical composition, specifically its inventory of light alloying elements such as sulfur.

The sulfur concentration is the primary parameter dictating the core's liquidus temperature, physical state, and mechanism of solidification \citep[e.g.,][]{hauck2006}. 
Whether Ganymede's present-day dynamo is powered by purely thermal convection or crystallization-driven compositional convection remains an active subject of debate, as the viable driving mechanism is intricately linked to the moon's uncertain internal structure. 
A previous parameterized thermal history model suggests that purely thermal convection is generally insufficient to sustain a core dynamo to the present day unless exceptional core cooling rates or unique tidal heating profiles are invoked \citep{Bland2008}, thereby underscoring the necessity of compositional buoyancy derived from core crystallization. 
Energetic analyses have since demonstrated the viability of such crystallization-driven regimes, providing quantitative scaling laws that show how the generated compositional buoyancy can theoretically produce the observed surface magnetic field \citep{Zhan2012,Christensen2015}.
Conversely, a more detailed heat transport model has demonstrated that purely thermal convection can successfully sustain a present-day dynamo without invoking compositional buoyancy, provided that specific structural and sulfur conditions, such as a large core radius and high initial temperatures, are satisfied \citep{kimura2009}. 
These distinct thermodynamic pathways dramatically alter the magnetic field's temporal evolution and lifespan \citep{Ruckriemen2018}. 
Therefore, any attempt to accurately resolve Ganymede's internal structure must intrinsically link the moon's present-day magnetic field strength to both the thermal evolution of the mantle and the specific solidification regime of its core.

Despite this established necessity, previous evolution models have struggled to comprehensively capture the coupled dynamics of Ganymede's interior, often isolating individual physical regimes. For instance, although the detailed heat transport model by Kimura et al. (2009) provided an advanced treatment of solid-state convection within the silicate mantle and the hydrosphere, it neglected the internal thermochemical evolution and complex solidification mechanics of the core itself. Conversely, numerical models focusing primarily on core solidification dynamics provided invaluable insights into top-down freezing processes \citep{hauck2006, Ruckriemen2015, Ruckriemen2018}. 
Yet, these core models typically utilized parameterized models to represent the overlying silicate mantle. 
While these parameterized models solve robust volumetric energy conservation equations and provide fundamental insights, capturing the precise transient heat flux across the core-mantle boundary across a broad compositional space remains challenging.
Although the thermodynamic and fluid dynamic behaviors of the FeS-layer and Fe-snow regimes have been treated in quantitative models \citep[e.g.,][]{Ruckriemen2015,Ruckriemen2018}, understanding their long-term bifurcation and evolution requires tracking them across a broad compositional space.
Most critically, none of these prior dynamic frameworks explicitly integrated cosmochemical constraints based on the accreting material.
While the exact final state depends on subsequent processes like metal-silicate element partitioning during differentiation, incorporating the bulk compositions of distinct chondrite parent bodies fundamentally reduces structural degeneracy.
This approach provides a plausible baseline that intrinsically links the potential ranges of mantle density, the total radiogenic heat inventory, and the core's elemental budget \citep[e.g.,][]{Warren2011}.

To lift the degeneracy of the gravity-derived MoI constraint and resolve the structural ambiguity of Ganymede, we employ a fully coupled thermal-chemical-magnetic evolution framework that incorporates diverse chondritic precursor compositions as cosmochemical constraints.
Consistent with previous evolutionary frameworks \citep[e.g.,][]{Ruckriemen2018}, we use the observed present-day surface magnetic field as an observational constraint. 
However, instead of varying individual thermochemical parameters, we apply this constraint to evaluate which interior structure models derived from specific parent-body bulk compositions can successfully reproduce the current dynamo activity.
Following the calculations by \citet{Bercovici2022}, we first anchor the internal layer dimensions and core sulfur inventories to their rigorous phase equilibrium results for a wide range of chondritic bulk compositions. 
By coupling these structurally constrained models with a solid-state mantle convection model, and by explicitly incorporating the thermochemical evolution of both the Fe-snow and FeS-layer regimes, we iteratively calculate the available dynamo power over 4.6 billion years. 
Ultimately, by isolating the specific subset of core sizes, sulfur concentrations, and chondritic bulk compositions capable of sustaining a magnetic field comparable to the observed $\sim$719~nT, this study demonstrates its primary novelty. 
Utilizing the hydrosphere density as an independent constraint breaks the structural degeneracy of Ganymede's primordial origin. 
This approach provides tightly constrained interior models essential for interpreting future geophysical data anticipated from the upcoming ESA JUICE mission.

The remainder of this paper is organized as follows. 
Section 2 describes our coupled evolutionary framework, outlining the gravitational and cosmochemical bulk composition constraints, the solid-state mantle convection model, and the thermochemical evolution of the core, which encompasses its comprehensive energy budget from an initially fully molten state and potential solidification pathways via the Fe-snow and FeS-layer regimes.
Section 3 presents our numerical results, detailing the temporal evolution of the thermal profiles within the mantle and core, the core-mantle boundary heat flux, the available dynamo power, and the resulting present-day magnetic field strength.
In Section 4, we examine the model limitations and sensitivities regarding core-mantle thermal coupling, hydrosphere simplifications, differentiation timing, and core thermal conductivity, while exploring how upcoming geodetic and magnetic field measurements from the JUICE mission can observationally discriminate between the proposed interior regimes. Finally, Section 5 summarizes our main conclusions and outlines their implications for the internal structure and primordial bulk composition of Ganymede.

\section{Methods}
\subsection{Model overview for the interior evolution}

We performed numerical simulations of Ganymede's global thermal evolution under a wide range of parameter conditions to investigate the state and the magnetic field generation in the core.
One of the major methods to calculate planetary thermal evolution is a parameterized convection model, which is a dimensionless energy balance model employing a scaling law between the Nusselt number (Nu) and the Rayleigh number (Ra).
Although a 3D numerical scheme is another robust approach, its high computing costs limit the number of calculations that may be performed.
In this study, we employed a 1D method based on the mixing length theory to estimate the convective heat flux in the sub-solidus regime. 
This method has been applied to terrestrial planets and icy satellites in previous studies \citep[e.g.,][]{Abe93,Senshu02,Wagner11,kamata19,kimura2020,kimura2024}. 
This approach possesses the distinct merit of producing a continuous temperature profile that incorporates conductive and convective contributions consistently, while maintaining significantly lower calculation costs compared to 3D modeling.
However, the mixing length theory remains an approximation. As highlighted by recent studies, accurately capturing absolute heat transport requires calibration of the mixing-length parameter against fully dynamic 3D convection simulations  \citep{wagner19,thiriet19}.
Despite this uncertainty in absolute thermal flux, the mixing length theory remains a robust and practical approach for our objective of systematically exploring a broad compositional parameter space to determine relative interior regime transitions. 
The qualitative impacts of this calibration uncertainty on the mantle heat extraction efficiency and the subsequent timing of core dynamo generation are discussed in Section 4.1.

In this study, we do not explicitly compute the heat transfer within the H$_{2}$O hydrosphere. 
Instead, we prescribed a constant temperature boundary of 300 K at the rocky mantle-hydrosphere boundary (RHB).
This simplification is physically justified by the significant contrast in thermal resistance between the high-viscosity silicate mantle and the low-viscosity hydrosphere. 
The heat transfer efficiency within the H$_{2}$O layer is orders of magnitude higher than that of the rocky mantle; thus, the overall cooling rate of the deep interior is predominantly governed by the convective heat flux through the rocky mantle.
The fixed temperature of $300\text{ K}$ is a representative value for the RHB under typical pressure conditions ($\sim 1.5\text{--}2\text{ GPa}$), consistent with previous interior models.
%
\subsection{Allowable range of interior structures under gravitational and compositional constraints}
In this study, we assume a spherically symmetric three-layer model for Ganymede consisting of a central metallic core, an overlying rocky mantle, and an outermost H$_{2}$O hydrosphere. 
To determine the permissible interior structures, we first establish the geophysically viable parameter space by solving the coupled equations for Ganymede's total mass ($M_G = 1.4819 \times 10^{23}\text{ kg}$) and normalized moment of inertia (MoI) factor of 0.3156 \citep{GomezCasajus2022}.
The core is assumed to be an Fe-FeS binary eutectic system \citep{Goldstein2009, Scott2002}. 
We systematically vary the average rocky mantle density within 3000 to 4000\,kg\,m$^{-3}$ and the core density within 4900 to 7100\,kg\,m$^{-3}$, which corresponds to a bulk sulfur concentration of 0 to 36 wt\%.
We evaluate two representative bulk hydrosphere densities of 1200 and 1300\,kg\,m$^{-3}$. 
This geophysical filtering delineates the broad continuous domain of compatible core radii, mantle densities, and layer thicknesses (shaded regions in Figure \ref{fig:1}). 
Because mass is conserved within the deep interior (the rocky mantle and the metallic core), these structural boundaries remain effectively static over 4.6 Gyr.
Even complete core stratification, such as segregation into a solid pure-Fe inner core and a liquid eutectic outer core, alters the normalized moment of inertia factor by only about 0.001, which is an order of magnitude smaller than the observational uncertainty ($\sim$0.012). 
Therefore, this deep structural framework derived from present-day constraints serves as a robust, invariant geometric baseline for our long-term evolutionary simulations, regardless of unmodeled phase changes in the overlying hydrosphere.

Within this geophysically permissible space, we next map the specific chemical constraints derived from the thermodynamic differentiation model of \citet{Bercovici2022} for various chondrite types (H, L, LL, CI, CM, CO, CV, and CK) which represent the bulk composition of the combined rocky mantle and metallic core. 
While carbonaceous, comet-like materials are traditionally considered the primary building blocks of Jovian satellites, the exact source and relative proportions of solids in the circum-Jovian disk remain major unresolved questions \citep{Nimmo2026}. 
It remains a viable possibility that the moons accreted a mixture of both outer solar system materials and thermally processed, dehydrated chondritic materials from the inner solar system \citep{Becker2024}. 
Consequently, we treat all chondrite classes as equally likely initial candidates, allowing the observed magnetic field to serve as an objective filter.
Although model of \citet{Bercovici2022} was originally formulated for planetesimals, this approach remains robust for Ganymede because the bulk partitioning budgets of iron and sulfur are primarily dictated by precursor chondritic ratios.
Based on these precursor ratios, the estimated core sulfur concentrations vary among the different bulk compositions because the sulfur partition coefficient during core-mantle differentiation is sensitive to temperature and pressure \citep{Bercovici2022}.
It should be noted that their model assumes anhydrous conditions. 
Qualitatively, the presence of substantial H$_{2}$O during early differentiation would increase the oxidation state, driving more iron into the silicate mantle as FeO and yielding a smaller metallic core with a relatively higher bulk sulfur concentration.
In this study, we adopt the anhydrous framework as a thermodynamically consistent first-order approximation, assuming that the majority of H$_{2}$O rapidly separated upward into the hydrosphere, leaving the deep segregation environment relatively depleted in volatiles.
This approach establishes baseline core sulfur budgets and structural ranges for various chondritic compositions, noting that future models incorporating complex hydration dynamics may further refine these estimates.
We utilize their framework solely to derive the initial core sulfur and iron mass budgets. 
In our models, the mantle density and the corresponding core radius are calculated independently by strictly satisfying the observational constraints of Ganymede's total mass and MoI.
By superimposing the resulting core-to-silicate radius ratios and core sulfur contents (Table \ref{tab:1}) onto the geophysically permitted domains, we identify the specific structural combinations for each chondritic model (colored bars in Figure \ref{fig:1}).

The resulting structural configurations depend strongly on the assumed bulk density of the H$_{2}$O hydrosphere.
The 1200\,kg\,m$^{-3}$ case is intended to represent a hydrosphere with a relatively thick subsurface ocean and a smaller volume fraction of high-pressure ice phases (ices V and VI).
In this case, the permissible core radius spans 550 to 850~km with a sulfur concentration of 14 to 36~wt.\% (Figure \ref{fig:1} left). 
The 1300\,kg\,m$^{-3}$ case is adopted as a high-density endmember, motivated by a larger portion of high-pressure ice in structures with thin or absent oceans, while allowing for uncertainties in pressure-temperature conditions and non-H$_{2}$O constituents.
This case yields a core radius of 600 to 850~km with a sulfur concentration of 10\,--\,22~wt.\% (Figure \ref{fig:1} right).
It should be noted that while the pure gravitational constraint dictates that a higher sulfur concentration necessitates a larger core radius, these quoted overall ranges represent the encompassing envelope of the discrete chondrite classes that remain viable under each specific hydrosphere density, rather than a continuous compositional variation.
These constrained structural and compositional parameters serve as the initial conditions for our subsequent 4.6-Gyr coupled thermochemical evolution simulations.
%
\begin{table}
    \centering
    \caption{
        Inferred radius ratio of the metallic core to the rocky-metallic interior and sulfur content for each chondrite type modeled by \citet{Bercovici2022}.
        }
    \label{tab:1}
        \begin{tabular}{cccc} \toprule
        Chondrite & Core radius \% & Sulfur \\
	type & to rocky interior & concentration (wt\%) \\ \midrule
        H  & $48.8\pm0.59$ & $9.4\pm3.5$ \\
        L  & $42.8\pm0.34$ & $16\pm3.1$ \\
        LL & $38.4\pm1.22$ & $22\pm5.3$ \\
        CI & $46.6\pm n/a$ & $35\pm n/a$ \\
        CM & $40.3\pm0.91$ & $34\pm2.3$ \\
        CO & $38.8\pm3.61$ & $21\pm6.5$ \\
        CV & $31.4\pm3.84$ & $22\pm8.3$ \\
        CK & $29.9\pm1.15$ & $24\pm1.4$ \\
        \bottomrule
        \end{tabular}
\end{table}

\subsection{Initial state}
We perform thermal evolution simulations over 4.6~Gyr until the present day within the ranges of internal structures determined in the previous section. 
We assume that Ganymede's interior is already differentiated into three layers at the onset of the simulation. 
For the calculations, the metallic core and the rocky mantle are discretized into concentric spherical shells. 
The initial temperature profile of the rocky mantle is assumed to be linear, extending from the core-mantle boundary (CMB) at its base to the boundary with the H$_{2}$O layer at its top. 
The core is assumed to be entirely molten at the onset of the simulation. Its initial temperature at the CMB is locked to its composition-dependent liquidus, and the internal core temperature follows an adiabatic gradient.

While the timing and duration of the differentiation process remain subjects of ongoing debate, we assume an initially differentiated three-layer structure as a baseline for our 4.6-Gyr thermal evolution model.
Recently, it has been proposed that Ganymede might have experienced protracted core formation starting from a cold interior \citep{Trinh2026}. 
However, our adoption of the conventional "hot start" scenario is supported by the fact that the gravitational potential energy released during rapid accretion, supplemented by the decay of short-lived radionuclides, is generally considered sufficient to trigger the melting of ice and the subsequent segregation of a metallic core \citep{Schubert2004, Monteux2014}. 
While Ganymede's low moment of inertia factor and intrinsic magnetic field are consistent with a fully differentiated interior, these present-day observations do not prohibit ongoing differentiation.
Both the early complete differentiation and the protracted core formation scenarios remain viable, and further observational evidence is needed to discriminate between them. 
In this study, we adopt the fully differentiated configuration as a standard benchmark for Ganymede in stark contrast to the partially differentiated state of its neighbor, Callisto  \citep{Nagel2004}.
By assuming a fully differentiated three-layer structure from the onset, we can isolate and evaluate the influence of core sulfur concentration and mantle heat transport on the evolution of magnetic field intensity, providing a necessary end-member comparison to delayed differentiation models.
%
\subsection{Numerical model for thermal and structural evolution}
The following is the general equation for heat transfer:
\begin{equation}
\rho_{m} C_{p,m}\frac{dT_{m}}{dt}=-\frac{1}{r^{2}}\frac{d}{dr}\left(r^{2}F_{cond}+r^{2}F_{conv}\right)+\rho_{m}Q_{rad}
\label{eq:heat_transfer}
\end{equation}
where $\rho$ is density, $C_{p}$ is specific heat, $T$ is temperature, and subscript $m$ represents the rocky mantle (Table \ref{tab:parameter}).
$t$ is time, $r$ is radial distance from the center, and $Q$ is heat production rate per unit mass.
The conductive and convective heat fluxes are given by
\begin{equation}
F_{cond}=-k_{cond,m}\frac{dT}{dr},
\end{equation}
\begin{equation}
F_{conv}=-k_{conv,m}\left\{\frac{dT}{dr}-\left(\frac{dT}{dr}\right)_{ad}\right\},
\end{equation}
where $k_{cond,m}$ is thermal conductivity, and $k_{conv,m}$ is {\it effective} thermal conductivity, which emulates the effect of thermal convection, is expressed as follows \citep[e.g.,][]{kimura2009}:
\begin{equation}
\!k_{conv,m}\!=\!\left\{
\begin{array}{ll}
\displaystyle{-\frac{\rho_{m}^{2}C_{p,m}\alpha_{m}g\ell^{4}}{18\eta_{m}}
\left\{\frac{\partial T}{\partial r}
\!-\!\left(\frac{\partial T}{\partial r}\right)_{ad}\right\}}
& \textrm{for }\;\displaystyle{\frac{\partial T}{\partial r}
\!<\!\left(\frac{\partial T}{\partial r}\right)_{ad}} \\
0 & \textrm{for }\;\displaystyle{\frac{\partial T}{\partial r}
\!>\!\left(\frac{\partial T}{\partial r}\right)_{ad}}
\end{array}
\right.
\end{equation}
where $g$ is the gravitational acceleration varying with radial depth, $\alpha$ is thermal expansion coefficient, and $\eta$ is viscosity.
$\ell$ denotes a characteristic length scale (mixing length), introduced to evaluate convective thermal transport. 
A more detailed explanation is provided later.
($dT/dr)_{ad}$ is the adiabatic temperature gradient given by
\begin{equation}
\left(\frac{dT}{dr}\right)_{ad}=-\frac{\alpha gT}{C_{p}}.
\end{equation}
Note that the both $dT/dr$ and $(dT/dr)_{ad}$ are negative; a positive convective heat flux appears only if the temperature gradient is steeper than the adiabatic temperature gradient.
$\ell$ linearly increases with depth to the peak value $b$ until it reaches the peak depth $a$, and then it decreases linearly with depth as follows:
\begin{equation}
\!\ell\!=\!\left\{
\begin{array}{ll}
\displaystyle{\frac{b}{a}(R_{m}-r)}
& \textrm{for }\;\displaystyle{r\!\ge\! R_{m}-aD_{mntl}} \\
\displaystyle{\frac{b}{1-a}(r-R_{c})}
& \textrm{for }\;\displaystyle{r\!\le\! R_{m}-aD_{mntl}}
\end{array}
\right.
\end{equation}
where $R_{m}$ and $R_{c}$ are the radii of the rocky mantle and the metallic core, respectively.
$D_{mntl}=R_{m}-R_{c}$ is the thickness of the rocky mantle.
$a=0.5$ and $b=0.5$, where $\ell$ is assumed to represent the distance to the nearest boundary of the layer to reproduce a $Nu\sim Ra^{1/3}$ relationship \citep[e.g.][]{kimura2009}.
As introduced in Section 2.1, we note that these theoretical baseline values are uncalibrated against 3D stagnant-lid convection simulations. 
The qualitative impact of using these uncalibrated parameters on the mantle heat transport efficiency and core cooling history is discussed in Section 4.1.
We consider the temperature--dependent olivine viscosity given by
\begin{equation}
\eta_{m}=\eta_{m,ref}\exp{\left[\frac{E_{m}}{R_{g}}\left(\frac{1}{T}-\frac{1}{T_{melt,m}}\right)\right]}
\end{equation}
where $\eta_{m,ref}$ is reference viscosity of the rocky mantle, $R_{g}$ is the gas constant, $E_{m}$ is the activation energy for diffusion creep in the rocky mantle (Table \ref{tab:parameter}).
$T_{melt,m}=(1.0\times10^{-7})P+1500$ is the melting temperature of the mantle where $P$ is the pressure \citep{zhang1994}.
Through this pressure dependence of the melting temperature, the viscosity also implicitly depends on depth.

The effective viscosity of silicate rock can vary by several orders of magnitude depending on poorly constrained factors, most notably the water content \citep{Hirth03}.
Although a higher water content initially reduces the rock viscosity, it concurrently lowers the mantle solidus \citep{Katz2003}.
This process induces partial melting and subsequent dehydration of the residual mantle under a stagnant lid regime, which efficiently removes water and eventually restores a higher viscosity \citep{Ruckriemen2018}.
Therefore, the assumption of a representative reference viscosity serves as a reasonable and robust approximation for evaluating the long-term thermal evolution of Ganymede.

The heat source term, $Q_{rad}$, in equation (\ref{eq:heat_transfer}) for the rocky mantle is then described as follows:
\begin{equation}
    Q_{\mathrm{rad}} = \sum_{\mathrm{i}}{c_{\mathrm{i}} H_{\mathrm{i}} \exp \left(- \lambda_{\mathrm{i}} t \right)}
\end{equation}
where $i$ represents isotopes, $c_{i}$ is initial concentration, $H_{i}$ is heat release, and $\lambda_{i}$ is the decay constant.
In this model, the radioisotope abundance is assumed to be the concentration rate according to assumed chondritic composition and included in the energy equation as the heat source amount per unit mass (Table \ref{tab:radiogenic_isotope}).
We assumed that the partitioning of heat-producing isotopes, such as $^{40}$K, into the metallic core is negligible due to the relatively low pressure of Ganymede's core (approximately 4 to 10~GPa). 
Consequently, all radiogenic heat sources are assumed to be entirely retained within the rocky mantle.
It should be noted that if early differentiation within the rocky interior caused heat-producing isotopes to concentrate near the upper part of the rocky mantle. This would likely accelerate the cooling of the deeper interior \citep{Morschhauser2011}.
%
\subsection{Evolution of the metallic core}

Our theoretical approach for modeling the thermochemical evolution of the core, including the Fe-snow regime, fundamentally follows the framework developed by \citet{Ruckriemen2015,Ruckriemen2018}.
We model the metallic core as an initially fully molten and well-mixed body, where the temperature profile follows an adiabatic gradient,
\begin{equation}
    \label{eq:adiabatic_temperature}
        T_{\mathrm{c}} = 
        T_{\mathrm{cmb}} 
        \exp\!
        \left\{
            \frac{\alpha_{\mathrm{c}}}{\rho_{\mathrm{c}} C_{\mathrm{p,c}}}
            \left( P - P_{\mathrm{cmb}} \right)
        \right\},
\end{equation}
where $T_{\mathrm{c}}$ and $T_{\mathrm{cmb}}$ are the temperature in the core, and the temperature at the CMB, respectively.
$\alpha_{\mathrm{c}}$ is the thermal expansivity of the core, $C_{\mathrm{p,c}}$ is the specific heat of the core (Table \ref{tab:parameter}), and $P_{\mathrm{cmb}}$ is the pressure at the CMB.
Unlike the core density, which is formulated later in this section as a function of local pressure, temperature, and sulfur concentration because its variation primarily dictates the compositional buoyancy driving the dynamo, we assume $\alpha_{\mathrm{c}}$, $C_{\mathrm{p,c}}$, and specific entropy of fusion ($\Delta S_{\mathrm{Fe}}$, $\Delta S_{\mathrm{FeS}}$) which will be described later, to be composition-independent constants.
Although recent experimental and thermodynamic studies have significantly advanced our understanding of pure iron properties \citep[e.g.,][]{Dorogokupets2017}, high-pressure melting relations in related iron-bearing systems \citep[e.g.,][]{Komabayashi2014}, and liquid Fe-S alloy properties \citep[e.g.,][]{Xu2021}, obtaining comprehensive and complete experimental thermodynamic data that simultaneously cover the entire pressure range of Ganymede's core (4 to 10\,GPa) across the wide range of sulfur concentrations (0 to 36 wt.\%) remains challenging.
Consequently, we adopt representative reference values following the approach of recent core evolution models \citep{Ruckriemen2018}.
The qualitative impacts of adopting these constant values on the core's thermal evolution and dynamo generation are discussed in detail in Section \ref{sec:core_thermal_parameters}.
Although a stable thermal stratification may develop at the top of the core due to early heating from the mantle or the general thermal conditions typical of small bodies like Ganymede \citep[e.g.,][]{Greenwood2021}, we assume an adiabatic profile throughout the fully molten state as a practical baseline.
While high thermal conductivity facilitates thermal diffusion, the degree of core stratification is modulated by the balance between this transport property and the heat flux extracted across the core-mantle boundary. Nevertheless, this adiabatic profile serves as a practical approximation with a negligible impact on the macroscopic energy balance during the subsequent cooling phase.
The time evolution of $T_{\mathrm{cmb}}$ is deduced by numerically integrating the following global energy balance equation over time:
\begin{equation}
    \label{eq:core_energy_conservation}
    \rho_{\mathrm{c}} C_{\mathrm{p,c}}
    \int_{V_{\mathrm{c}}} \frac{\partial T_{\mathrm{cmb}}}{\partial t} \, dV
    = Q_{\mathrm{grav}} - 4 \pi R_{\mathrm{c}}^{2} F_{\mathrm{cmb}},
\end{equation}
where $t$ is the time, $R_{\mathrm{c}}$ is the core radius, $V_{\mathrm{c}}$ is the core volume, $\rho_{\mathrm{c}}$ is the core density, and $F_{\mathrm{cmb}}$ is the heat flux through the CMB.
The heating due to gravitational potential release in the core, $Q_{\mathrm{grav}}$, is given as follows,
\begin{equation}
    Q_{\mathrm{grav}} = - \frac{dE_{\mathrm{grav}}}{dt},
\label{eq:qgrav}
\end{equation}
\begin{equation}
    E_{\mathrm{grav}}
    = - 4 \pi G \int_{0}^{R_c} M(r) \rho(r) r dr,
\end{equation}
where $E_{\mathrm{grav}}$ is the gravitational potential energy, $G$ is the gravitational constant, and $M(r)$ is the mass of the core within the radius $r$.
In case of the fully molten state of the core, solidification and compositional gradient can not be created in the core, thus $Q_{\mathrm{grav}}$ will be negligible because it will be sufficiently small compared with the secular cooling due to the rocky mantle.
Therefore, the cooling from the rocky mantle controls the core temperature before starting solidification in the core.
Liquidus in the core $T_{\mathrm{l,c}}$ is assumed linear relationship to the sulfur concentration, $x_{S}$, which is based on \citet{fei1995,fei1997,Boehler1992,Boehler1996,hauck2006},
\begin{equation}
    T_{\mathrm{l,c}} = 
    T_{\mathrm{l,end}} +
    \frac{ T_{\mathrm{l,end}} - T_{\mathrm{l,eut}}}{x_{\mathrm{S,eut}} - x_{\mathrm{S,end}}}
        \left(x_{\mathrm{S}} - x_{\mathrm{S,end}} \right),
\end{equation}
where $T_{\mathrm{l,eut}}$ and $T_{\mathrm{l,end}}$ is the eutectic point for the Fe--FeS system and the liquidus for the endmember composition of Fe--FeS system, respectively.
$x_{\mathrm{S,eut}}$ and $x_{\mathrm{S,end}}$ is the sulfur mass fraction for the eutectic composition and for the endmember composition, respectively.
Here, the eutectic sulfur mass fraction, $x_{\mathrm{S,eut}}$, strongly depends on the local pressure. To account for this pressure dependence as illustrated in Figure 2, we dynamically calculate $x_{\mathrm{S,eut}}$ using the following relation \citep{Rivoldini2011}:
\begin{equation}
x_{\mathrm{S,eut}} = 0.11 + 0.187 \exp(-0.065 \cdot 10^{-9} P).
\end{equation}
In case that its average value is smaller than the eutectic value, we consider Fe as an endmember, in case that its average value is larger than the eutectic value, we consider FeS as an endmember.
Thus, $x_{\mathrm{S,end}}$ is given as follows,
\begin{equation}
    x_{\mathrm{S,end}} =
    \begin{cases}
        0
        & \left\langle x_{\mathrm{S}} \right\rangle < x_{\mathrm{S,eut}} \\[6pt]
        0.36
        & \left\langle x_{\mathrm{S}} \right\rangle > x_{\mathrm{S,eut}},
    \end{cases}
\end{equation}
$T_{\mathrm{l,end}}$ and $T_{\mathrm{l,eut}}$ are given as follows,
\begin{equation}
    T_{\mathrm{l,end}} =
    \begin{cases}
        1809 \left(1 + 1.99 \times 10^{-11} P - 1.84 \times 10^{-22} P^{2} \right)
        & \left\langle x_{\mathrm{S}} \right\rangle < x_{\mathrm{S,eut}} \\[6pt]
        1495 \left(1 + 1.97 \times 10^{-11} P - 1.43 \times 10^{-22} P^{2} \right)
        & \left\langle x_{\mathrm{S}} \right\rangle > x_{\mathrm{S,eut}},
    \end{cases}
\end{equation}
\begin{equation}
    T_{\mathrm{l},\mathrm{eut}} = 
    1261 \left(1 - 6.91 \times 10^{-12} P \right).
\end{equation}
These expressions are derived by empirically fitting the data from \citet{Boehler1992,fei1995,Boehler1996,fei1997,fei2000}.

The pressure dependence of the solid Fe and FeS densities is calculated with the Murnaghan equation of state,
\begin{equation}
    \rho_{\mathrm{Fe,sol}}=\rho_{0}\left(1+\frac{K_{\mathrm{T,0}}^{\prime \mathrm{Fe,sol}}}{K_{\mathrm{T,0}}^{\mathrm{Fe,sol}}}(P-P_{0})\right)^{1/K_{\mathrm{T,0}}^{\prime \mathrm{Fe,sol}}},
\label{eq:rho_fes_sol}
\end{equation}
\begin{equation}
    \rho_{\mathrm{FeS,sol}}=\rho_{0}\left(1+\frac{K_{\mathrm{T,0}}^{\prime \mathrm{FeS,sol}}}{K_{\mathrm{T,0}}^{\mathrm{FeS,sol}}}(P-P_{0})\right)^{1/K_{\mathrm{T,0}}^{\prime \mathrm{FeS,sol}}}.
\label{eq:rho_c_liq}
\end{equation}
The fixed parameter values are given in Table \ref{tab:parameter}.

Assuming the isothermal bulk modulus in a liquid phase $K_{\mathrm{T,P}}^{\mathrm{liq}}$ linearly depends on the pressure, the density of the liquid phase can be calculated using the following relationship,
\begin{equation}
    \label{eq:rho_linear}
        \rho_{\mathrm{c}}^{\mathrm{liq}} = 
        \rho_{\mathrm{c,0}}^{\mathrm{liq}} \left(1 + \frac{K_{\mathrm{T,0}}^{\prime \mathrm{liq}}}{K_{\mathrm{T,0}}^{\mathrm{liq}}} \left(P - P_{\mathrm{0}} \right) \right)
            ^{\frac{1}{K_{\mathrm{T,0}}^{\prime \mathrm{liq}}}},
\end{equation}
where isothermal bulk modulus $K_{\mathrm{T,0}}^{\mathrm{liq}}$ at the reference pressure $P_{0}$ and $K_{\mathrm{T,0}}^{\prime \mathrm{liq}}$ is its pressure derivative.
$K_{\mathrm{T,0}}^{\mathrm{liq}}$ is given as follows \citep{Ruckriemen2018,Sanloup2000,Hixson1990},%
\begin{equation}
    K_{\mathrm{T,0}}^{\mathrm{liq}} =
    83.4 - 412 x_{\mathrm{S}} + 592 x_{\mathrm{S}}^{2}.
\end{equation}
Density of liquid Fe--FeS at atmospheric pressure $\rho_{\mathrm{c,0}}^{\mathrm{liq}}$ is given by the following relationship assuming that liquid Fe and FeS form an ideal solution,
\begin{equation}
    \rho_{\mathrm{c,0}}^{\mathrm{liq}} = 
    \left(
        \frac{1 - x_{\mathrm{FeS}}^{\mathrm{liq}}}{\rho_{\mathrm{Fe,0}}^{\mathrm{liq}}} + 
        \frac{x_{\mathrm{FeS}}^{\mathrm{liq}}}{\rho_{\mathrm{FeS,0}}^{\mathrm{liq}}} 
    \right)^{-1},
\end{equation}
where $x_{\mathrm{FeS}}^{\mathrm{liq}}$ is the mass fraction for FeS in a liquid phase.
$x_{\mathrm{S}}$ converts to $x_{\mathrm{FeS}}$ through $x_{\mathrm{FeS}} = x_{\mathrm{S}} / 0.36$
.
Densities of liquid Fe $\rho_{\mathrm{Fe,0}}^{\mathrm{liq}}$ and FeS $\rho_{\mathrm{FeS,0}}^{\mathrm{liq}}$ at the reference pressure $P_{\mathrm{0}}$ of 10$^{5}$\,Pa are given as follows \citep{Rivoldini2009,AndersonW1994,Kaiura1979},
\begin{equation}
    \rho_{\mathrm{Fe,0}}^{\mathrm{liq}}
    = 7034.96 - 0.926\left(T_{\mathrm{c}} - 1811 \right),
\end{equation}
\begin{equation}
    \rho_{\mathrm{FeS,0}}^{\mathrm{liq}} 
    = 4598 - 0.58\left(T_{\mathrm{c}} - 273 \right),
\end{equation}

It has been reported that the electrical resistivity of the Fe-FeS system can vary by more than an order of magnitude with increasing sulfur content \citep{Pommier2018b}.
Assuming that the electrical resistivity is linear with respect to sulfur concentration and temperature, it is given by the following equation extrapolated from measurements at 4.5\,GPa \citep{Pommier2018b}:
\begin{equation}
    \rho_{\mathrm{el,c}} =
    \begin{cases}
        \left(0.50 T_{\mathrm{c}} + 2700\right) x_{\mathrm{S}} + \left(0.0125  T_{\mathrm{c}} + 55\right)
        & 0.0 \le x_{\mathrm{S}} \le 0.05 \\[6pt]
        \left(0.83 T_{\mathrm{c}} + 1067\right) x_{\mathrm{S}} + \left(0.167 T_{\mathrm{c}} + 350\right)
        & 0.05 < x_{\mathrm{S}} < 0.20 \\[6pt]
        \left(T_{\mathrm{c}} + 2813\right) x_{\mathrm{S}} + 800
        & 0.20 \le x_{\mathrm{S}} \le 0.36 \\[6pt]
    \end{cases}
\label{eq:rho_elc}
\end{equation}
The thermal conductivity $k_{cond,c}$ can be calculated by the following Wiedemann-Franz law,
\begin{equation}
    k_{cond,c} = \frac{L_{0} T_{c}}{\rho_{el,c}},
\label{eq:kcond}
\end{equation}
where $L_0$ the Sommerfeld value of the Lorenz number.
Based on this equation, the thermal conductivity of the liquid core typically ranges from approximately 1.5 to 2.2 W m$^{-1}$K$^{-1}$ for sulfur contents between 36 and 20 wt.\% at 1500~K.
These remarkably low values directly reflect the high electrical resistivity of the sulfur-rich liquid phase demonstrated by \citet{Pommier2018b}

The position of solid precipitation during the cooling and solidification of a fully molten metallic core is determined by the relationship between the pressure gradients of the adiabatic temperature, $dT_{ad,c}/dP$, and the liquidus temperature, $dT_{l,c}/dP$. 
In the Earth's core, the liquidus temperature is more sensitive to pressure ($dT_{ad,c}/dP < dT_{l,c}/dP$), causing the core temperature to reach the liquidus first at the center and leading to the growth of a solid inner core. 
However, in smaller bodies like Ganymede where core pressures are low (approximately 5 to 10\,GPa), the solidification process differs. 
Experimental studies indicate that the pressure dependence of the liquidus temperature in the Fe-FeS system is significantly smaller or even negative under these pressure conditions \citep{fei1995,fei1997,Boehler1992,Boehler1996}. 
Consequently, the adiabatic gradient exceeds the liquidus gradient ($dT_{ad}/dP > dT_{l}/dP$), resulting in solid precipitation initiating at the core-mantle boundary rather than at the center \citep{hauck2006,breuer2015}.
In this iron snow regime, solid Fe particles forming at the top of the core sink toward the center and subsequently remelt in the hotter, deeper regions. 
This configuration consists of an iron snow zone where solid Fe falls through the liquid Fe-FeS while maintaining local thermodynamic equilibrium, overlying a fully molten interior. 
Regarding the potential for magnetic field generation, \citet{Bland2008} suggest that a snowing core cannot sustain an active dynamo. 
Furthermore, experimental work on particle settling indicates that individual particles tend to settle independently without initiating large-scale convection \citep{Blanchette2005}. 
Numerical modeling by Zhan and Schubert (2012) also demonstrated that magnetic field generation within the snow zone tends to produce multipolar solutions, which is in contrast to the observed dipolarity of Ganymede's field \citep{Kivelson2002}. 
Instead, the primary driver of the dipole dynamo is considered to be the compositional convection occurring in the liquid region below the snow zone. 
\citet{Christensen2015} successfully modeled the magnetic field associated with such sub-snow zone convection, finding that a dipolar field is not only feasible but also produces an estimated strength comparable with the measured field of Ganymede. 
As cooling proceeds, the depth of remelting increases until the entire core is occupied by the iron snow zone. 
At this stage, global stable stratification occurs and the dynamo action ceases. Therefore, the simulations in this study are terminated when the snow zone reaches the center of the core.

It should be noted that our model assumes a fully adiabatic profile throughout the liquid core. If a thermally stratified, conductive layer developed at the top of the core \citep[e.g.,][]{Greenwood2021}, it would act as a thermal blanket, reducing the efficiency of heat extraction. 
Under such conditions, the deep core would cool more slowly, potentially delaying the overall cooling history and the onset of core solidification.

In the Fe-snow region, which is in a two-phase coexistence state of solid and liquid phases, the temperature always coincides with the liquidus temperature.
Therefore, the sulfur mass fraction in the liquid Fe-FeS, $x_{liq}^S$, is determined by
\begin{equation}
    T_{\mathrm{melt,c}}\left(x_{\mathrm{S}}^{\mathrm{liq}} \right)=T_{\mathrm{c}}
\end{equation}
Because this region becomes compositionally and thermally stably stratified, large-scale convection is effectively suppressed. 
The temperature in the Fe-snow region is thus given by the following heat conduction equation:
\begin{equation}
    \label{eq:conductive_temperature}
\rho_{\mathrm{c}} C_{\mathrm{p,c}} \frac{\partial T_{\mathrm{c}}}{\partial t} =
k_{cond,c}\frac{1}{r^{2}}\frac{d}{dr}\left(r^{2}\frac{dT_{\mathrm{c}}}{dr}\right)
+ q_{\mathrm{grav}}^{+} + q_{\mathrm{lat}}^{+}
\end{equation}
where $q_{grav}^+$ is the heat generation per unit volume due to the release of gravitational potential in the Fe-snow region, and $q_{lat}^+$ is the latent heat released per unit volume associated with the precipitation of solid Fe.
The heat generation due to the release of gravitational potential for the entire core, $Q_{grav}$, is distributed uniformly throughout the core, and latent heat is generated by the precipitation of solid Fe, as given by the following equations, respectively \citep{Ruckriemen2018}:
\begin{equation}
    q_{\mathrm{grav}}^{+} =
    \frac{Q_{\mathrm{grav}}^{+}}{V_{\mathrm{c}}^{+}} = 
    \frac{Q_{\mathrm{grav}}}{V_{\mathrm{c}}}
\end{equation}
\begin{equation}
    q_{\mathrm{lat}}^{+} =
    \frac{1}{V_{\mathrm{c}}^{+}}
    \frac{\partial m_{\mathrm{sol}}}{\partial t} \Delta S_{\mathrm{Fe}} T_{\mathrm{c}}
\end{equation}
where $Q_{\mathrm{grav}}$ is the heat generation due to the release of gravitational potential for the entire core (Eq.\ref{eq:qgrav}), $V_{\mathrm{c}}^{+}$ is the volume of the Fe-snow region, $m_{\mathrm{sol}}$ is the total bulk mass of solid Fe produced across the entire Fe-snow region, and $\Delta S_{\mathrm{Fe}}$ is the change in specific entropy of fusion for the phase change (Table \ref{tab:parameter}).
By dividing this bulk term by $V_{\mathrm{c}}^{+}$, we assume a spatially uniform mass precipitation rate per unit volume. 
However, because $T_{\mathrm{c}}$ is a radial profile, the resulting latent heat term $q_{\mathrm{lat}}^{+}$ acts as a local profile evaluated at each grid point.
Assuming thermodynamic equilibrium, the ratio of the solid phase mass to the combined mass of the liquid and solid phases, $f_{sol}$, is given by the following equation:
\begin{equation}
    f_{\mathrm{sol}} = 
    \frac{x_{\mathrm{FeS}}^{\mathrm{liq+sol}} - x_{\mathrm{FeS}}^{\mathrm{liq}}}{x_{\mathrm{FeS}}^{\mathrm{sol}} - x_{\mathrm{FeS}}^{\mathrm{liq}}}
    = 1 - \frac{x_{\mathrm{FeS}}^{\mathrm{liq+sol}}}{x_{\mathrm{FeS}}^{\mathrm{liq}}}
\end{equation}
where $x_{FeS}^{liq+sol}$ is the mass fraction of FeS in the combined liquid and solid phases, $x_{FeS}^{sol}$ is the mass fraction of FeS in the solid phase, and $x_{FeS}^{liq}$ is the mass fraction of FeS in the liquid phase.
The mass fraction of FeS and the mass fraction of S are converted as $x_{FeS} = x_S / 0.36$.
Since the solid phase is pure Fe, $x_{FeS}^{sol} = 0$.
It is assumed that the precipitation, sinking, and remelting of solid Fe occur on a timescale shorter than the discrete step $\Delta t = 100$ yr in the numerical calculations of this study.
While the exact size of iron snow particles is unconstrained, a simple estimate using Stokes' law \citep[e.g.,][]{Ruckriemen2015} shows that the maximum settling time ranges from $\sim$16,000 years for 10~$\mu$m particles to $\sim$1.6 years for 1 mm particles. Given this broad physically plausible range and assuming a continuous macroscopic precipitation flux, a discrete time step of $\Delta t = 100$ yr provides a sufficiently resolved approximation for the long-term energy balance.
Consequently, $f_{sol}$ can be regarded as the mass fraction of solid Fe precipitated from the liquid phase during $\Delta t$.
Therefore, the Fe precipitation rate $\partial m_{sol}/\partial t$ is given by the following equation \citep{Ruckriemen2018}:
\begin{equation}
    \frac{\partial m_{\mathrm{sol}}}{\partial t}
    = \frac{f_{\mathrm{sol}}}{\Delta t} m_{\mathrm{c}}^{+}
\end{equation}
where $m_c^+$ is the mass of the snow region.
The fully molten region below the Fe-snow region is assumed to be well-mixed, and its temperature is given by the adiabatic temperature gradient.
\begin{equation}
    \label{deeper_core_adiabatic_temperature}
        T_{\mathrm{c}} = 
        T_{\mathrm{ub}} 
        \exp\!
        \left\{
            \frac{\alpha_{\mathrm{c}}}{\rho_{\mathrm{c}} C_{\mathrm{p,c}}}
            \left( P - P_{\mathrm{ub}} \right)
        \right\}
\end{equation}
$T_{ub}$ and $P_{ub}$ are the temperature and pressure at the boundary between the fully molten region and the Fe-snow region, respectively.
$T_{ub}$ is given as the temperature that satisfies the energy balance of the fully molten region \citep{Ruckriemen2015}.
\begin{equation}
    \label{deeper_core_energy_conservation}
    \rho_{\mathrm{c}} C_{\mathrm{p,c}}
    \int_{V_{\mathrm{c}}^{-}} \frac{\partial T_{\mathrm{ub}}}{\partial t} \, dV
    = Q_{\mathrm{grav}}^{-} + Q_{\mathrm{lat}}^{-} - 4 \pi R_{\mathrm{c}}^{2} F_{\mathrm{ub}}
\end{equation}
where $V_{\mathrm{c}}^-$ is the volume of the fully molten region, and $F_{\mathrm{ub}}$ is the heat flux through the upper boundary of the fully molten region.
$Q_{\mathrm{lat}}^-$ is the latent heat absorbed when Fe remelts in the upper part of the fully molten region, and $Q_{\mathrm{grav}}^-$ is the heat generation due to the release of gravitational potential in the fully molten region, given by the following equation:
\begin{equation}
    Q_{\mathrm{grav}}^{-} = 
    Q_{\mathrm{grav}} \frac{V_{\mathrm{c}}^{-}}{V_{\mathrm{c}}}
\end{equation}
\begin{equation}
    Q_{\mathrm{lat}}^{-} = 
    - \frac{\partial m_{\mathrm{sol}}}{\partial t} \Delta S_{\mathrm{Fe}} T_{\mathrm{ub}}
\end{equation}
where $\partial m_{melt}/\partial t = -\partial m_{sol}/\partial t$ holds because the timescale from Fe precipitation to remelting in the Fe-snow region is assumed to be sufficiently short relative to the assumed time step $\Delta t$.
The sulfur mass fraction in the fully molten region, $x_S^-$, is given by mass conservation:
\begin{equation}
    \label{deeper_core_sulfer_concentration}
    x_{\mathrm{S}}^{-} = 
    \frac{1}{M_{\mathrm{c}}^{-}} 
    \left(\left\langle x_{\mathrm{S}} \right\rangle M_{\mathrm{c}} - 
            \int_{V_{\mathrm{c}}^{+}} x_{\mathrm{S}}^{+} \rho_{\mathrm{Fe,sol}}dV\right)
\end{equation}
where $M_c$ is the total mass of the core and $M_c^-$ is the mass of the fully molten region .

In the FeS-layer regime, when solid FeS precipitates at the top of the core, it does not sink because its density is lower than that of the surrounding liquid, forming a solid FeS outer core at the top of the core.
Calculations using the Eq.(\ref{eq:rho_fes_sol}) and (\ref{eq:rho_c_liq}) confirm that the bulk density of the liquid Fe-FeS ($\rho_{\mathrm{c}}^{\mathrm{liq}}$) remains significantly higher ($6500\,-\,6800~\mathrm{kg\,m^{-3}}$) than that of the crystallized solid FeS ($\rho_{\mathrm{FeS,sol}}=5500\,-\,5700~\mathrm{kg\,m^{-3}}$) under Ganymede's core conditions, robustly supporting this buoyancy-driven formation.
This solidification process is known as the FeS-layer regime and consists of a solid FeS outer core growing from the core-mantle boundary toward the deep interior and a fully molten region beneath it .
In the solid FeS outer core, heat is transported only by conduction, and the temperature is given by Equation (\ref{eq:conductive_temperature}).
Unlike the Fe-snow regime where Fe precipitates throughout the Fe-snow region, FeS precipitates at the boundary between the solid FeS outer core and the deep fully molten region, thus $q_{lat}^+$ is given as follows \citep{Ruckriemen2018}:
\begin{equation}
    q_{\mathrm{lat}}^{+} =
    \begin{cases}
        \frac{1}{\Delta V_{\mathrm{FeS,sol}}} \frac{\partial m_{\mathrm{sol}}}{\partial t} \Delta S_{\mathrm{FeS}} T_{\mathrm{ub}} & \text{at the FeS solidification front,} \\
        0 & \text{otherwise.}
    \end{cases}
\end{equation}
where $\Delta V_{\mathrm{FeS,sol}}$ is the volume of the solid FeS outer core increased by solidification, and $\Delta S_{\mathrm{FeS}}$ is the change in specific entropy of fusion for the phase change from solid to liquid FeS (Table \ref{tab:parameter}).
The temperature of the fully molten region is given by Equations (\ref{deeper_core_adiabatic_temperature}) and (\ref{deeper_core_energy_conservation}).
However, in the FeS-layer regime, $Q_{\mathrm{lat}}^-$ is zero because no solidification or melting occurs within the fully molten region.
The sulfur mass fraction of the fully molten region is given by mass conservation in Equation (\ref{deeper_core_sulfer_concentration}).
%
\subsection{Magnetic field generation in the metallic core}

To estimate the total power available for the magnetic dynamo, we employ the thermodynamic energy balance approach based on \citet{Buffett1996} and previously applied to Ganymede by \citet{Ruckriemen2018}.
The total available power $\Phi_{\mathrm{all}}$ is given by:

\begin{equation}
    \Phi_{\mathrm{all}} = \Phi_{\mathrm{s}} - \Phi_{\mathrm{ad}} + \Phi_{\mathrm{lat}} + \Phi_{\mathrm{C}}
\end{equation}
where $\Phi_{\mathrm{s}}$ is the power available for the dynamo derived from secular cooling at the top of the fully molten region, $\Phi_{\mathrm{ad}}$ is the power required to sustain convection while maintaining an adiabatic state, $\Phi_{\mathrm{lat}}$ is the power derived from latent heat at the top of the fully molten region, and $\Phi_{\mathrm{C}}$ is the power available due to compositional gradients.
Note that while basic thermodynamic constants (e.g., specific heat and thermal expansivity) are fixed, the transport and structural properties governing these dynamo calculations, such as core density, electrical resistivity, and thermal conductivity, dynamically evolve over time as functions of temperature, pressure, and sulfur concentration.
In an entirely molten core, the viability and intensity of the dynamo are primarily governed by core cooling via heat transport in the silicate mantle.
When the heat flux at the core-mantle boundary (CMB) exceeds the adiabatic heat flux, thermal convection is initiated, facilitating global mixing. 
The source of the dynamo power is the gravitational potential energy released as convective mixing resolves gravitational instabilities. 
Specifically, the dynamo power is defined by the difference in gravitational potential between the unstable and fully mixed states of the core. 
Consequently, the power available from secular cooling and that required to sustain convection are given by \citet{Buffett1996} and \citet{Ruckriemen2015},
\begin{equation}
    \label{eq:thermal_dynamo_power_fully_molten}
    \Phi_{\mathrm{s}} =
    \frac{\alpha_{\mathrm{c}}}{C_{\mathrm{p,c}}} 4 \pi R_{\mathrm{c}}^{2} F_{\mathrm{cmb}} \Delta \psi
\end{equation}
\begin{equation}
    \label{eq:thermal_dynamo_power_fully_molten}
    \Phi_{\mathrm{ad}} =
    \frac{\alpha_{\mathrm{c}}}{C_{\mathrm{p,c}}} 4 \pi R_{\mathrm{c}}^{2} F_{\mathrm{ad}} \Delta \psi
\end{equation}
Here, $F_{\mathrm{ad}} = \frac{k_{\mathrm{cond,c}}\alpha_{\mathrm{c}}T_{\mathrm{c}} g}{C_{\mathrm{p,c}}}$ is the adiabatic heat flux, and $\Delta \Psi$ is the difference in gravitational potential between the surface and the mean interior of the fully molten region. 
In a fully molten core, $\Phi_{\mathrm{lat}}$ and $\Phi_{\mathrm{C}}$ are zero because there is no latent heat release or compositional inhomogeneity.

As core cooling progresses and solidification begins, the processes driving the dynamo also change.
When the sulfur content in the core is below the eutectic composition (e.g., $\sim$25 wt\%) and solidification occurs in the iron (Fe) snow regime, the fully molten region is cooled by thermal conduction through the overlying Fe-snow zone. 
Furthermore, the endothermic effect associated with the remelting of Fe-snow contributes to the cooling of the upper part of the molten region. 
Consequently, the dynamo power in the Fe-snow regime is given by the following equations \citep{Ruckriemen2018}:
\begin{equation}
    \label{eq:dynamo_power_secular_fesnow}
    \Phi_{\mathrm{s}} =
    \frac{\alpha_{\mathrm{c}}}{C_{\mathrm{p,c}}} 4 \pi R_{\mathrm{ub}}^{2} F_{\mathrm{ub}} \Delta \psi
\end{equation}
\begin{equation}
    \label{eq:dynamo_power_adiabatic_fesnow}
    \Phi_{\mathrm{ad}} =
    \frac{\alpha_{\mathrm{c}}}{C_{\mathrm{p,c}}} 4 \pi R_{\mathrm{ub}}^{2} F_{\mathrm{ad}} \Delta \psi
\end{equation}
\begin{equation}
    \label{eq:dynamo_power_latent_fesnow}
    \Phi_{\mathrm{lat}} =
    \frac{\alpha_{\mathrm{c}}}{C_{\mathrm{p,c}}} Q_{\mathrm{lat}} \Delta \psi
\end{equation}
where $R_{\mathrm{ub}}$ is the radius of the fully molten region. 
Note that $Q_{\mathrm{lat}}$ represents the absolute value of the heat absorbed during remelting. 
The remelting of Fe-snow enriches the upper part of the fully molten region with Fe, leading to an increase in density. 
The dynamo power derived from the compositional convection that arises to resolve this gravitational instability is given by the following equation:
\begin{equation}
    \label{eq:dynamo_power_chemical_fesnow}
    \Phi_{\mathrm{C}} =
    \int_{V_{\mathrm{c}}^{-}}
        \frac{\partial \left\langle \rho_{\mathrm{c}^{-}} \right\rangle}
        {\partial t} \Delta \psi dV
\end{equation}
When the sulfur content is relatively high and solidification occurs in the FeS-layer regime, a solid FeS-layer forms at the top of the core. 
This regime differs from the Fe-snow regime in that the layer is completely solidified and remains at the top rather than sinking into the deep core. 
The crystallization and upward accretion of solid FeS occurs at the boundary between the fully molten region and the FeS-layer that has formed above it, accompanied by the release of latent heat.
Physically, the upward conductive heat transport into the solid FeS layer is balanced by the sum of this released latent heat and the heat flux extracted from the underlying liquid core ($F_{\mathrm{ub}}$). Because the latent heat acts as a local heat source at the boundary, it reduces the demand on $F_{\mathrm{ub}}$ to supply the upward conduction.
This mitigates the cooling of the upper molten region and acts to suppress the initiation of thermal convection. 
In the calculations of this study, this latent heat effect is not explicitly represented as a $Q_{\mathrm{lat}}$ term in the dynamo power equation; instead, it is incorporated as a reduction in the heat flux $F_{\mathrm{ub}}$ from the surface of the fully molten region. 
Consequently, $\Phi_{\mathrm{lat}}$ is zero. 
Furthermore, during the crystallization of FeS, the Fe component that is not incorporated into the solid phase is expelled into the liquid phase. 
Similar to the remelting of Fe in the Fe-snow regime, this process enriches the upper molten region with Fe, potentially creating a gravitationally unstable state. 
Therefore, the dynamo power derived from compositional convection is given by equation (\ref{eq:dynamo_power_chemical_fesnow}), as in the Fe-snow regime. 
$\Phi_{\mathrm{s}}$ and $\Phi_{\mathrm{ad}}$ are similarly given by equations (\ref{eq:dynamo_power_secular_fesnow}) and (\ref{eq:dynamo_power_adiabatic_fesnow}).
The intensity of the generated magnetic field is estimated using the resulting dynamo power. 
Following scaling laws derived from previous dynamo simulations based on magnetohydrodynamic calculations, the root-mean-square magnetic field strength within the dynamo region, $B_{\mathrm{rms}}$, can be described as follows \citep{Christensen2006,Aubert2009}:
\begin{equation}
    B_{\mathrm{rms}} =
    c_{\mathrm{ub}} f_{\mathrm{ohm}}^{\frac{1}{2}} 
    \left(\frac{\Phi}{V_{\mathrm{c}}^{-} \left\langle \rho_{\mathrm{c}}^{-} \right\rangle \Omega R_{\mathrm{ub}}^{2}} \right)^{\frac{1}{3}}
    \left(\left\langle \rho_{\mathrm{c}}^{-} \right\rangle \mu_{0} \right)^{\frac{1}{2}}
    \Omega R_{\mathrm{ub}}
\label{eq:b_rms}
\end{equation}
where $\mu_0$ is the vacuum magnetic permeability, $\Omega$ is the angular velocity of Ganymede's rotation, $c_{\mathrm{ub}}$ is a coefficient determined by the scaling law fitting, $f_{\mathrm{ohm}}$ is the fraction of energy dissipated by Ohmic heating, $\langle \bar{\rho_c} \rangle$ is the mean density of the fully molten region, and $R_{\mathrm{ub}}$ is the radius of the fully molten region. 
Furthermore, the dipole magnetic field strength at the surface of Ganymede, $B_{\mathrm{dip}}^{\mathrm{surf}}$, is given by
\begin{equation}
    B_{\mathrm{dip}}^{\mathrm{surf}} =
    \frac{B_{\mathrm{rms}}}{b_{\mathrm{dip}}}
    \left(\frac{R_{\mathrm{ub}}}{R_{\mathrm{p}}} \right)^{3},
\end{equation}
where $b_{\mathrm{dip}}$ is the ratio of the total magnetic field strength to its dipole component and $R_{\mathrm{p}}$ is the radius of Ganymede.
Since dynamo simulations indicate that in models with a dipole-dominated field, $f_{\mathrm{ohm}}$ is greater than 0.6 and $b_{\mathrm{dip}}$ ranges from 7 to 17, this study adopts $f_{\mathrm{ohm}} = 1$ and $b_{\mathrm{dip}} = 10$ \citep{Christensen2015}.
Note that while the un-normalized total dynamo power ($\Phi$) is strictly used in Eq.(\ref{eq:b_rms}) to calculate the magnetic field strength, we evaluate the intrinsic physical contribution of each driving process using the area power density $\phi$ (in mW\,m$^{-2}$), which is the dynamo power normalized by the surface area of the dynamo region. 
This approach is intended to remove the purely geometric influence of the shrinking size of the dynamo region, allowing us to accurately track the true efficiencies of the thermal and compositional drivers over time. 
Geometric scale differences in the dynamo region are then explicitly reflected only in the resulting magnetic field strength \citep{Christensen2006,Christensen2009}.
\newpage
\begin{table}
    \centering
    \caption{Parameters using for calculation}
    \label{tab:parameter}
    \begin{center}
        \begin{tabular}{llll} \toprule
        Symbol & Description & Value & Reference\\ \midrule

        $R_{\mathrm{m}}$        & Mantle radius            & 1600\,--\,1850 km & \\
        $\left\langle \rho_{\mathrm{m}} \right\rangle$ & Mantle average density & 3000\,--\,4000 kg\,m$^{-3}$ & \\
        $C_{\mathrm{p,m}}$      & Mantle specific heat & 900 J\,kg$^{-1}$\,K$^{-1}$ & \citet{Kirk1987} \\
        $\alpha_{\mathrm{m}}$   & Mantle thermal expansivity & 2.4 $\times$ $10^{-5}$ K$^{-1}$ & \citet{Kirk1987} \\
        $k_{\mathrm{cond,m}}$   & Mantle thermal conductivity & 3.0 W\,m$^{-1}$\,K$^{-1}$& \citet{Kirk1987} \\

        $E_{\mathrm{m}}$ & Activation energy of the rocky mantle & 240 $\times$10$^{3}$ J\,mol$^{-1}$ & \citet{Karato1993} \\ 
        $\eta_{\mathrm{ref}}$ & Reference mantle viscosity & 1.0 $\times$ $10^{19}$ Pa\,s & \citet{Ruckriemen2018} \\

        $R_{\mathrm{c}}$        & Core radius              & 530\,--\,840 km & \\
        $\left\langle \rho_{\mathrm{c}} \right\rangle$ & Core average density & 4900\,--\,7100 kg\,m$^{-3}$ & \\
        $C_{p,c}$               & Core specific heat & 800 J\,mol$^{-1}$\,K$^{-1}$ & \citet{Buffett1996}\\
        $\alpha_{\mathrm{c}}$   & Core thermal expansivity & 8.2 $\times$ $10^{-5}$ K$^{-1}$ & \citet{Nasch1995}\\
        $\left\langle x_{\mathrm{S}} \right\rangle$ & Core average sulfur content & 0.094\,--\,0.36 & \\

        $K_{\mathrm{T,0}}^{\mathrm{Fe,sol}}$        & Isothermal bulk modulus of solid Fe                               & 103 GPa & \citet{Ahrens2002} \\
        $K_{\mathrm{T,0}}^{\prime \mathrm{Fe,sol}}$ & First pressure derivative of $K_{\mathrm{T,0}}^{\mathrm{Fe,sol}}$ & 5.0                   & \citet{Ahrens2002} \\
        $K_{\mathrm{T,0}}^{\mathrm{FeS,sol}}$       & Isothermal bulk modulus of solid FeS                              & 54.3 GPa & \citet{Urakawa2004} \\
        $K_{\mathrm{T,0}}^{\prime \mathrm{FeS,sol}}$& First pressure derivative of $K_{\mathrm{T,0}}^{\mathrm{FeS,sol}}$& 4.0                   & \citet{Urakawa2004} \\
        $K_{\mathrm{T,0}}^{\prime \mathrm{liq}}$    & First pressure derivative of $K_{\mathrm{T,0}}^{\mathrm{liq}}$    & 4.66                   & \citet{AndersonW1994} \\

        $\Delta S_{Fe}$         & Specific entropy of fusion for Fe & 136 J\,kg$^{-1}$\,K$^{-1}$ & \citet{Barin1993} \\
        $\Delta S_{FeS}$        & Specific entropy of fusion for FeS & 247 J\,kg$^{-1}$\,K$^{-1}$ & \citet{Barin1993} \\
        $L_{0}$                 & Sommerfeld value         & 2.445 $\times$ 10$^{-8}$ W\,$\Omega$\,K$^{-2}$ &  \\

        $\Omega$           & Ganymede's rotation rate & 1.02 $\times$ $10^{-5}$  & \\
        $c_{\mathrm{ub}}$   & Scaling prefactor                   & 1.17 & \citet{Aubert2009} \\
        $f_{\mathrm{ohm}}$ & Fraction of ohmic dissipation       & 1    & \citet{Christensen2015} \\
        $b_{\mathrm{dip}}$ & Attenuation factor for dipole field & 10   & \citet{Christensen2015} \\
        \bottomrule
        \end{tabular}
    \end{center}
\end{table}
\vspace{2\baselineskip}
\begin{table}
    \caption{The values of long-lived radioactive isotopes for various chondritic compositions}
    \label{tab:radiogenic_isotope}
    \begin{center}
        \begin{tabular}{lllllllllll} \toprule
            & $H_{i}$$^{a}$ & $\lambda_{i}$$^{a}$ & $c_{i}$(H)$^{b}$ & $c_{i}$(L)$^{b}$ & $c_{i}$(LL)$^{b}$ & $c_{i}$(CI)$^{b}$ & $c_{i}$(CM)$^{b}$ & $c_{i}$(CO)$^{b}$ & $c_{i}$(CV)$^{b}$ & $c_{i}$(CK)$^{b}$ \\ \midrule
            ${}^{238}$U & 9.465 & 4468 & 26.21 & 30.24 & 30.24 & 16.13 & 24.19 & 36.29 & 34.28 & 30.24 \\
            ${}^{235}$U & 56.87 & 703.81 & 8.41 & 9.70 & 9.70 & 5.17 & 7.76 & 11.64 & 10.99 & 9.70 \\
            ${}^{232}$Th & 2.638 & 14030 & 47.62 & 52.63 & 58.90 & 36.34 & 51.38 & 100.25 & 72.68 & 72.68 \\
            ${}^{40}$K & 2.917 & 1277 & 1086.0 & 1283.99 & 1228.17 & 767.60 & 516.39 & 502.43 & 502.43 & 404.74 \\ \bottomrule
        \end{tabular}
    \end{center}
    \vspace{.05cm}
    $^{a}$\citet{Robuchon2011}\\
    $^{b}$\citet{Lodders1998}
\end{table}
\clearpage
\begin{figure}
	\centering
	\includegraphics[width=1.0\textwidth]{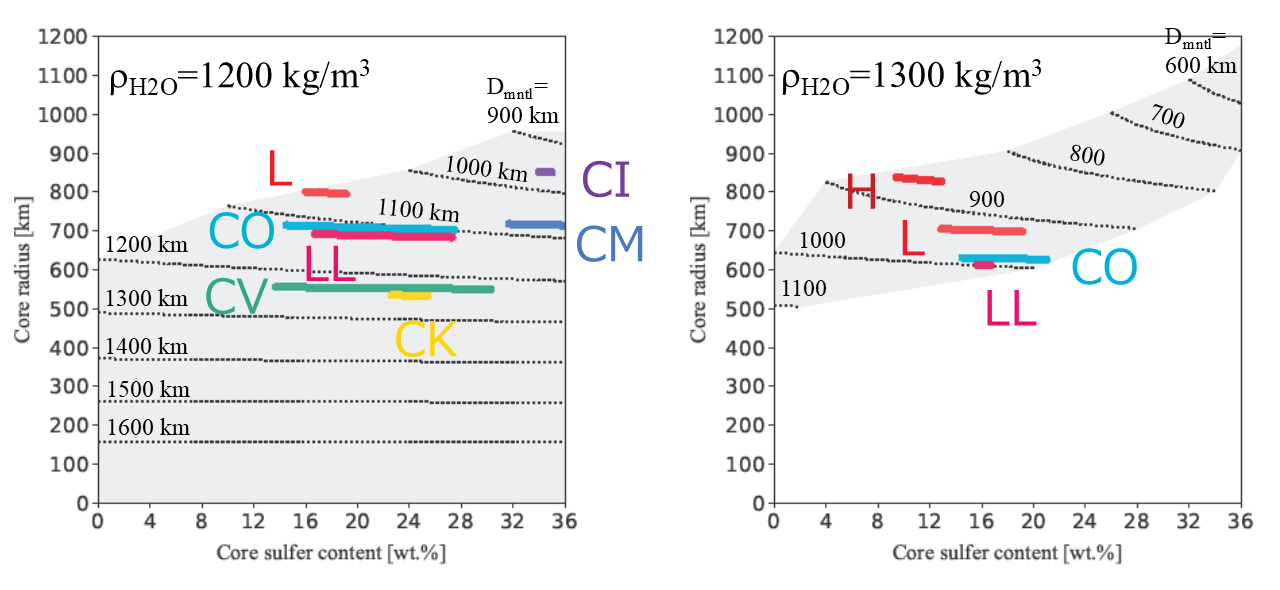}
	\caption{Structural ranges of Ganymede's interior consistent with the moment of inertia factor of 0.3156 (shaded region), assuming various chondritic compositions (H, L, LL, CI, CM, CO, CV, and CK) for the bulk rock-metal fraction. 
Dotted lines indicate contours of rocky mantle thickness (D$_{\mathrm{mntl}}$). Colored lines denote the specific viable ranges for each assumed chondrite composition.
Chondritic models whose structural parameters fall outside the permissible MoI range (shaded region) under a given hydrosphere density are incompatible and thus omitted from the respective panel.
}
	\label{fig:1}
\end{figure}
\begin{figure}
	\centering
	\includegraphics[width=0.8\textwidth]{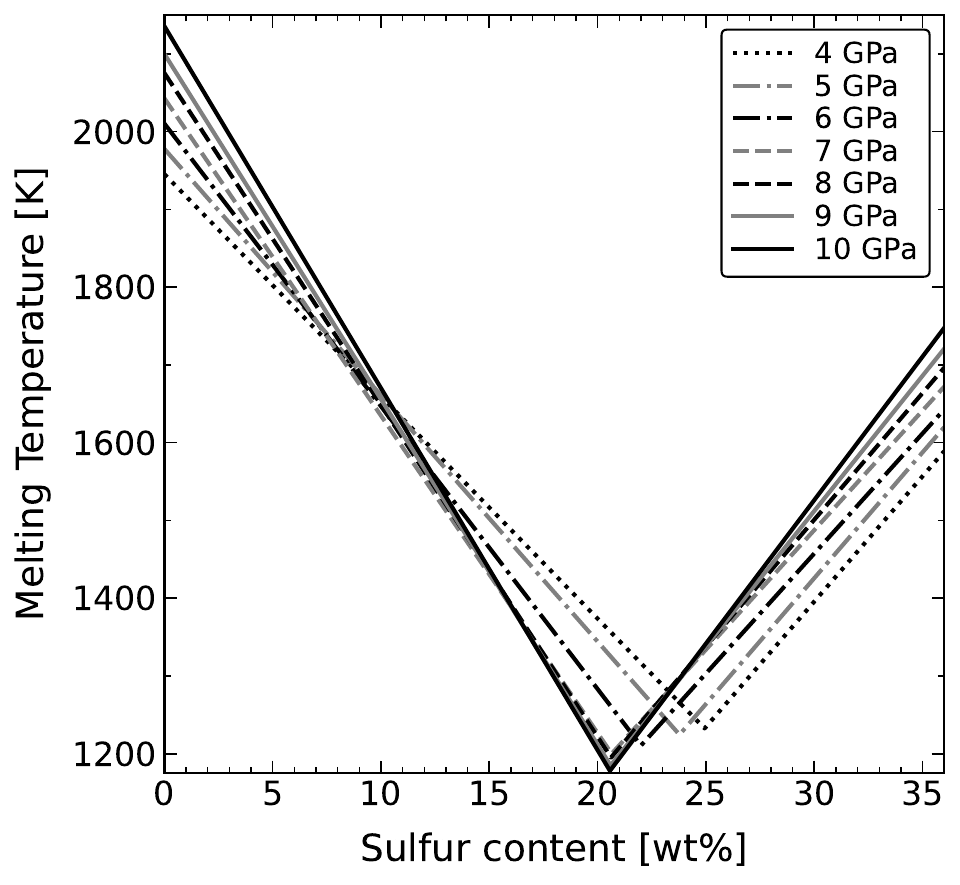}
	\caption{
Model melting temperatures as a function of bulk core sulfur content in the Fe-FeS for pressures from 4 to 10~GPa. 
Melting temperatures of pure Fe, pure FeS, and eutectic compositions match available data from the literature \citep{Boehler1992,fei1995,Boehler1996,fei1997,fei2000}.
Temperatures are the result of linear interpolation between pure end-members and eutectic values at intermediate compositions.}
	\label{fig:fefes}
\end{figure}
\section{Results}

\begin{figure}[htbp]
  \centering
  \begin{minipage}{0.48\textwidth}
    \centering
    \includegraphics[width=\textwidth]{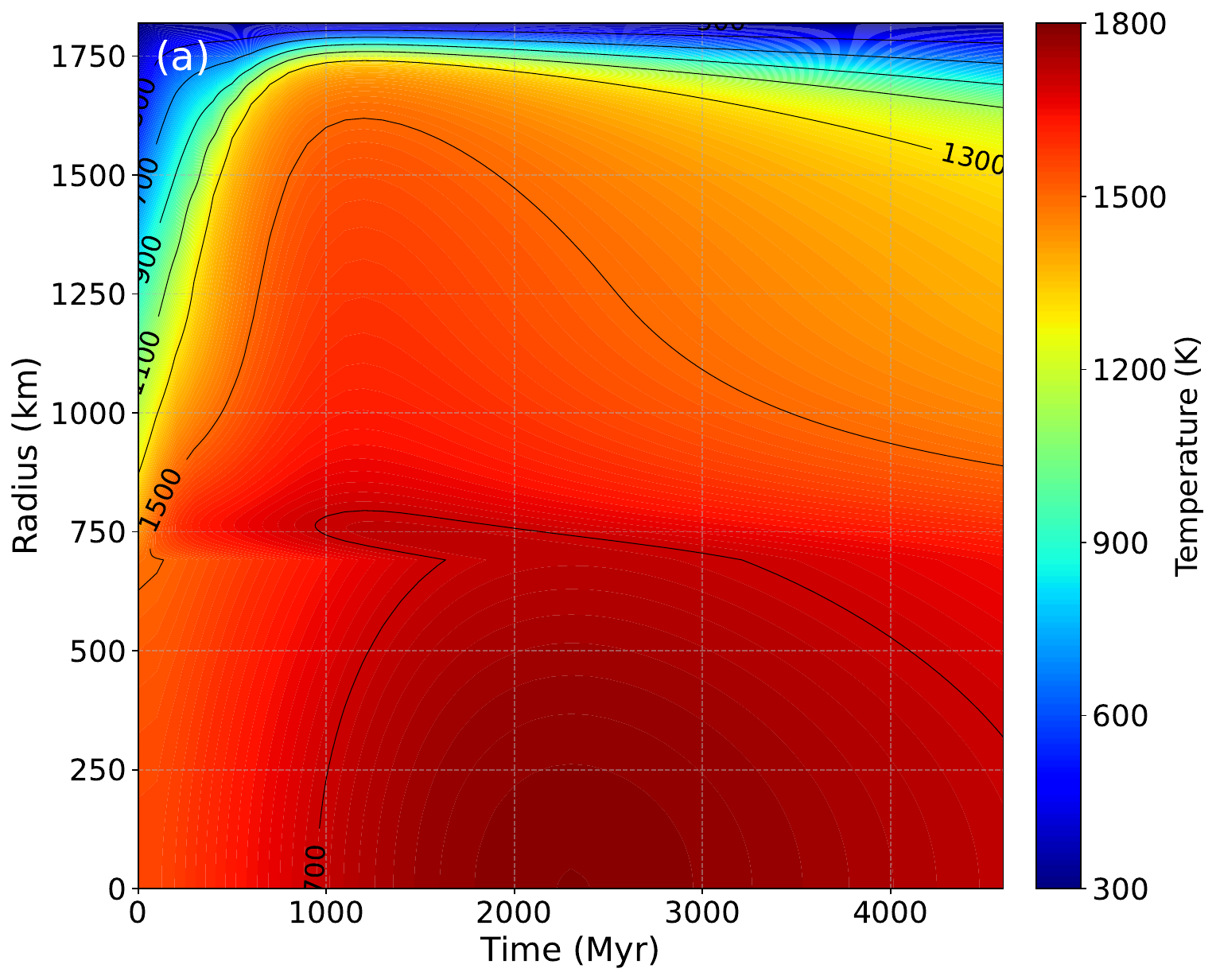}
  \end{minipage}
  \hfill
  \begin{minipage}{0.48\textwidth}
    \centering
    \includegraphics[width=\textwidth]{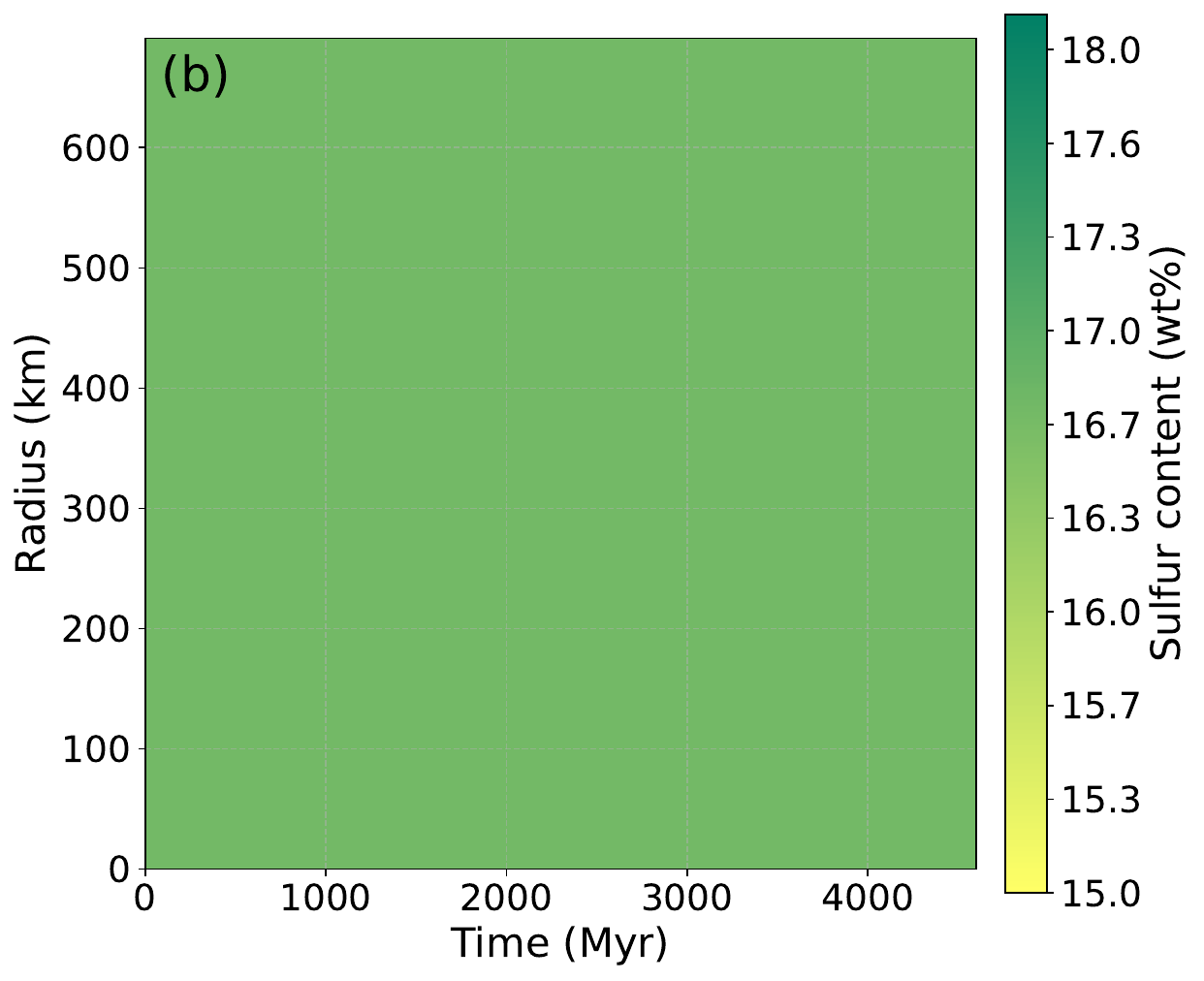}
  \end{minipage}

  \vspace{2ex} %

  \begin{minipage}{0.48\textwidth}
    \centering
    \includegraphics[width=\textwidth]{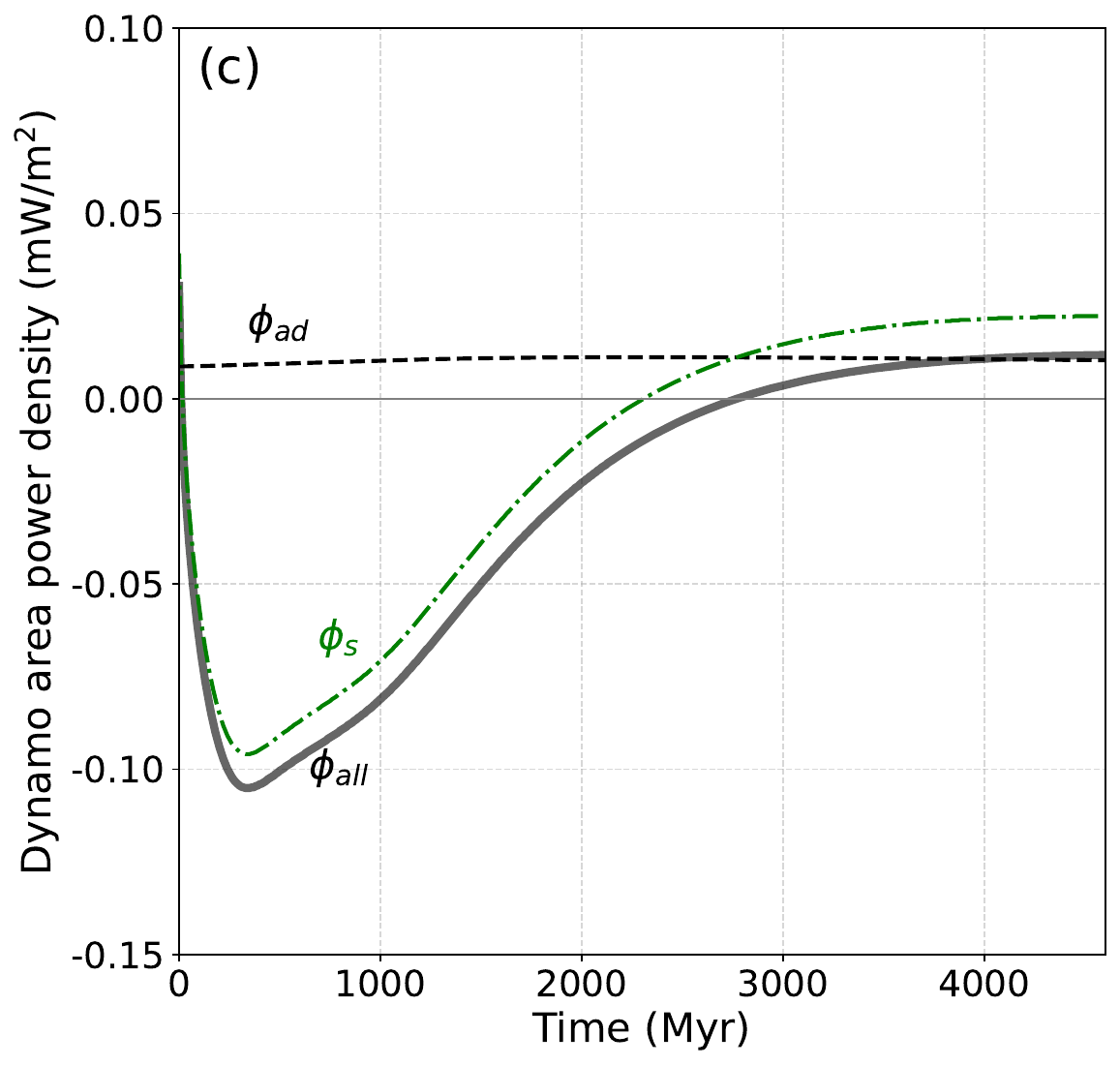}
  \end{minipage}
  \hfill
  \begin{minipage}{0.48\textwidth}
    \centering
    \includegraphics[width=\textwidth]{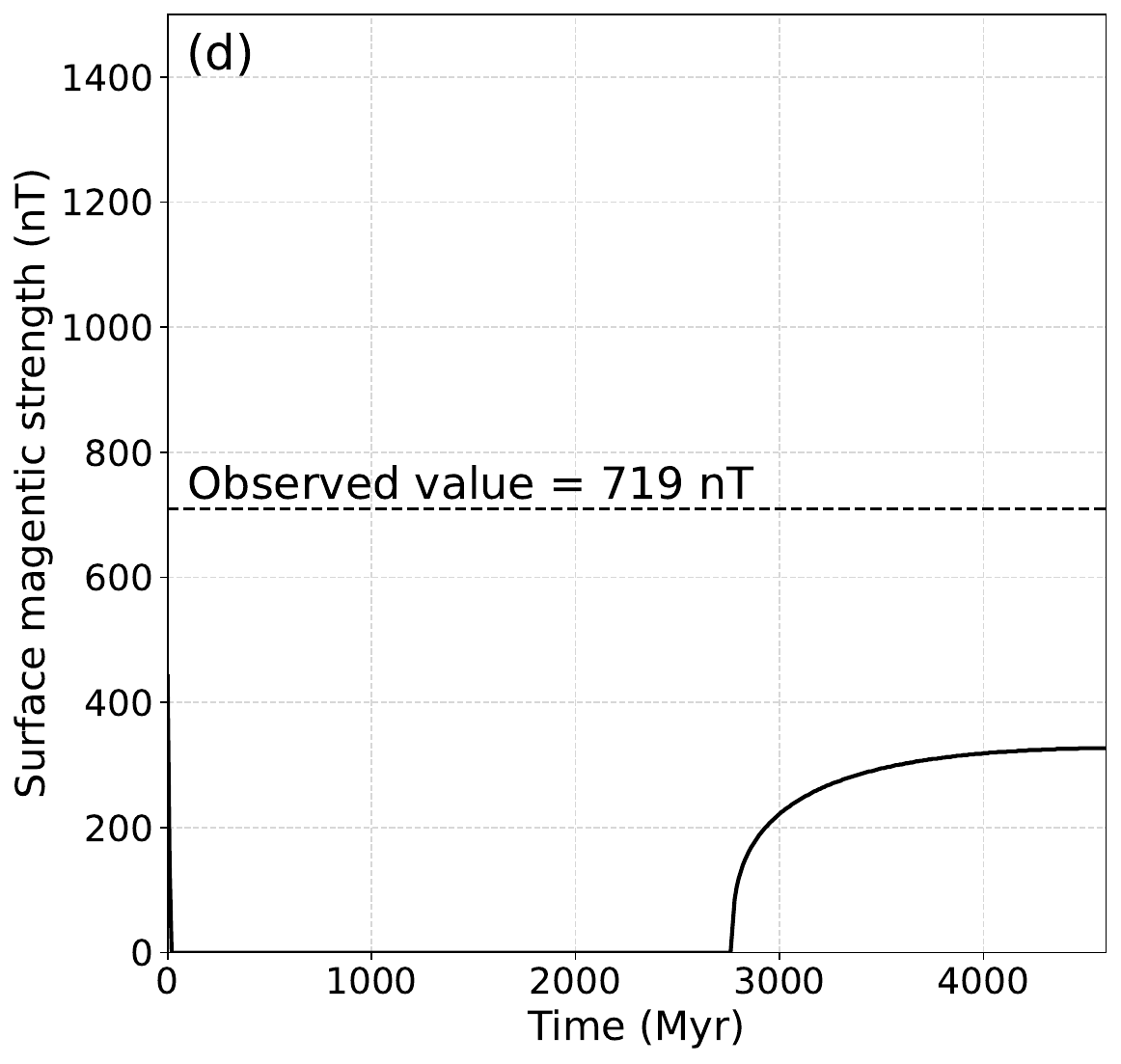}
  \end{minipage}
  \caption{Representative thermo-chemical evolution model of (a) temperature profile changing with time of the rocky mantle and the metallic core, (b) sulfur content changing with time of the core, (c) the area power density available for the dynamo $\phi_{\mathrm{all}}$ and the contributions of $\phi_{\mathrm{ad}}$ and $\phi_{\mathrm{s}}$, and (d) the dipole surface magnetic field strength with the observed value of 719\,nT \citep{Kivelson2002}.
In this model, assumed radius of the metallic core is 690\,km, the bulk sulfur content is 16.7\,wt\%, the bulk density of the hydrosphere is 1,200\,kg\,m$^{-3}$, and the LL chondritic composition is assumed. Note that in panel (b), the uniform color distribution represents an entirely molten regime without core solidification.}
\label{fig:LLmin}
\end{figure}

\begin{figure}[htbp]
\centering
    \includegraphics[width=0.6\textwidth]{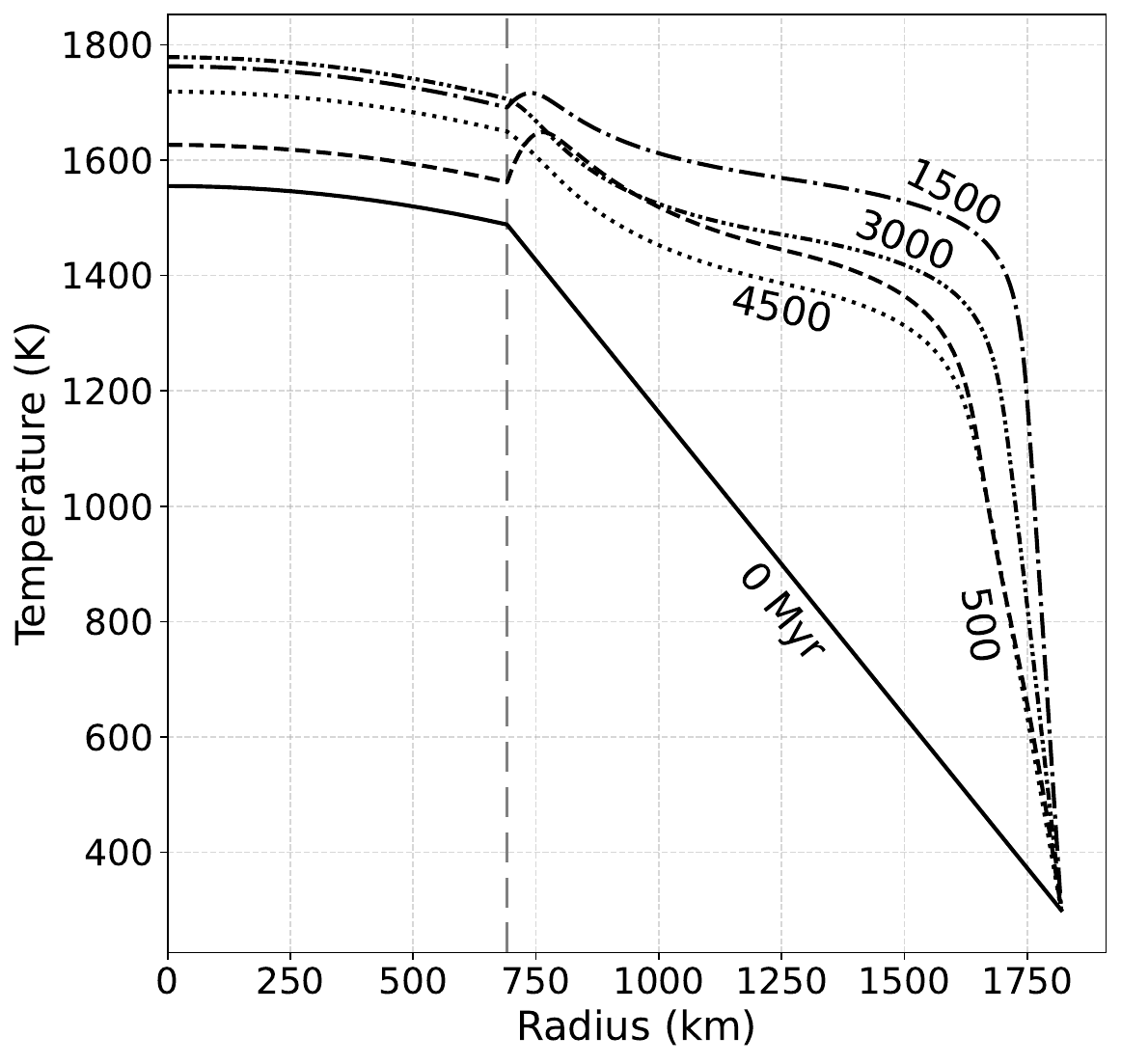}
  \caption{Temperature profiles at 0, 500, 1500, 3000, and 4500 Myr for the case shown in Figure \ref{fig:LLmin}. 
The vertical gray line indicates the location of the CMB.}
\label{fig:LLmin_temperature}
\end{figure}

\begin{figure}[htbp]
  \centering
  \begin{minipage}{0.48\textwidth}
    \centering
    \includegraphics[width=\textwidth]{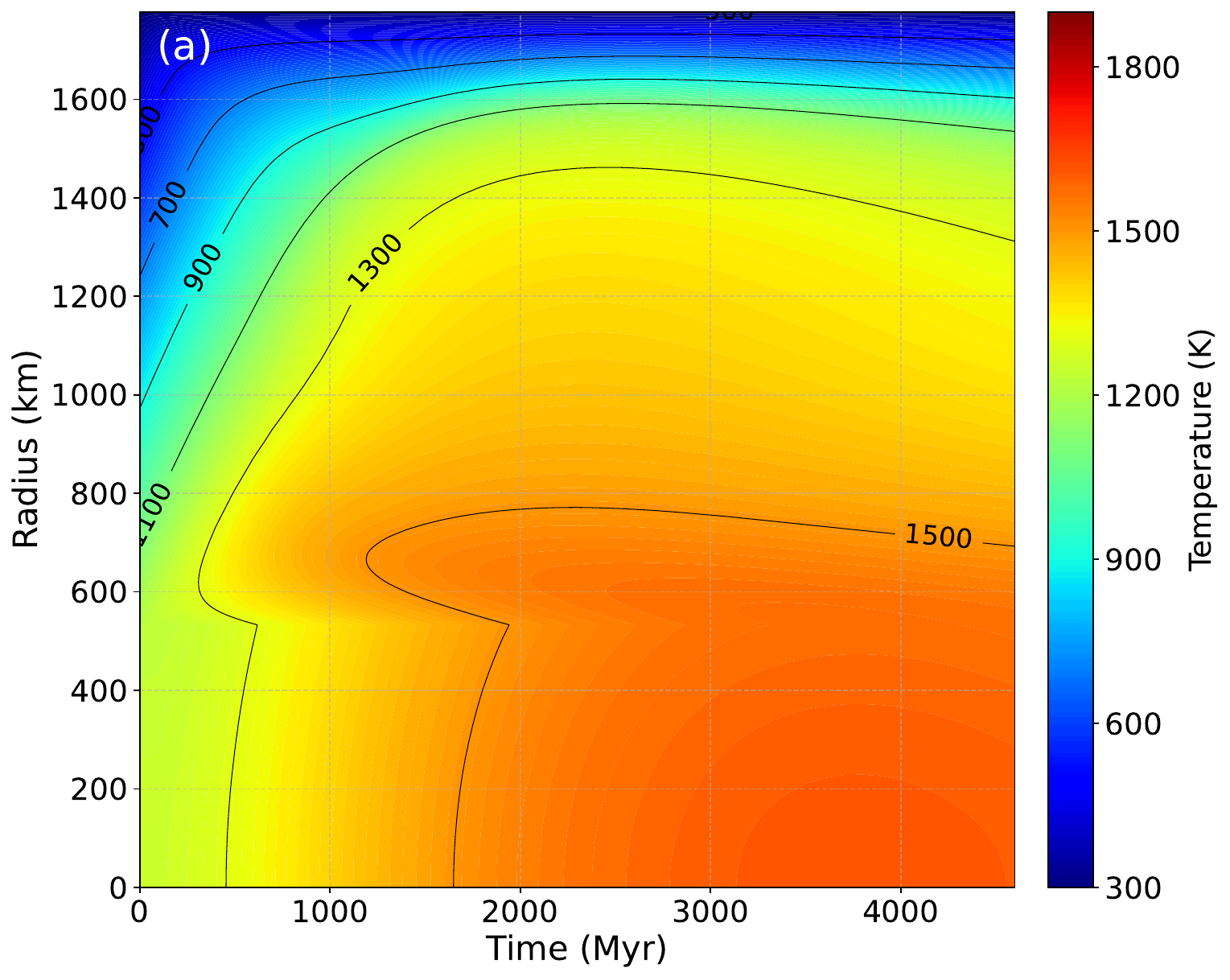}
  \end{minipage}
  \hfill
  \begin{minipage}{0.48\textwidth}
    \centering
    \includegraphics[width=\textwidth]{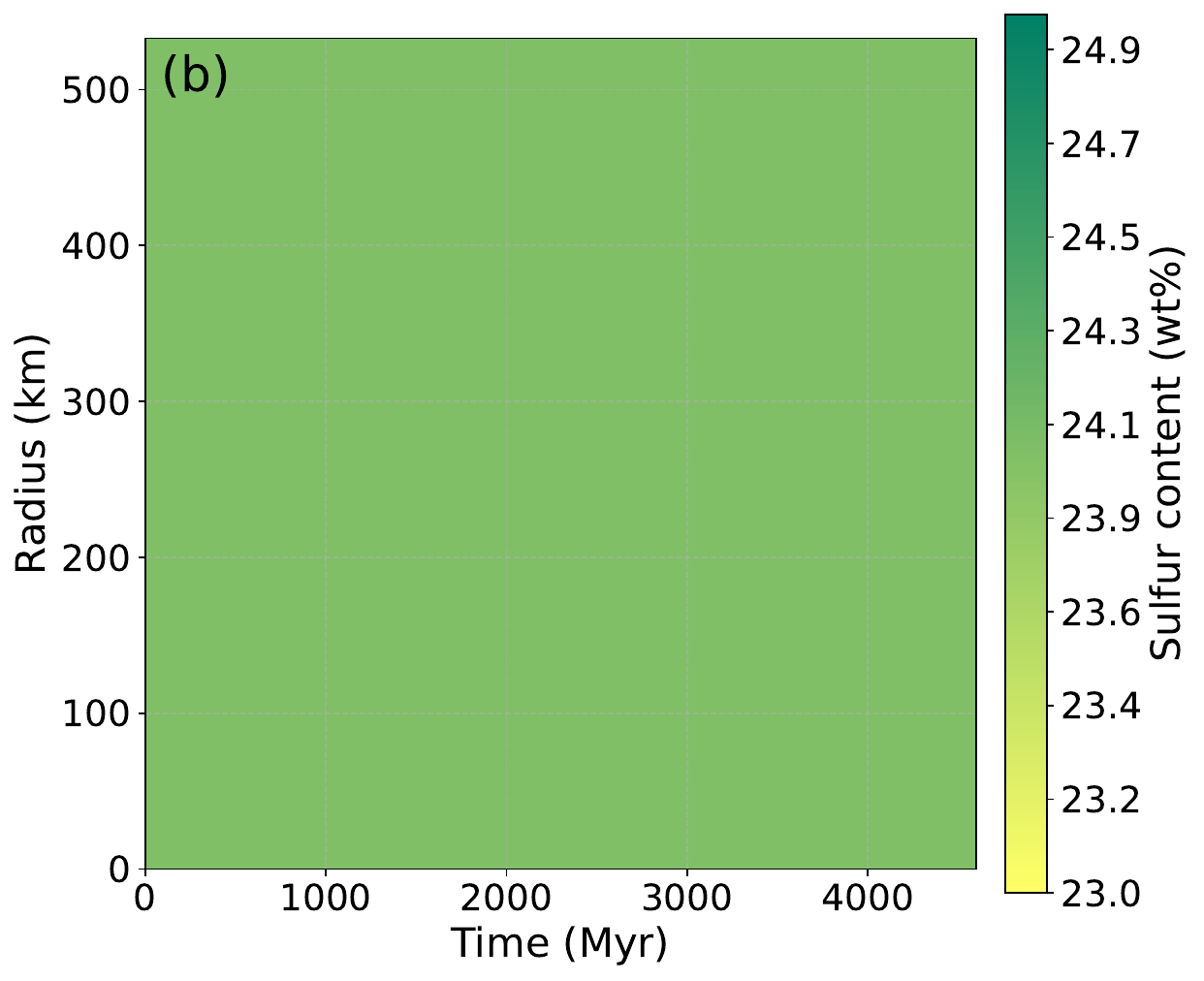}
  \end{minipage}

  \vspace{2ex} %

  \begin{minipage}{0.48\textwidth}
    \centering
    \includegraphics[width=\textwidth]{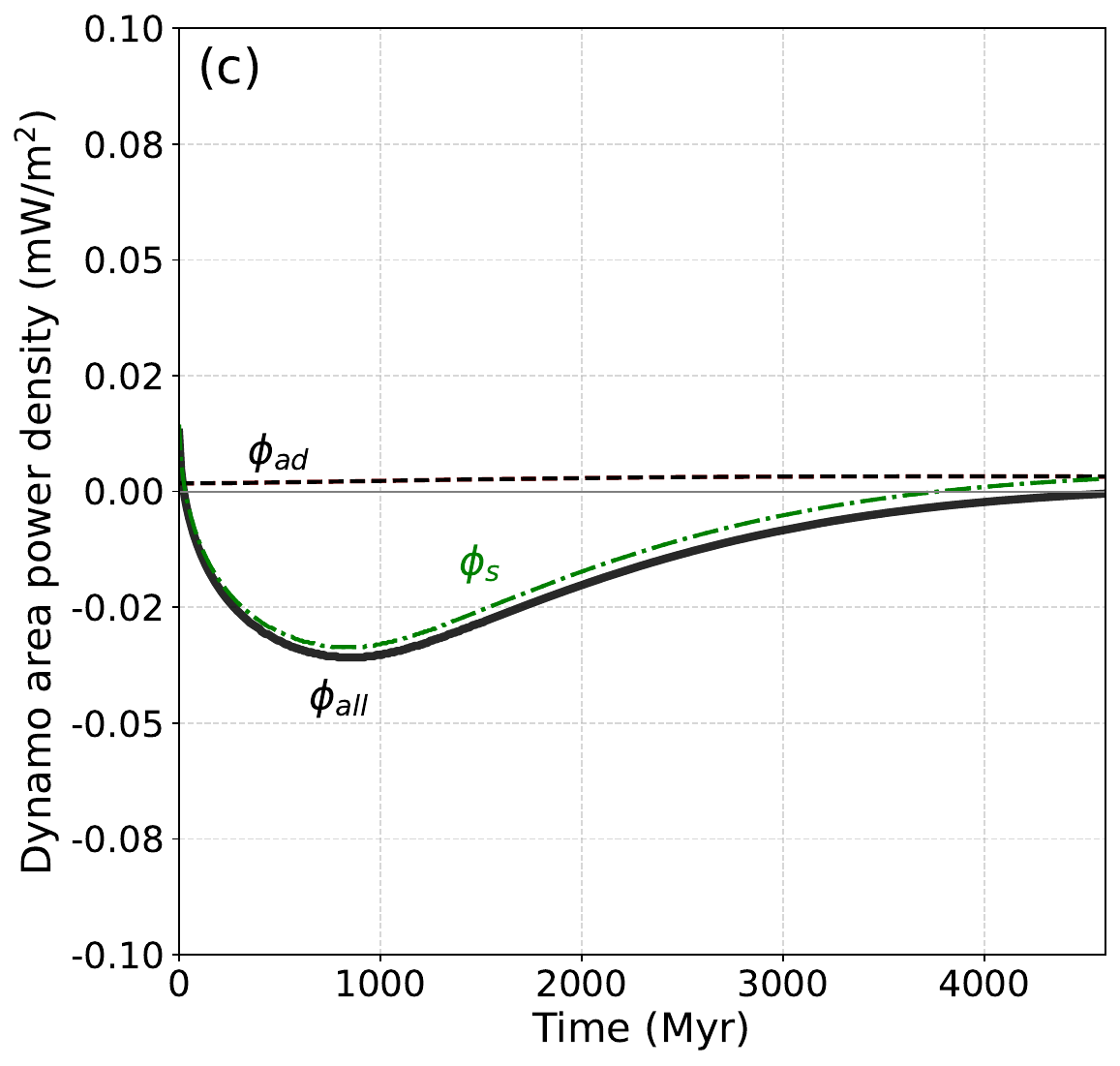}
  \end{minipage}
  \hfill
  \begin{minipage}{0.48\textwidth}
    \centering
    \includegraphics[width=\textwidth]{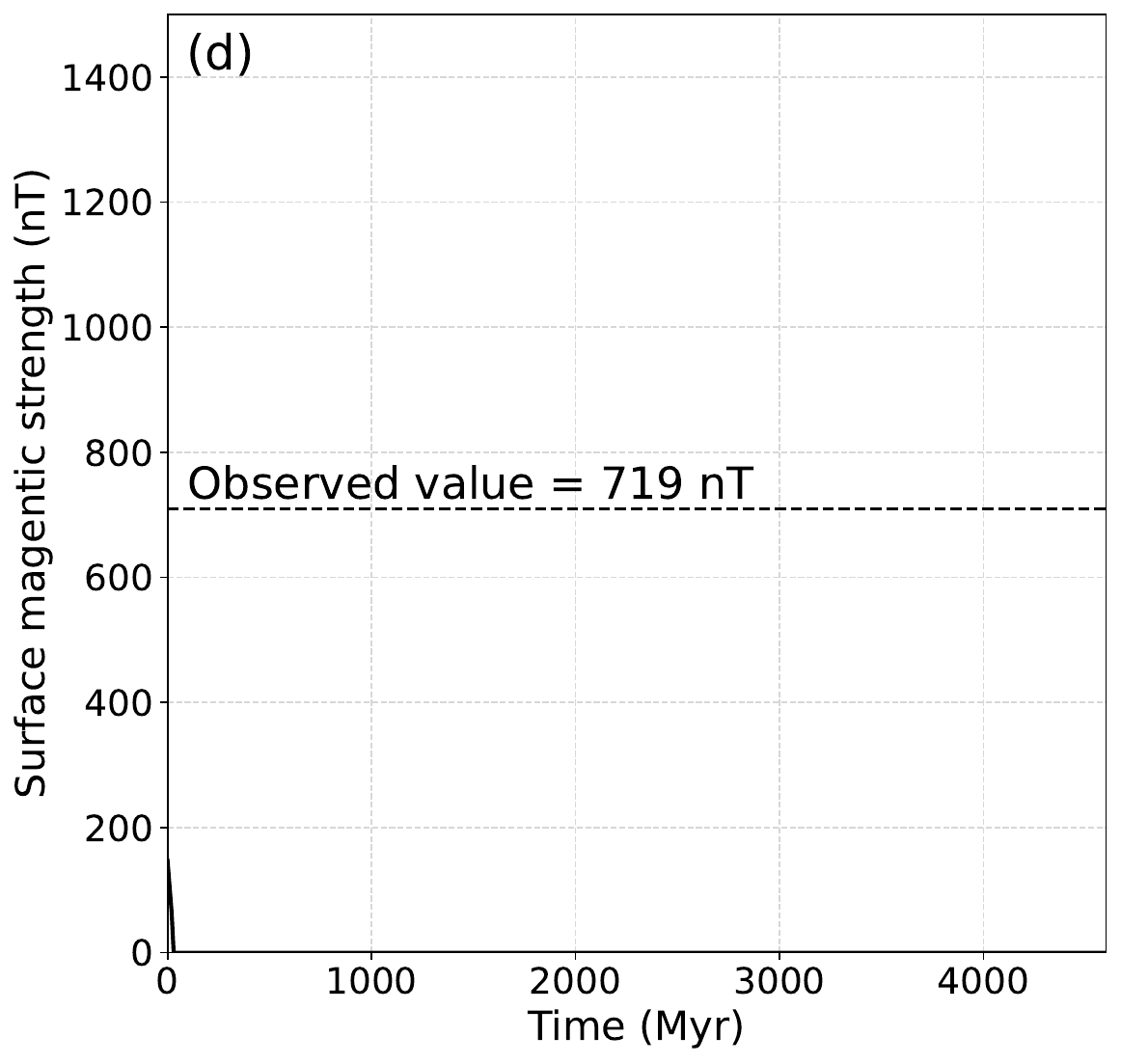}
  \end{minipage}
  
  \caption{Same as Figure \ref{fig:LLmin} but assumed radius of the metallic core is 530\,km, the bulk sulfur content is 24.0\,wt\%, the bulk density of the hydrosphere is 1,200\,kg\,m$^{-3}$, and the CK chondritic composition is assumed. This case also follows an entirely molten regime but fails to generate a magnetic field.}
\label{fig:CKmax}
\end{figure}

\begin{figure}[htbp]
  \centering
  \begin{minipage}{0.48\textwidth}
    \centering
    \includegraphics[width=\textwidth]{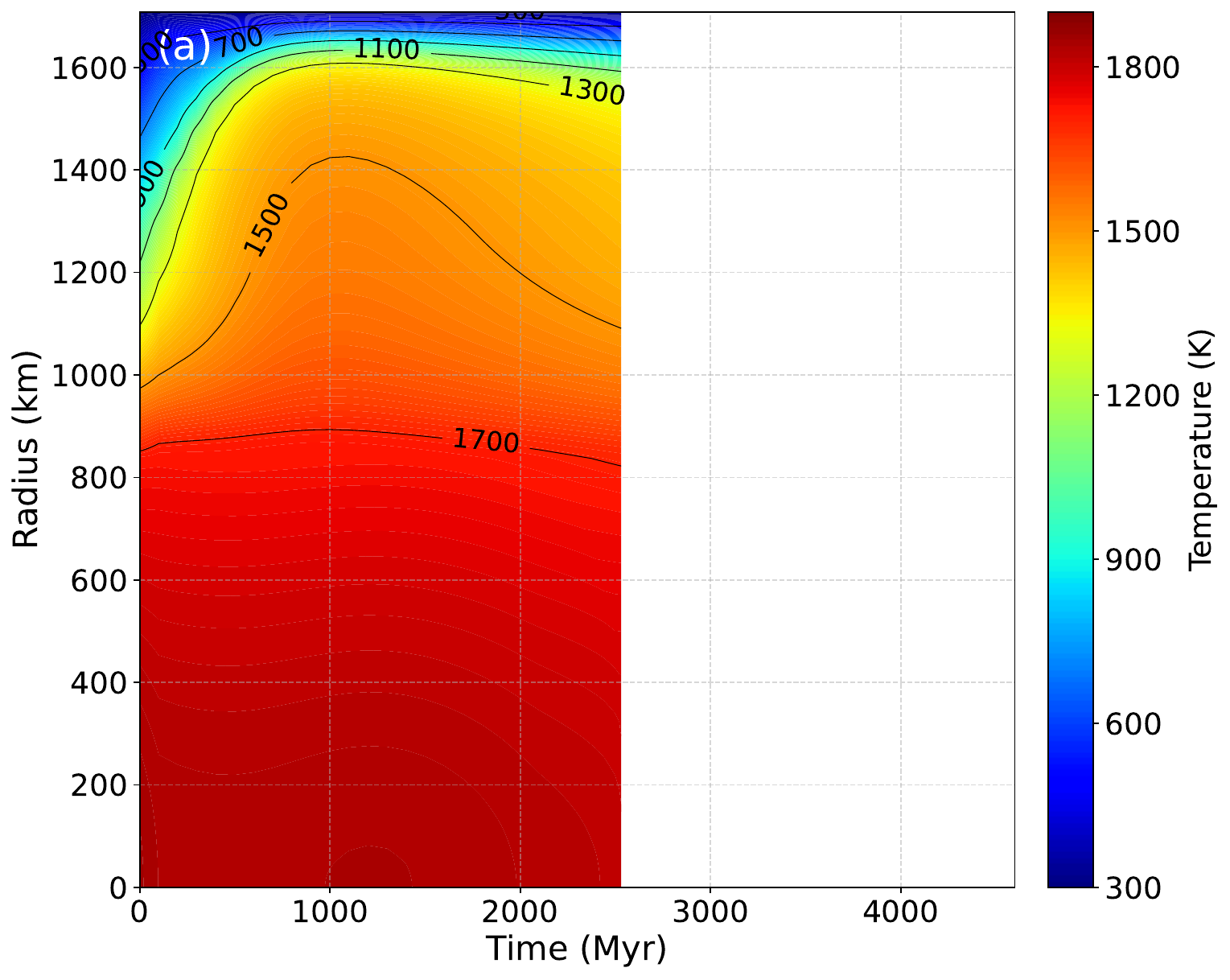}
  \end{minipage}
  \hfill
  \begin{minipage}{0.48\textwidth}
    \centering
    \includegraphics[width=\textwidth]{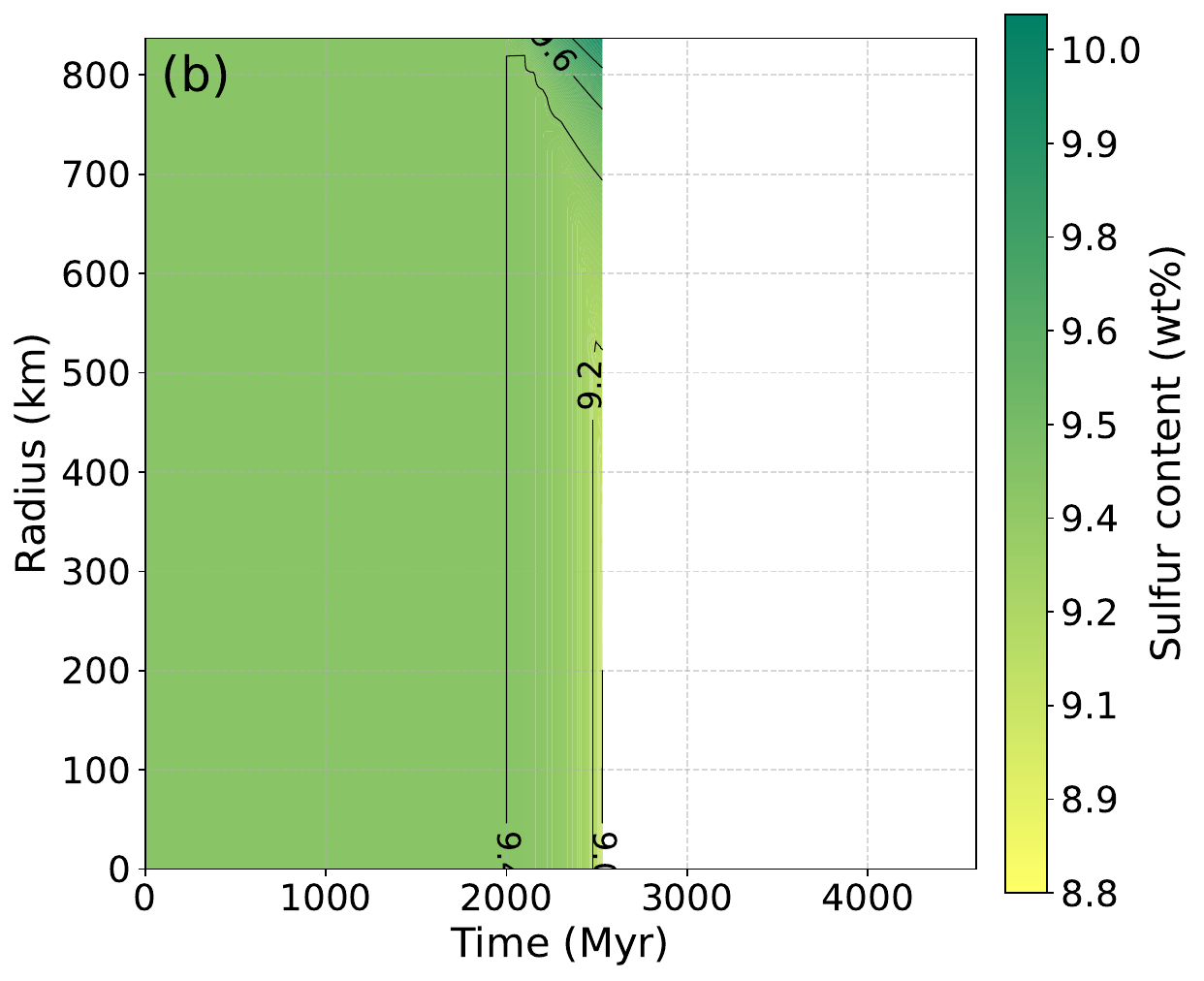}
  \end{minipage}

  \vspace{2ex} %

  \begin{minipage}{0.48\textwidth}
    \centering
    \includegraphics[width=\textwidth]{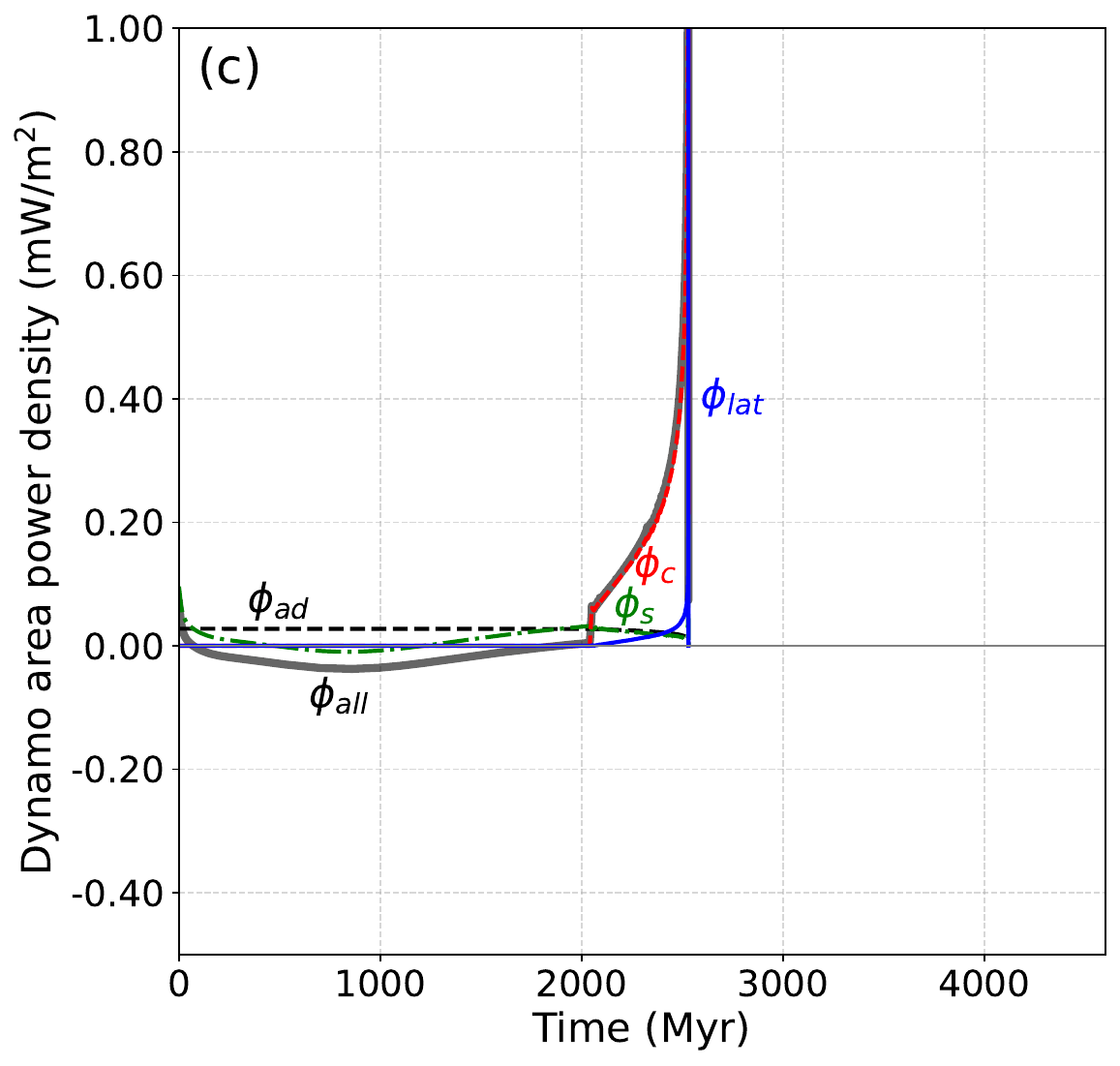}
  \end{minipage}
  \hfill
  \begin{minipage}{0.48\textwidth}
    \centering
    \includegraphics[width=\textwidth]{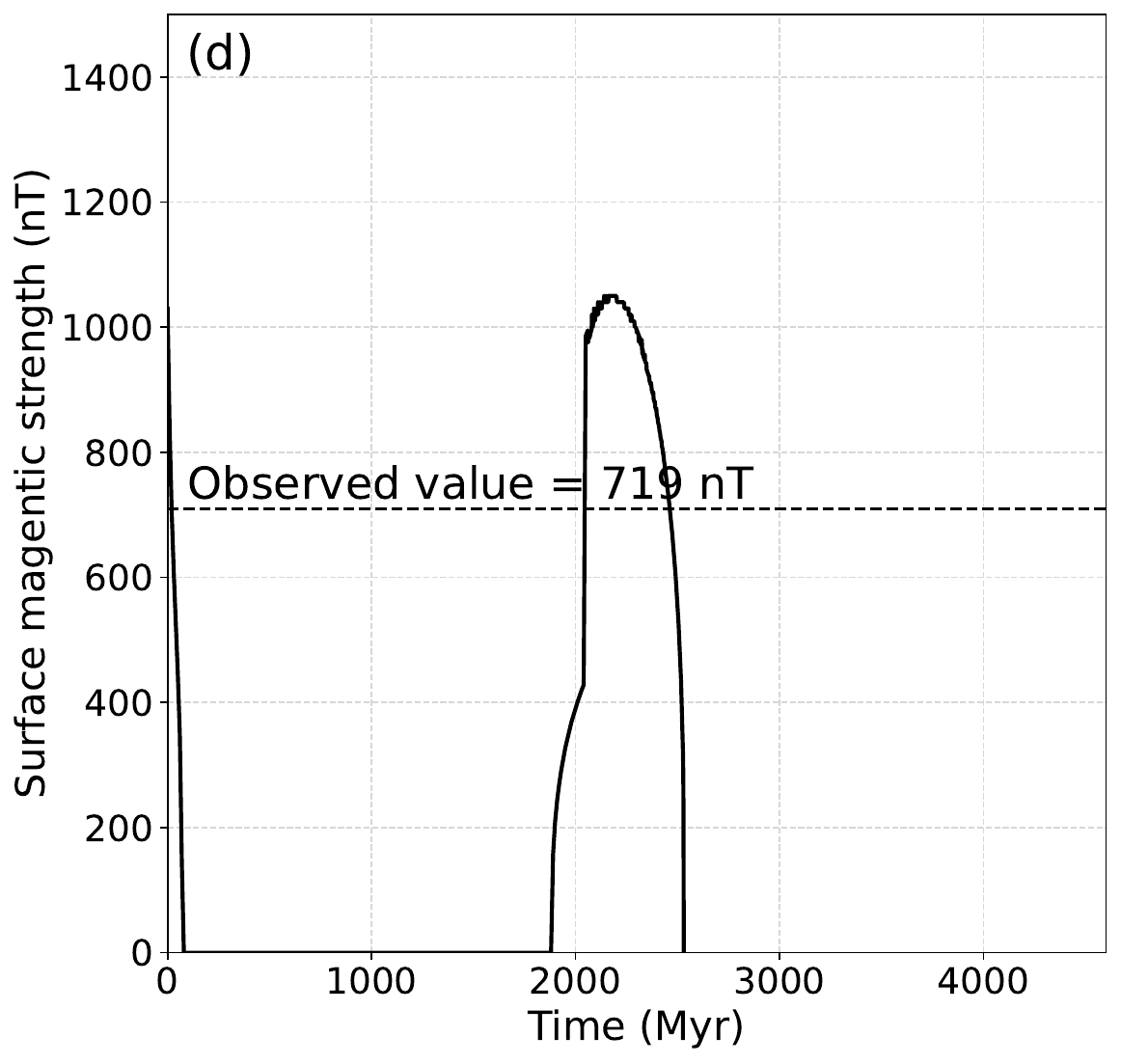}
  \end{minipage}
  
  \caption{Same as Figure \ref{fig:LLmin} but assumed radius of the metallic core is 840\,km, the bulk sulfur content is 9.4\,wt\%, the bulk density of the hydrosphere is 1,300\,kg\,m$^{-3}$, and the H chondritic composition is assumed, representing the Fe-snow regime where the dynamo ultimately ceases. $\phi_{\mathrm{C}}$ and $\phi_{\mathrm{lat}}$ denote the area power density available for the dynamo from contributions of the compositional convection and the heat absorbed during remelting, respectively.}
\label{fig:Hmin}
\end{figure}

\begin{figure}[htbp]
  \centering
  \begin{minipage}{0.48\textwidth}
    \centering
    \includegraphics[width=\textwidth]{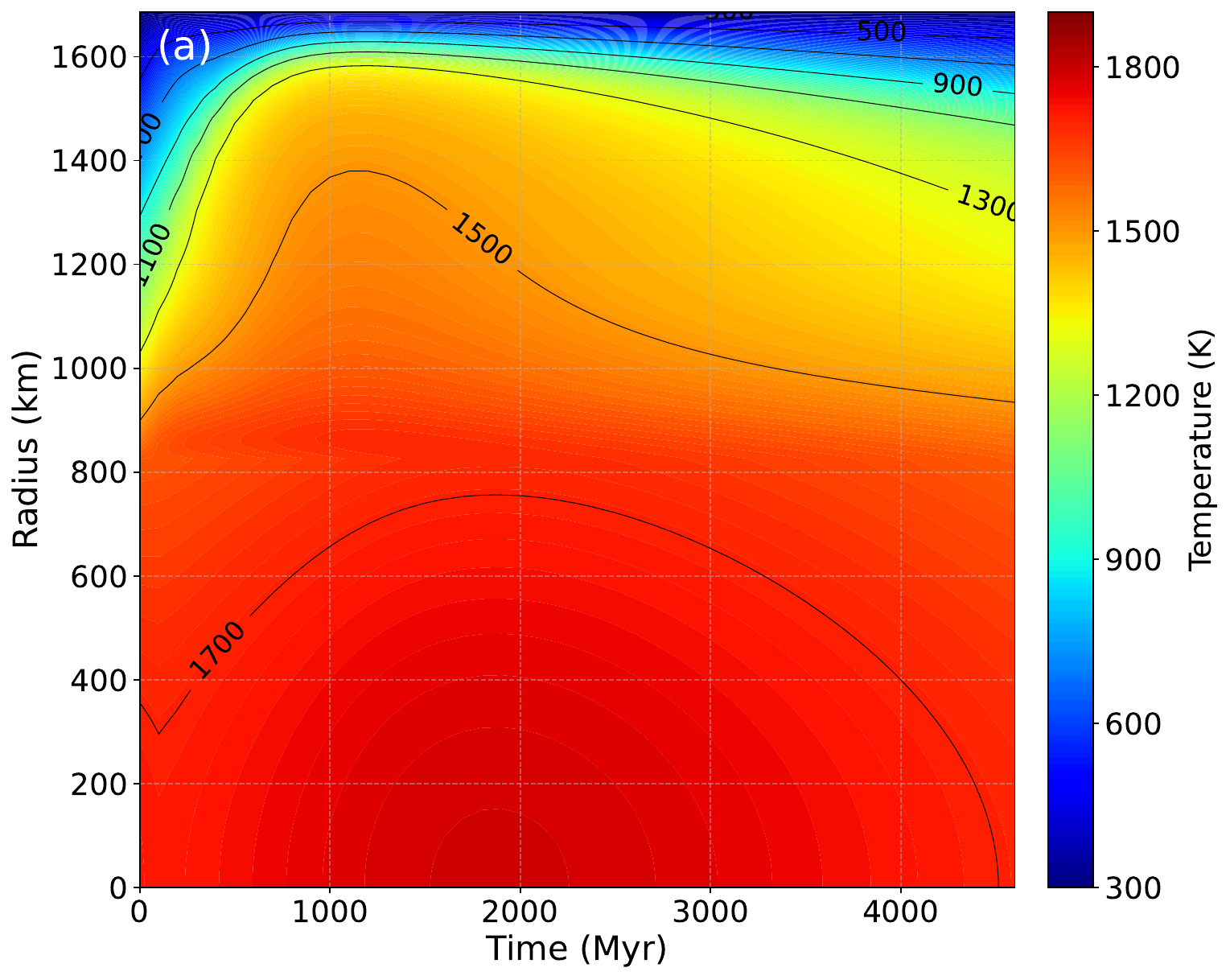}
  \end{minipage}
  \hfill
  \begin{minipage}{0.50\textwidth}
    \centering
    \includegraphics[width=\textwidth]{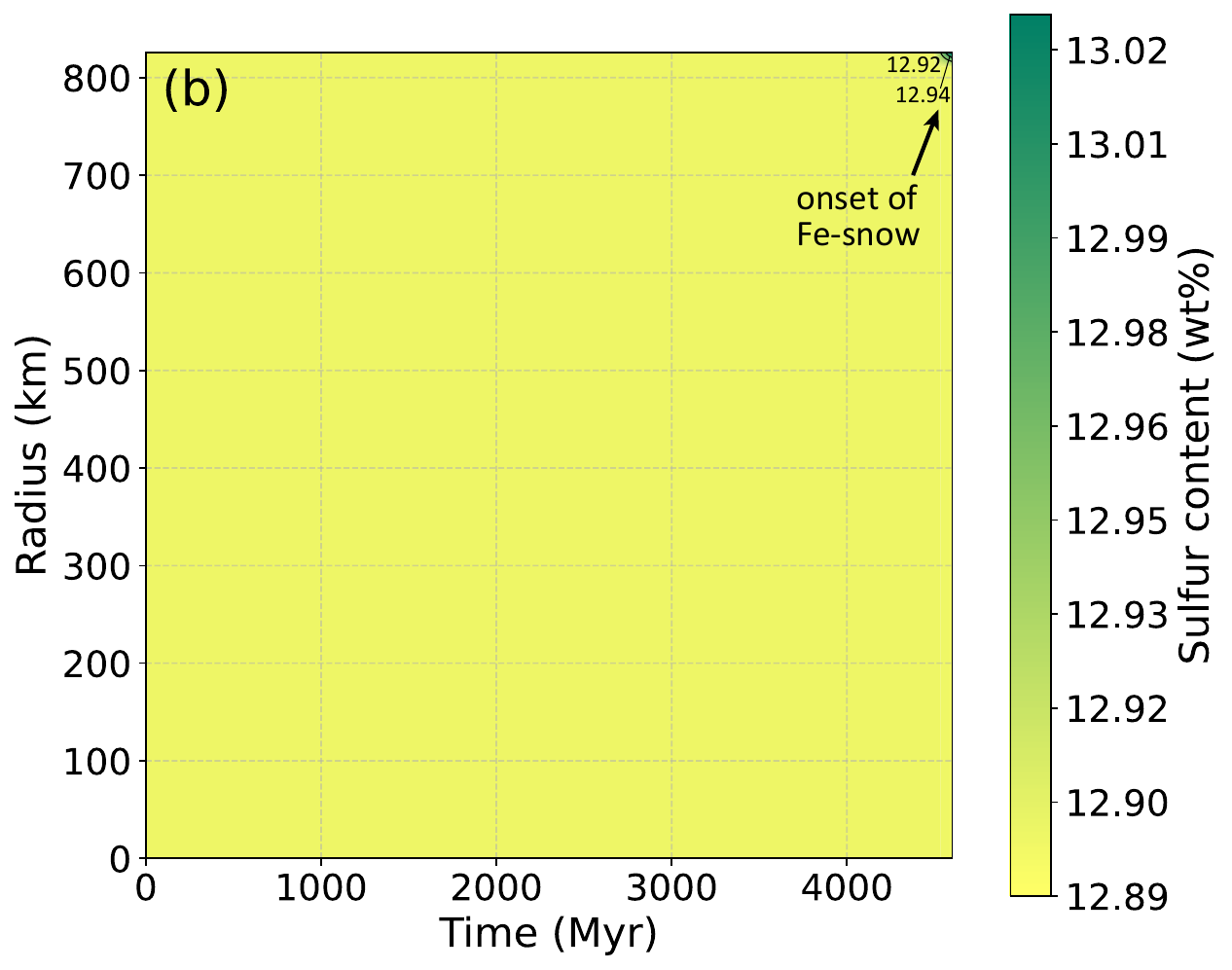}
  \end{minipage}

  \vspace{2ex} %

  \begin{minipage}{0.48\textwidth}
    \centering
    \includegraphics[width=\textwidth]{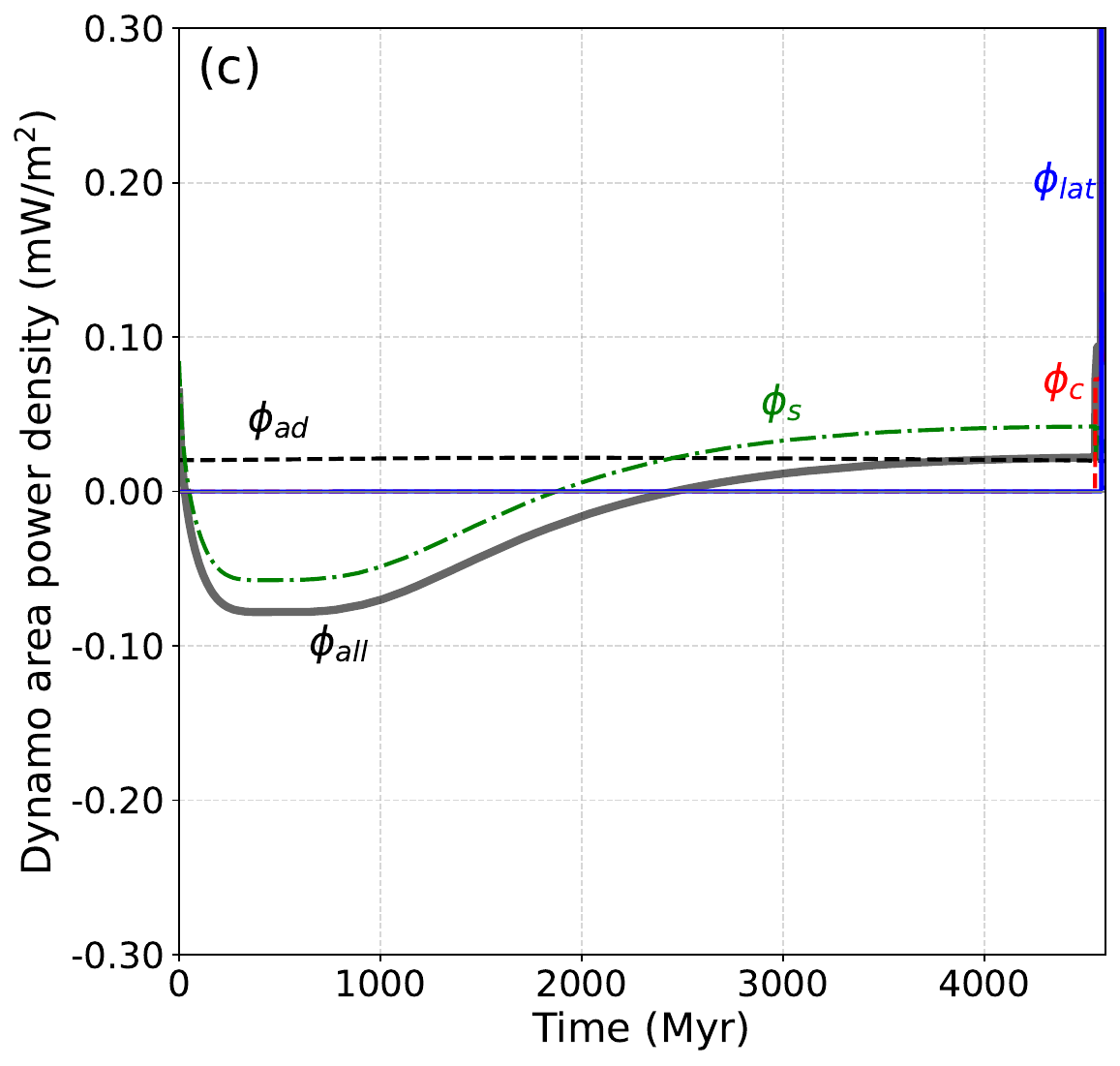}
  \end{minipage}
  \hfill
  \begin{minipage}{0.48\textwidth}
    \centering
    \includegraphics[width=\textwidth]{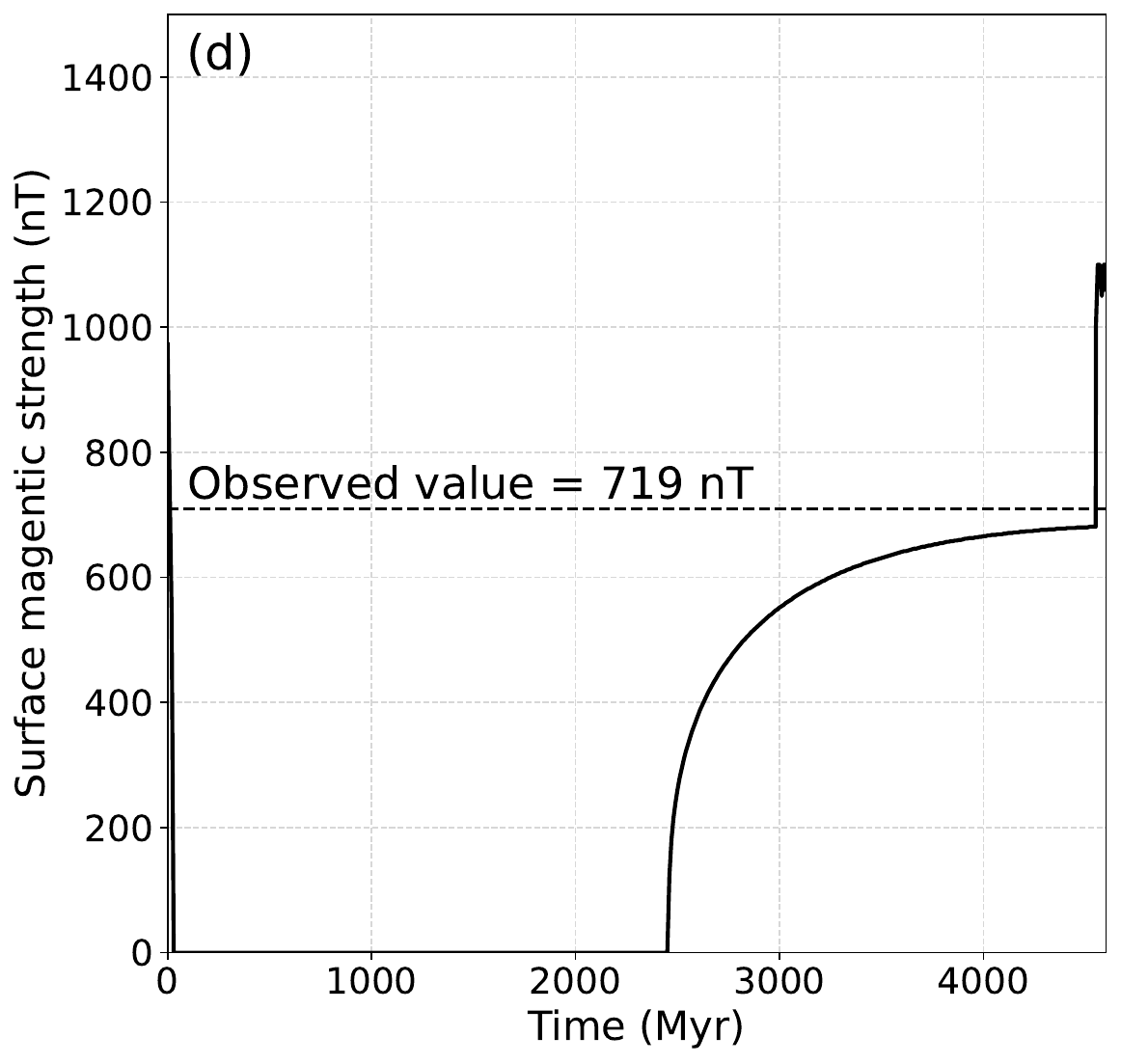}
  \end{minipage}
  
  \caption{Same as Figure \ref{fig:LLmin} but assumed radius of the metallic core is 830\,km, the bulk sulfur content is 12.9\,wt\%, the bulk density of the hydrosphere is 1,300\,kg\,m$^{-3}$, and the H chondritic composition is assumed, representing the Fe-snow regime that sustains an active present-day dynamo.}
\label{fig:Hmax}
\end{figure}

\begin{figure}[htbp]
\centering
\includegraphics[width=0.6\textwidth]{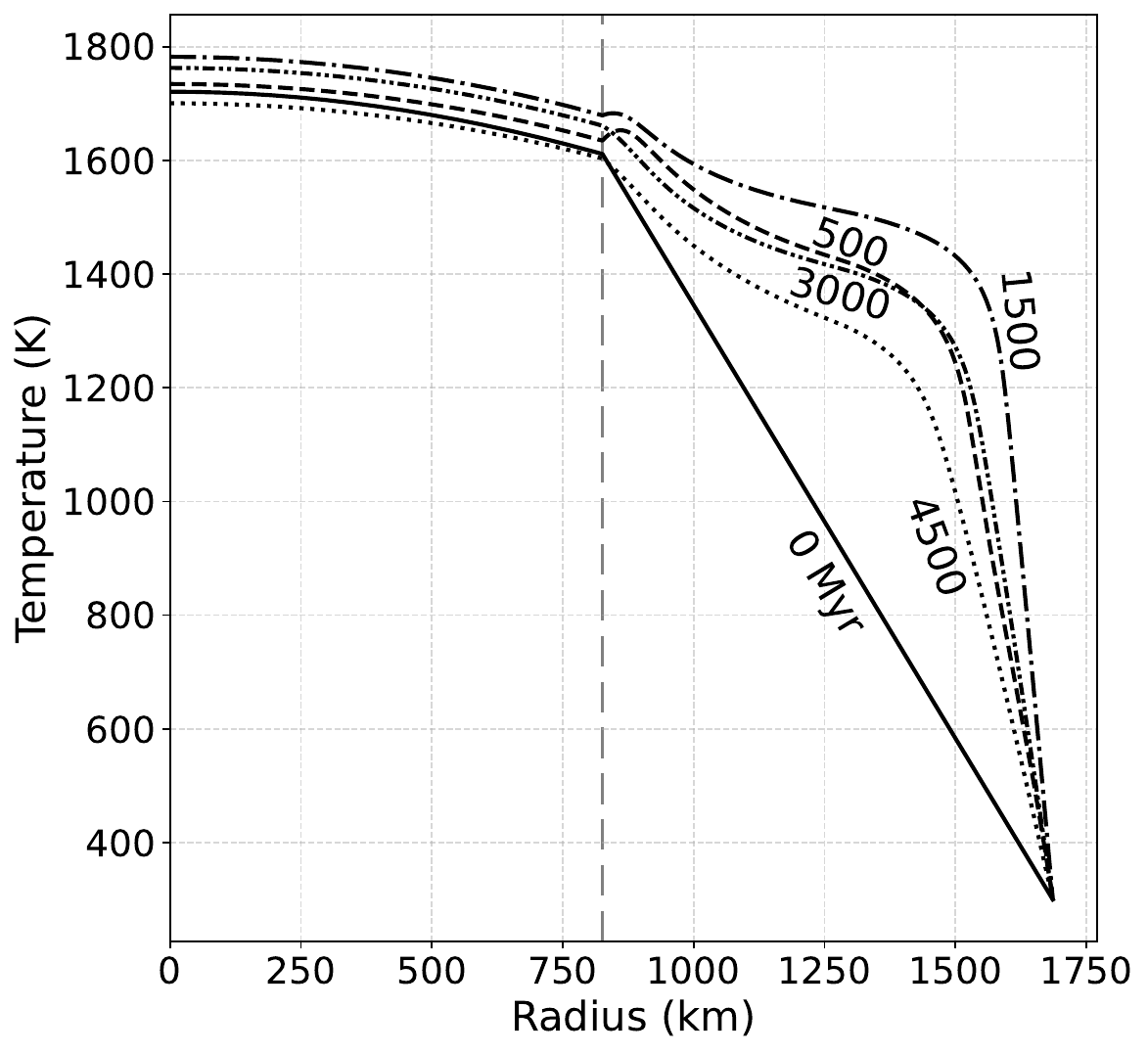}

\caption{Temperature profiles at 0, 500, 1500, 3000, and 4500 Myr for the case shown in Figure \ref{fig:Hmax}. 
The vertical gray line indicates the location of the CMB.}
\label{fig:Hmax_temperature}
\end{figure}

\begin{figure}[htbp]
  \centering
  \begin{minipage}{0.48\textwidth}
    \centering
    \includegraphics[width=\textwidth]{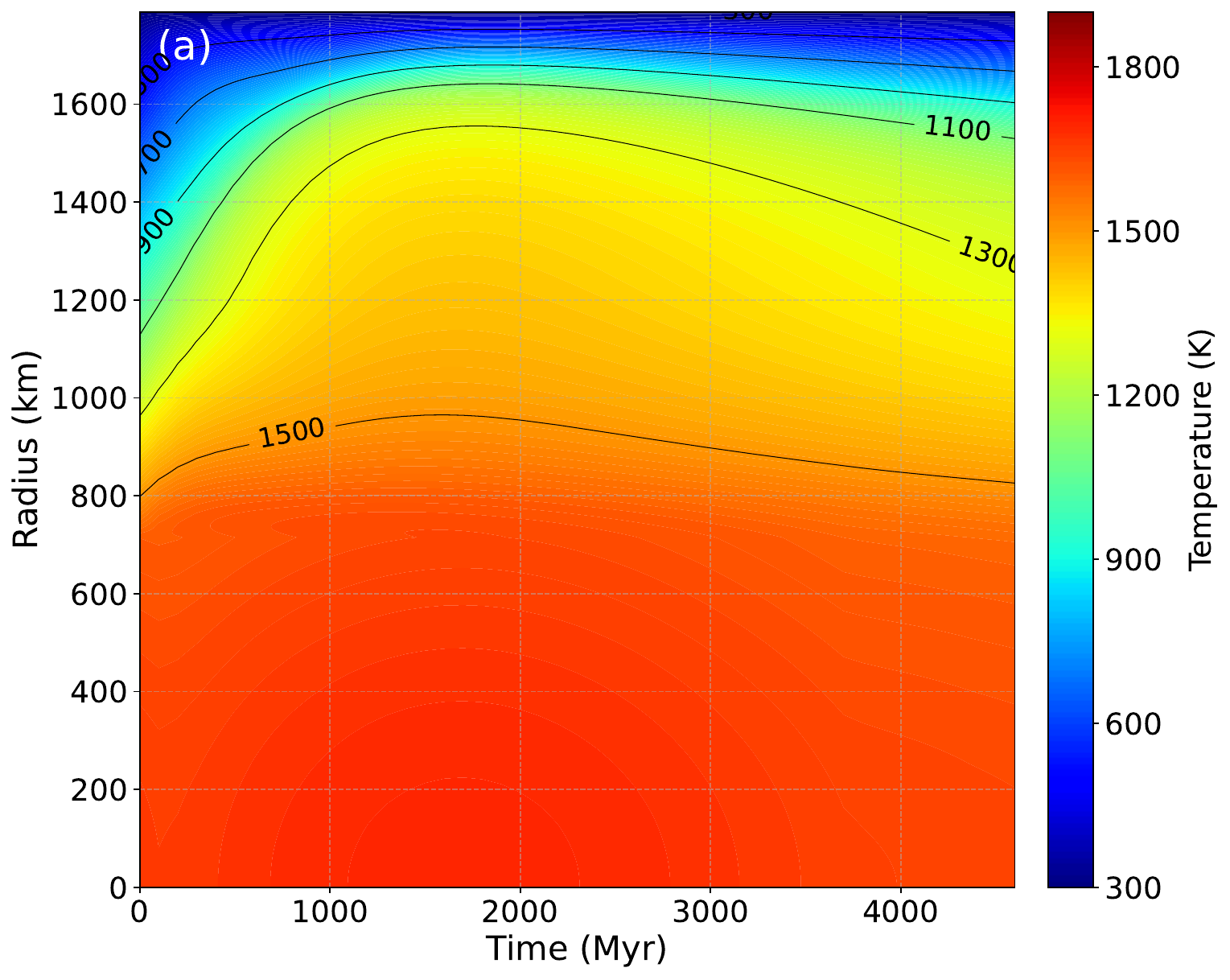}
  \end{minipage}
  \hfill
  \begin{minipage}{0.48\textwidth}
    \centering
    \includegraphics[width=\textwidth]{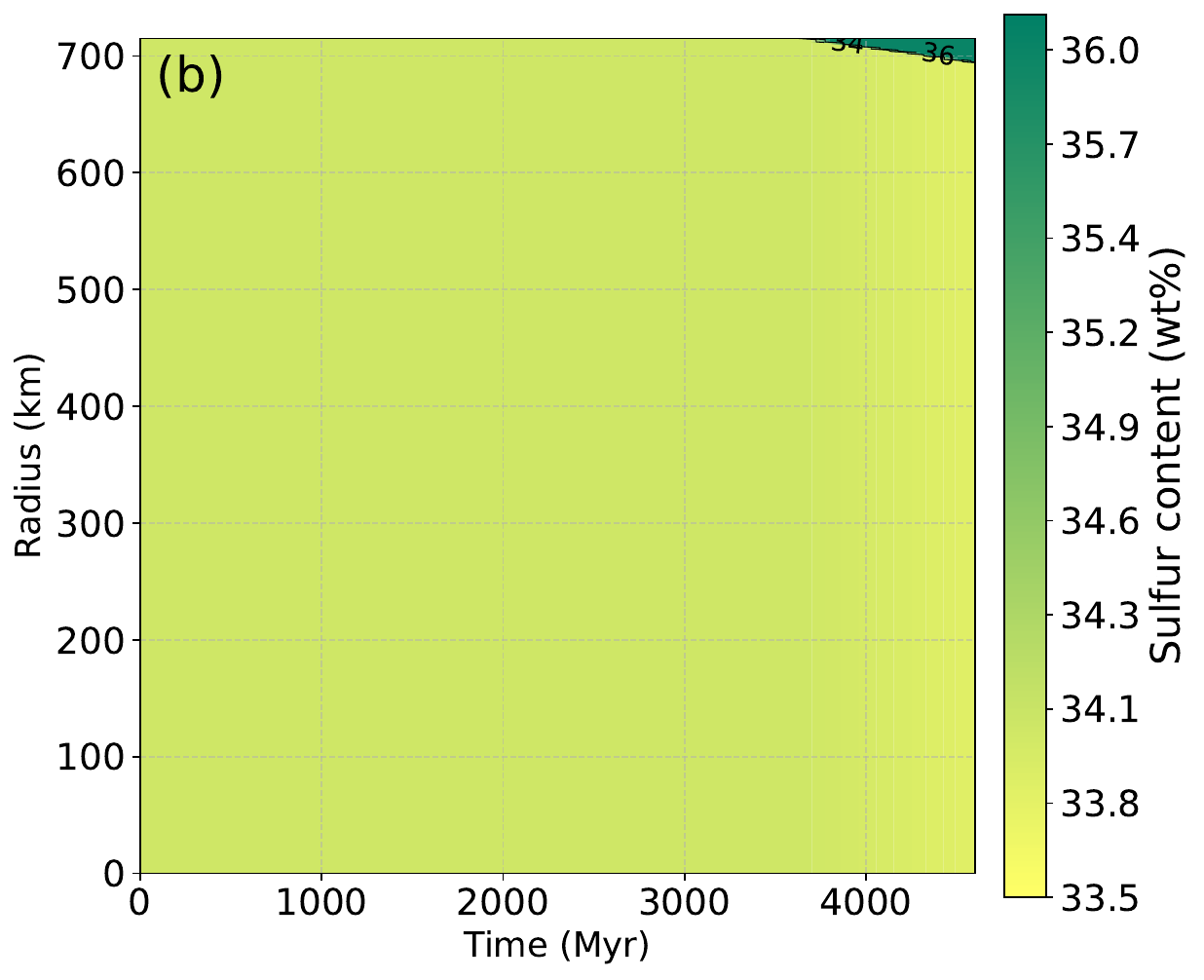}
  \end{minipage}

  \vspace{2ex} %

  \begin{minipage}{0.48\textwidth}
    \centering
    \includegraphics[width=\textwidth]{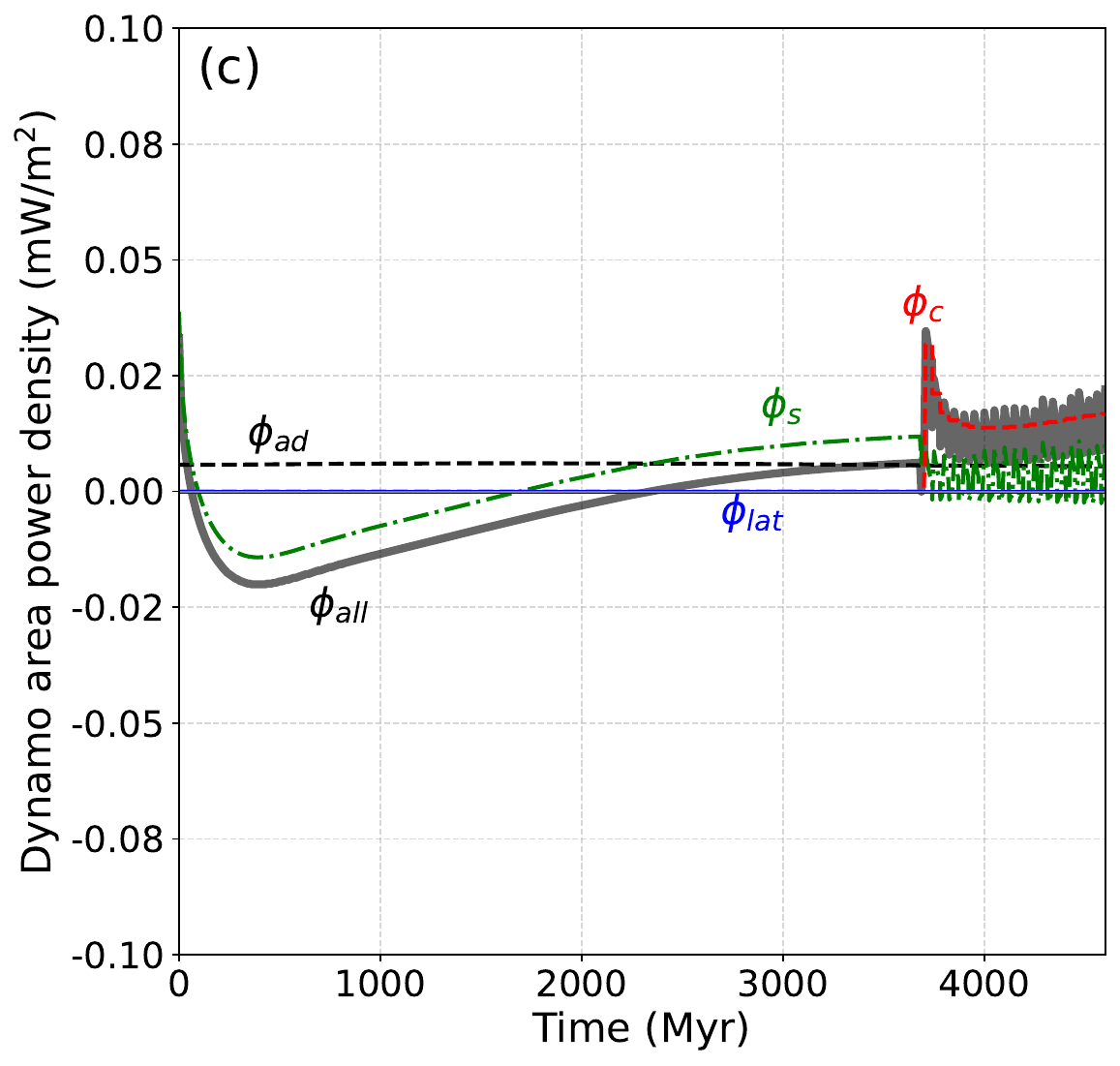}
  \end{minipage}
  \hfill
  \begin{minipage}{0.48\textwidth}
    \centering
    \includegraphics[width=\textwidth]{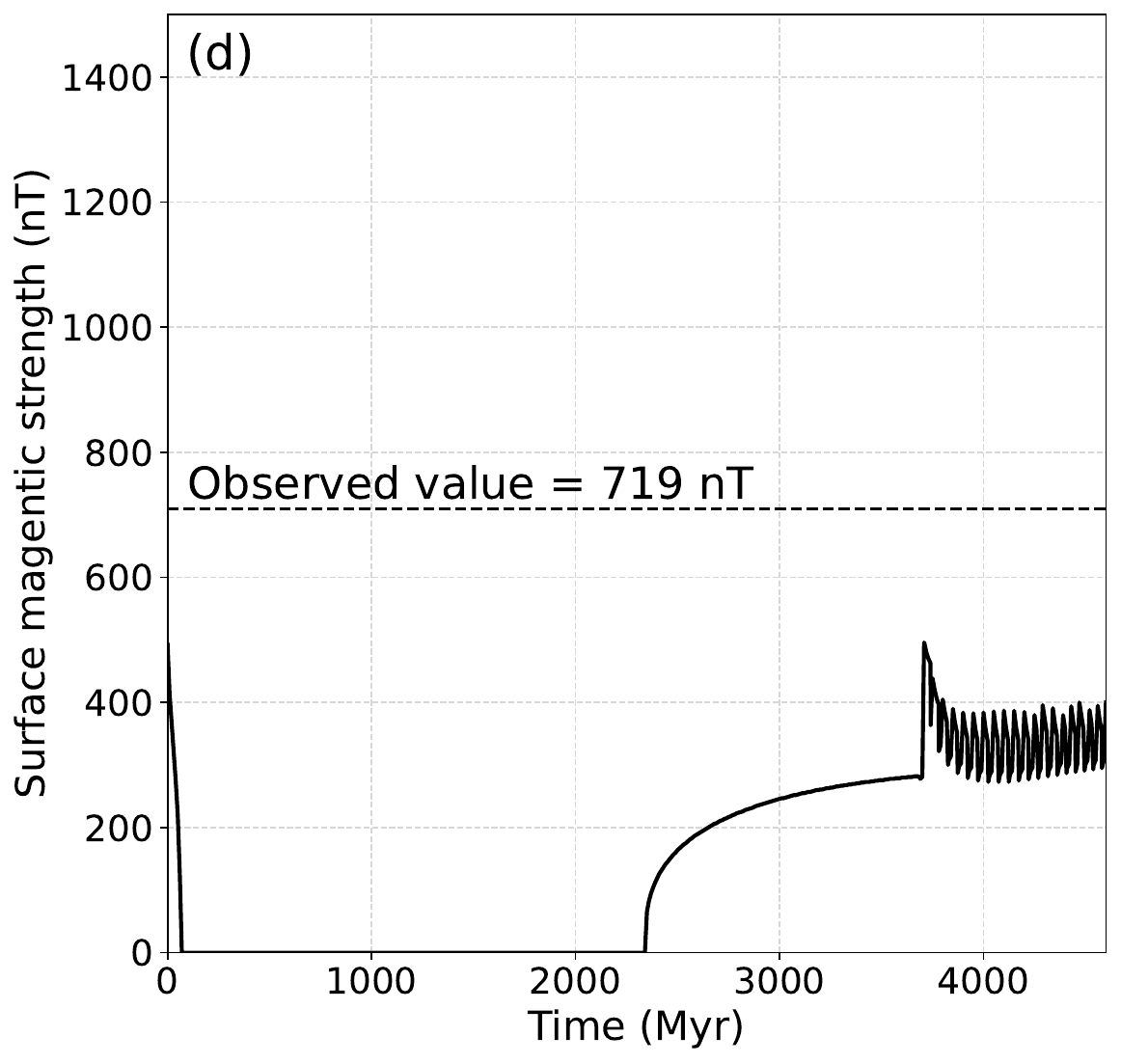}
  \end{minipage}
  
  \caption{Same as Figure \ref{fig:LLmin} but assumed radius of the metallic core is 720\,km, the bulk sulfur content is 34.0\,wt\%, the bulk density of the hydrosphere is 1,200\,kg\,m$^{-3}$, and the CM chondritic composition is assumed, representing the FeS-layer solidification regime.}
\label{fig:CMmid}
\end{figure}

We describe the results of our calculations for the interior thermal evolution of Ganymede, particularly the evolution of the rocky mantle and the metallic core, the sulfur content in the core, the dynamo driving power in the core, and resulting strength of the magnetic field at the Ganymede's surface for 4.6\,Gyrs, within a range of interior structures and bulk composition.
To provide a comprehensive overview of our results, Table \ref{tab:simulation_summary} summarizes the full set of simulations performed in this study. 
From this broad parameter space, we have selected five representative cases to discuss in detail (Figures \ref{fig:LLmin} to \ref{fig:CMmid}). These specific cases were chosen to illustrate the distinct core evolution trajectories and magnetic field generation mechanisms identified in our model: the entirely molten regime with and without an active dynamo (Figures \ref{fig:LLmin} and \ref{fig:CKmax}), the Fe-snow regime leading to dynamo cessation or sustained activity (Figures \ref{fig:Hmin} and \ref{fig:Hmax}), and the FeS-layer regime (Figure \ref{fig:CMmid}).

As a general trend in our calculations, the core evolves from an initially fully molten state in response to the thermal history of the overlying mantle. The decay energy of radioactive isotopes warms the rocky mantle, and the core subsequently follows this warming trend via thermal coupling across the core-mantle boundary (CMB).
As the radiogenic heat sources in the mantle deplete, progressive mantle cooling drives the continuous cooling of the core, eventually inducing the segregation of solid phases depending on the core composition.

Figure\,\ref{fig:LLmin} depicts the typical results for the time evolution of Ganymede's interior, the sulfur content in the metallic core, the dynamo driving power, and the magnetic field strength at Ganymede's surface for the LL chondritic bulk composition and the bulk density of the hydrosphere of 1,200\,kg\,m$^{-3}$ (Figure\,\ref{fig:1}).
The radius and the initial sulfur content of the metallic core are 690\,km and 16.7\,wt.\%, respectively, corresponding to an initial CMB temperature (i.e., the composition-dependent core liquidus) of 1,488~K.
The mantle temperature reaches its peak at approximately 1.0\,Gyr (Figure\,\ref{fig:LLmin}a). 
Due to efficient convective heat transport within the mantle, the core temperature peaks at approximately 1,790\,K at around 2.5\,Gyr, after which the core enters a cooling phase.
Throughout the cooling phase, the core remains entirely molten without initiating solidification, preventing compositional segregation and thus maintaining a uniform sulfur distribution under sustained thermal convection  (Figure\,\ref{fig:LLmin}b).
Throughout this cooling phase, the adiabatic heat flux at the core-mantle boundary ($F_{\mathrm{ad}}$) ranges from approximately 0.70 to 0.88~mW\,m$^{-2}$.
This regime enhances the power available for the dynamo derived from secular cooling at the top of the fully molten region (Figure\,\ref{fig:LLmin}c).
Consequently, this secular cooling activates a dynamo sustained within the entirely molten core, yielding a present-day surface magnetic field strength equivalent to approximately half of the observed value (Figure\,\ref{fig:LLmin}d).
Temporal evolution of the temperature profiles is provided in Figure \ref{fig:LLmin_temperature}, demonstrating that the mantle requires less than 0.5~Gyr to fully relax from the initial linear profile and establish a quasi-steady convective state. 
While the early thermal behavior during this initial relaxation phase is influenced by the initial conditions, such as the initial core temperature set at the liquidus, the subsequent thermal trajectory of the core over the long term is largely insensitive to the exact initial mantle temperature distribution. 
Further implications of different initial thermal states and the detailed sensitivities of the initial core temperature are discussed in Section 4.1.

Figure\,\ref{fig:CKmax} depicts the corresponding results for the CK chondritic bulk composition with a hydrosphere bulk density of 1,200 kg\,m$^{-3}$.
The radius and the initial sulfur content of the metallic core are 530 km and 24.0 wt.\%, respectively.
Initially, the decay energy of radioactive isotopes within the rocky mantle elevates the temperature at the mantle base. Consistent with the general trend, once this basal temperature exceeds that of the core, the heat flux reverses, causing the core to subsequently heat up (Figure\,\ref{fig:CKmax}a).
The core temperature continuously rises to a peak of approximately 1,600 K at around 4.0 Gyr, after which the core enters a cooling phase.
This prolonged core warming until relatively recently is attributed to the thick mantle containing a large total inventory of radioisotopes, combined with low initial temperatures that delay the onset of solid-state mantle convection, resulting in inefficient heat transport until the mantle warms sufficiently.
In this interior structure model, the core remains entirely molten without initiating solidification, preventing compositional segregation and thus maintaining a constant core sulfur concentration (Figure\,\ref{fig:CKmax}b).
Although the core cools after 4.0 Gyr, the dynamo driving power supplied by this secular cooling remains below the critical threshold required to initiate convection within the core ($\phi_{\text{ad}}$); consequently, neither a dynamo nor a magnetic field is generated (Figure\,\ref{fig:CKmax}c and \ref{fig:CKmax}d).
The brief cooling and associated magnetic field generation immediately after the start of the simulation likely stem from the assumed initial linear temperature profile in the mantle.
Figure \ref{fig:Hmin} shows the results for the H chondritic bulk composition with a hydrosphere bulk density of 1,300 kg m$^{-3}$.
The radius and the initial sulfur content of the metallic core are 840 km and 9.4 wt.\%, respectively.
The core enters a cooling phase at approximately 1.0 Gyr, and core solidification initiates at around 2.0 Gyr, transitioning into the Fe-snow regime (Figure \ref{fig:Hmin}a).
This behavior is attributed to a thin rocky mantle with a small total inventory of radioisotopes, combined with a high initial core temperature ($\sim$1,723~K at CMB) resulting from the low core sulfur concentration, which lowers the basal mantle viscosity and facilitates efficient heat extraction.
Approximately at 2.5\,Gyr, the Fe-snow zone extends the entire core, causing the simulation to terminate (Figure \ref{fig:Hmin}b and \ref{fig:Hmin}c).
This result is consistent with previous estimates indicating that the lifetime of a dynamo in the Fe-snow regime ranges from 0.1 to 1.0 Gyr \citep{hauck2006,Ruckriemen2015,Ruckriemen2018}.
Following the onset of the Fe-snow regime, the sulfur concentration within the Fe-snow zone increases over time, whereas that in the fully molten region decreases (Figure \ref{fig:Hmin}b).
Furthermore, a gravitationally stable stratification develops, where the liquid phase in the upper portion of the Fe-snow zone becomes relatively enriched in sulfur, while the lower portion becomes slfur-depleted.
Upon entering the Fe-snow regime, the dynamo driving power derived from the compositional gradient and the latent heat absorbed during the remelting of solid Fe is generated, increasing sharply over time (Figure \ref{fig:Hmin}c).
This sharp increase occurs because the expansion of the Fe-snow zone into deeper regions increases the total mass of solid Fe that precipitates and remelts.
Consequently, the total area power density $\phi_{\text{all}}$ (and the compositional contribution $\phi_{\text{c}}$) reaches a maximum value of 66 mW m$^{-2}$, while the latent heat contribution $\phi_{\text{lat}}$ peaks at 3.5 mW m$^{-2}$, just before the Fe-snow zone attains the center of the core. 
Additionally, the dynamo driving power resulting from the compositional gradient is more than an order of magnitude larger than that driven by other processes.
The surface magnetic field strength, which was 450\,nT driven by thermal convection due to secular cooling immediately prior to the Fe-snow regime, rises to 1,040\,nT following its onset due to the increased dynamo driving power (Figure \ref{fig:Hmin}d).

Subsequently, however, the effect of the shrinking dynamo region becomes dominant, causing the magnetic field strength to decrease.
Because the magnetic field completely disappears 2.0 Gyr ago in this interior structure model, it fails to reproduce the observed field.

Figure \ref{fig:Hmax} shows the results for the H chondritic bulk composition with a hydrosphere bulk density of 1,300\,kg\,m$^{-3}$.
The radius and the initial sulfur content of the metallic core are 830\,km and 12.9\,wt.\%, respectively.
The core transitions into the Fe-snow regime at approximately 4.5\,Gyr, where the remelting of Fe-snow enhances the dynamo driving power, consequently increasing the surface magnetic field strength to 1,095\,nT.
Within the uncertainty range of the magnetic field calculation parameters, this interior structure model can successfully reproduce Ganymede's observed magnetic field.
However, it must be emphasized that this successful reproduction with the H-chondrite composition is strictly conditional upon a precise temporal coincidence. 
Because the dynamo driven by the Fe-snow regime is inherently short-lived (spanning only a few hundred million years, shown in Figure \ref{fig:Hmin}), the model requires the onset of this regime to have occurred fortuitously late in Ganymede's history to be actively sustained at the present day.
Temporal evolution of the temperature profiles is provided in Figure \ref{fig:Hmax_temperature}.
As in Figure 4, the mantle reaches a convective state within 500 Myr from the initial state. Similar trends are observed in cases with other internal structures or chondritic compositions.

Figure \ref{fig:CMmid} shows the results for the CM chondritic bulk composition with a hydrosphere bulk density of 1,200\,kg\,m$^{-3}$.
The radius and the initial sulfur content of the metallic core are 720\,km and 34\,wt.\%, respectively.
The core temperature reaches its peak at approximately 2.0\,Gyr, and the core initiates solidification at around 4.0\,Gyr, transitioning into the FeS-layer regime (Figure \ref{fig:CMmid}a).
In this regime, the crystallization of solid FeS expels Fe into the underlying liquid phase, shifting the melt composition toward the eutectic point. This compositional shift effectively lowers the liquidus temperature of the remaining melt. 
Consequently, the freezing process is decelerated, allowing the dynamo to potentially operate much longer (until the eutectic temperature is reached). 
This contrasts sharply with the Fe-snow regime, where remelting raises the liquidus and causes the dynamo to cease considerably earlier.
During this process, the sulfur concentration in the fully molten region decreases (Figure \ref{fig:CMmid}b).
Upon the onset of FeS solidification, the dynamo driving power temporarily increases but quickly returns to its baseline level (Figure \ref{fig:CMmid}c). 
The thermodynamic interactions during this phase are highly intertwined. 
Latent heat released at the FeS solidification front contributes to the heat conducted through the overlying solid layer, reducing the heat that must be extracted from the underlying liquid core for a given outward conductive flux (Section 2.6).
After the onset of FeS crystallization, the secular-cooling contribution ($\phi_s$) decreases, while Fe enrichment of the underlying liquid produces a compositional contribution ($\phi_C$) (Figure \ref{fig:CMmid}c). The combined contributions yield a surface magnetic field whose baseline remains comparable to that before solidification (Figure \ref{fig:CMmid}d).
The latent heat released at the boundary between the solid FeS layer and the deeper molten core does not simply suppress the overall secular cooling ($\phi_s$), but it maintains the boundary at the liquidus temperature. 
Due to these complex feedbacks between the latent heat and the compositional liquidus shift, the surface magnetic field strength does not change significantly before or after the onset of core solidification.
The estimated present-day surface magnetic field strength is 406\,nT as a maximum, and this interior structure model fails to reproduce Ganymede's observed magnetic field within the uncertainty range of the magnetic field calculation parameters (Figure \ref{fig:CMmid}d).
Note that the strong oscillatory behavior observed in the dynamo driving power and the resulting magnetic field strength (Figures \ref{fig:CMmid}c and \ref{fig:CMmid}d) is likely a numerical effect associated with the spatially discrete treatment of the FeS solidification front and latent heat release.
Reducing the time step from 100 to 10 years leaves the oscillation amplitudes, long-term evolution, and present-day surface magnetic field strength nearly unchanged. 
The macroscopic baseline trend clearly demonstrates that the generated magnetic field strength does not change significantly across the onset of FeS crystallization and remains below the observed value.

\begin{figure}
\centering
\includegraphics[width=1.0\textwidth]{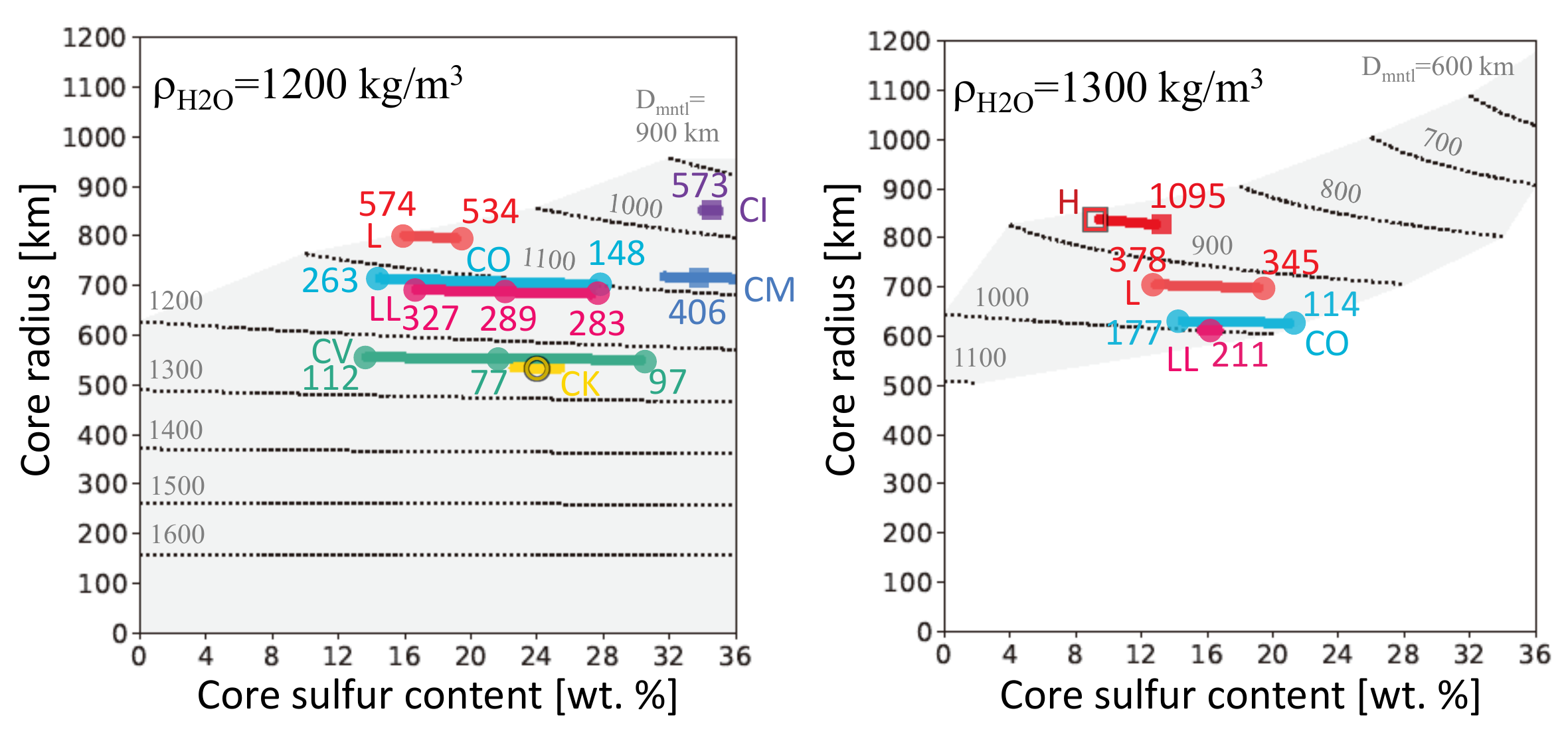}
\caption{
Estimated present-day surface magnetic field strength (nT) of Ganymede, indicated by the numbers adjacent to the symbols, plotted on the viable interior structure ranges shown in Figure \ref{fig:1}.
The grey shaded region represents the structural ranges consistent with the moment of inertia factor of 0.3156 for various chondritic compositions, and the dotted lines indicate contours of rocky mantle thickness.
The symbols represent the following core and dynamo regimes:
open circle ($\bigcirc$) denotes structures with no core solidification and no magnetic field generation.
Open square ($\square$) indicates structures where the dynamo region has disappeared, failing to sustain a field.
Filled circles (\textbullet) represent structures with an entirely molten core that successfully generates a field.
Filled squares ($\blacksquare$) indicate structures with core solidification and active magnetic field generation.}
	\label{fig:interior_mf}
\end{figure}

The present-day surface magnetic field strength (4.6\,Gyr after the start of the calculation) is shown in Figure \ref{fig:interior_mf} for various interior structure models permitted under the constraints of the moment of inertia and bulk composition.
Based on these results, the calculations can be broadly classified into four regimes according to the thermo-chemical evolution of the core and the magnetic field generation.
First, in the regime where no core solidification occurs and no magnetic field is generated (open circles in Figure \ref{fig:interior_mf}; see also Figure \ref{fig:CKmax}), the core sulfur concentration is close to the eutectic composition. 
Consequently, under the present simulation setup where the initial temperature is set to the melting point of the core composition, the initial temperatures of both the mantle and the core are low, delaying the onset of the core cooling phase. 
Due to this delayed cooling and the inefficient heat extraction through the thick rocky mantle, the power derived from secular cooling ($\phi_{\text{s}}$) remains permanently below the adiabatic threshold ($\phi_{\text{ad}}$). Because the core neither initiates thermal convection nor produces compositional buoyancy via solidification, magnetic field generation completely fails.
This regime was identified exclusively for the CK chondritic bulk composition.
Second, in the regime where a magnetic field is generated while the core remains entirely molten without initiating solidification (filled circles in Figure \ref{fig:interior_mf}; see also Figure \ref{fig:LLmin}), the resulting surface magnetic field strength ranges from several tens of nT to over 500\,nT. 
As the core sulfur concentration deviates further from the eutectic composition, the higher initial temperatures of the core and mantle enhance the efficiency of heat removal from both regions via solid-state mantle convection, thereby increasing the generated magnetic field strength. 
Furthermore, while the abundances of heat-producing elements vary among chondrite classes (e.g., higher $^{238}$U and $^{232}$Th in CO, CV, and CK; higher $^{40}$K in H, L, and LL; see Table \ref{tab:radiogenic_isotope}), our models indicate that these concentration differences are ultimately secondary to structural variations. 
A larger core radius (corresponding to a thinner mantle) reduces the total inventory of heat sources within the mantle, an effect that dominates over the specific elemental abundances. 
This smaller total heat inventory facilitates core cooling and thus leads to a stronger magnetic field.
Although none of the calculated magnetic field strengths in this regime reach the observed value of\,719 nT, some are compatible with the observation when the uncertainty in \(b_{\mathrm{dip}}\) is considered. For fields calculated with the adopted \(b_{\mathrm{dip}}=10\), Eq. (47) gives a compatible range from 503 to 1222\,nT when \(b_{\mathrm{dip}}\) varies from 7 to 17.
Third, in the regime where a magnetic field is generated during partial core solidification (filled squares in Figure \ref{fig:interior_mf}; see also Figure \ref{fig:Hmax} and \ref{fig:CMmid}), the interior structure models with the H chondritic composition experience the Fe-snow regime because the sulfur concentration is lower than the eutectic composition. 
Among these models, those with higher sulfur concentrations still generate a magnetic field stronger than the observed value at present. 
For the CI and CM chondritic compositions, the sulfur concentration is higher than the eutectic composition, resulting in a magnetic field driven by the FeS-layer regime.
In this regime, the slow growth of the solid FeS outer core allows the dynamo region to persist for a long duration, which is considered to prolong the period of magnetic field generation.
Fourth, in the regime where the dynamo region disappears, failing to sustain a magnetic field (open squares in Figure \ref{fig:interior_mf}; see also Figure \ref{fig:Hmin}), the cessation occurs exclusively within the Fe-snow regime. 
As the temperature decreases, the expansion of the Fe-snow zone reaches the center of the core, causing the dynamo region to vanish and magnetic field generation to cease. 

Overall, interior structure models capable of generating a magnetic field with the currently observed strength preferably possess a relatively large core radius (approximately 750 km or larger) and a bulk core sulfur concentration distinct from the eutectic composition. 
As shown in Figure \ref{fig:interior_mf}, models with extremely low sulfur concentrations (e.g., H chondrites, core sulfur concentration: 9.4 wt.\%) or those close to the eutectic composition (CK chondrite, 24.0 wt.\%) are more susceptible to dynamo failure under specific structural configurations. 
However, a comparison across different chondritic compositions indicates that this behavior is not solely governed by sulfur content. Instead, the success or failure of a present-day dynamo depends on the combined effects of sulfur-controlled solidification regimes and core-mantle structural dimensions that dictate the core cooling efficiency.
In contrast, compositions with intermediate sulfur concentrations successfully maintain an active dynamo ($\phi_{\text{all}} > 0$) across a wide range of hydrosphere densities and core radii.
Furthermore, under a low hydrosphere density scenario, other specific groups such as CV, CI, and CM compositions can also maintain an active dynamo depending on their structural parameters.
To further evaluate the viability of these successful models, we consider whether they can quantitatively explain the present-day magnetic field strength observed by Galileo ($\sim$719~nT). 
We therefore identify models as compatible with the observed field if their present-day surface field strengths calculated with \(b_{\text{dip}}=10\) fall within 503\,--\,1222\,nT.
Under this criterion, L and CI chondrites yield favorable parameter spaces at a hydrosphere density of 1,200~kg\,m$^{-3}$ (generating 534\,--\,574\,nT), whereas the H chondrite composition satisfies this condition under a high-density scenario of 1,300~kg\,m$^{-3}$ (reaching 1,095\,nT).
Furthermore, the model results reveal a systematic physical trend where the predicted magnetic field strength decreases systematically with increasing hydrosphere density for a given chondritic composition due to shifts in core-mantle structural partitioning. 
The inferred compositional constraints are conditional on the adopted range of \(b_{\mathrm{dip}}\).

While this broad trend is robust, the exact boundary of the compatible set exhibits sensitivity to the chosen threshold. 
Raising the criterion toward 600\,nT restricts the acceptable parameter space primarily to high-field models such as the H chondrite case, whereas lowering it toward 400\,nT broadens the viable window to include L chondrite models under the higher hydrosphere density. 
Although the uncertainty in $b_{\text{dip}}$ expands the window of compatible field strengths and threshold variations affect the exact boundary, the resulting viable ranges for the interior structure and bulk composition remain consistent with this systematic variation.
Ultimately, while our results do not uniquely isolate a single bulk composition, chondritic models with moderate core sulfur concentrations provide the most robust conditions for maintaining Ganymede's strong present-day magnetic field.
%
\begin{table*}[t]
\centering
\caption{Summary of simulated interior models, their core evolution regimes, and present-day magnetic field strengths.}
\label{tab:simulation_summary}
\begin{tabular}{ccccccc}
\hline
Chondrite & $\rho_{\mathrm{H_2O}}$ & Core Radius & Core Sulfur & Solidification & Present-day & Detailed \\
Type & (kg m$^{-3}$) & (km) & (wt.\%) & Regime & Magnetic Field (nT) & Results \\
\hline
CK & 1200 & 530 & 24.0 & Entirely Molten & No & Fig.\ref{fig:CKmax} \\
CV & 1200 & 541 & 13.7 & Entirely Molten & Yes (112) &  \\
CV & 1200 & 539 & 22.0 & Entirely Molten & Yes (77) &  \\
CV & 1200 & 540 & 30.3 & Entirely Molten & Yes (97) &  \\
LL & 1200 & 690 & 16.7 & Entirely Molten & Yes (327) & Fig.\ref{fig:LLmin} and \ref{fig:LLmin_temperature} \\
LL & 1200 & 670 & 22.0 & Entirely Molten & Yes (289) &  \\
LL & 1200 & 667 & 27.3 & Entirely Molten & Yes (283) &  \\
CO & 1200 & 698 & 14.5 & Entirely Molten & Yes (263) &  \\
CO & 1200 & 687 & 27.5 & Entirely Molten & Yes (148) &  \\
CM & 1200 & 720 & 34.0 & FeS-Layer & Yes (406) & Fig.\ref{fig:CMmid} \\
L  & 1200 & 784 & 16.0 & Entirely Molten & Yes (574) &  \\
L  & 1200 & 778 & 19.1 & Entirely Molten & Yes (534) &  \\
CI & 1200 & 844 & 35.0 & FeS-Layer & Yes (573) &  \\
LL & 1300 & 595 & 16.7 & Entirely Molten & Yes (211) & \\
CO & 1300 & 613 & 14.5 & Entirely Molten & Yes (177) &  \\
CO & 1300 & 608 & 21.0 & Entirely Molten & Yes (114) &  \\
L & 1300 & 686 & 12.9 & Entirely Molten & Yes (378) & \\
L & 1300 & 679 & 19.1 & Entirely Molten & Yes (345) & \\
H  & 1300 & 840 & 9.4  & Fe-Snow & No (Ceased) & Fig.\ref{fig:Hmin} \\
H  & 1300 & 830 & 12.9 & Fe-Snow & Yes (1095) & Fig.\ref{fig:Hmax} and \ref{fig:Hmax_temperature} \\
\hline
\end{tabular}
\end{table*}

%
%
%
%
%
\clearpage
\section{Discussions}

\subsection{Model Limitations and Robustness of Core-Mantle Thermal Coupling}
In this study, we employed a one-dimensional numerical scheme based on the mixing length theory (MLT) to evaluate sub-solidus convective heat flux within Ganymede's rocky mantle. 
This approach successfully avoids the prohibitive computational costs of full three-dimensional simulations, making it uniquely suited for our extensive parameter space exploration.
However, this 1D formulation inherently neglects three-dimensional lateral heterogeneities in the core-mantle boundary heat flux.
In a realistic 3D mantle dynamic regime, localized downwellings and mantle plumes induce spatial variations in thermal extraction \citep[e.g.][]{amit2012}. 
Such lateral thermal anomalies can distort fluid motions in the outer core and introduce non-axisymmetric magnetic field components, which cannot be captured by our 1D formulation.

Furthermore, the MLT formulation implemented in our study carries inherent uncertainties regarding the absolute efficiency of vertical heat transport. 
As demonstrated by \citet{wagner19} and \citet{thiriet19}, the mixing-length parameters in a stagnant-lid regime are expected to be substantially lower than those assumed in conventional uncalibrated formulations. 
Consequently, our 1D approach may overestimate the convective heat transport efficiency of the mantle. 
If corrected through rigorous 3D calibration, the reduced heat extraction would theoretically result in a warmer deep interior, thereby delaying the overall cooling of the core. 
Such a delay would shift the onset of core crystallization and the generation of compositionally driven dynamos to a more recent epoch. 
While this absolute timing shift might favorably extend the lifespan of otherwise short-lived dynamo models, it does not fundamentally alter the underlying thermodynamic mechanisms. 
The macroscopic structural conclusions of this study, specifically the bifurcation between the Fe-snow and FeS-layer regimes based on bulk sulfur content and the relative dependence of core cooling on mantle thickness, remain physically robust regardless of this calibration uncertainty.

The thermal and structural evolution of Ganymede's core is predominantly governed by the detailed thermodynamic properties of the implemented Fe-FeS phase diagram. 
At low sulfur concentrations below approximately 10 wt.\% sulfur, the liquidus temperature generally increases with pressure. 
Conversely, within the intermediate composition range from approximately 7 to 13 wt\% sulfur, the liquidus curves exhibit highly complex behavior where the relationships between pressure and melting temperature intersect and invert.
As discussed by \citet{Xu2021}, the specific conditions where the core adiabat and liquidus intersect within Ganymede's core pressure range (4 to 10~GPa) depend strongly on the chosen equations of state and liquidus profile. 
Therefore, our specific finding that the Fe-snow regime occurs for sulfur concentrations below approximately 9 wt.\% is a direct consequence of the thermodynamic parameterization adopted in this study, and applying different equations of state could shift this transition boundary.
Crucially, Ganymede's core radius of approximately 840 km dictates an internal pressure range where the outer region near the core-mantle boundary falls directly within this complex 4 to 7~GPa transition zone.
Our dynamic simulations demonstrated that even for the H-chondrite model, which possesses the lowest initial sulfur concentration of approximately 9 wt.\% sulfur and thus sits within this sensitive regime, the core undergoes top-down crystallization, specifically the Fe-snow regime, rather than simple bottom-up core solidification. 
This outcome arises because the temperature gradient of the cooling core dynamically intersects the complex liquidus profile from the lower-pressure CMB side first. 
This result highlights that Ganymede's unique spatial and pressure scale physically enables top-down buoyancy-driven dynamos even under low-sulfur conditions.

This finding aligns with the comprehensive parameter sweep by \citet{Ruckriemen2018}, which demonstrated that successful dynamo models are viable across almost the entire range of sulfur concentrations. 
Building upon this theoretical foundation, our study moves beyond arbitrary parameter explorations by explicitly anchoring the core's initial sulfur inventory and structural dimensions to rigorous cosmochemical bulk compositions. 
By capturing the precise dynamic intersection between the core adiabat and the complex liquidus topology within our coupled 1D MLT framework, we demonstrate that sustained dynamos naturally emerge even when strictly bounded by these meteoritic precursor constraints. 
This result is a direct consequence of coupling the detailed core thermodynamic and transport properties with the mantle cooling history. 
Consequently, this coupled approach allows us to utilize the observed present-day magnetic field to effectively map viable evolutionary pathways directly to distinct planetary bulk origins (such as L, CI, or H chondrites).

Furthermore, the specific preference for non-eutectic bulk sulfur concentrations in sustaining the observed 719 nT field strength is directly linked to a physical feedback through the temperature-dependent mantle viscosity. Locking the initial core temperature to its composition-dependent liquidus inherently influences the overlying mantle dynamics from the onset of the evolution. Near the eutectic composition, the lower initial core temperature yields a lower temperature boundary condition at the core-mantle boundary (CMB), which exponentially increases the basal mantle viscosity and diminishes the CMB heat flux, thereby suppressing thermal dynamo activity. Conversely, the higher initial temperatures required by non-eutectic bulk sulfur concentrations lower the mantle viscosity and accelerate core cooling. This core-mantle thermal coupling explains why non-eutectic compositions preferentially facilitate the generation and sustenance of the observed field strength within our modeling framework.

Regarding the initial thermal state, our model assumes a simple linear temperature profile in the mantle following an early, complete differentiation. 
One might alternatively consider a hotter, fully convective initial mantle thermal profile. 
However, due to the strongly temperature-dependent viscosity of the mantle rock, an initially hotter mantle would exhibit drastically lower viscosity and highly efficient convective heat transport. 
This would cause the initial heat loss to temporarily outpace the temperature rise driven by radiogenic heating, leading to rapid cooling until the viscosity increases sufficiently to establish a thermal balance. 
Because of this self-regulating viscosity thermostat, the mantle would converge to a similar quasi-steady convective state within a timescale comparable to the initial heating phase of our linear model (less than 0.5~Gyr). 
Therefore, the long-term thermal evolution of the core and the conditions for present-day dynamo generation remain robust against uncertainties in the initial mantle temperature profile.

High-pressure experimental data for the Fe-FeS system carry non-negligible uncertainties that introduce both qualitative and quantitative sensitivities to our model \citep{hauck2006,Ruckriemen2018}. 
Qualitatively, the exact sulfur concentration at which the liquidus slope inverts is subject to experimental variations. 
Because the initial sulfur content of the H-chondrite model ($\sim$9.4\,wt.\% sulfur) lies directly within the 7-13 wt.\% sulfur uncertainty window, a shift in this crossover boundary toward higher sulfur concentrations would prevent the core from maintaining the criteria for top-down crystallization. 
In such a scenario, the crystallization mode would invert to a standard bottom-up core growth, shifting the primary buoyancy source from Fe-snow to classical compositional convection. 
While classical inner-core growth could also explain the present-day field if it initiated recently, this shift fundamentally alters the evolutionary timeline required to sustain the observed surface magnetic field strength of approximately 719~nT. 
Consequently, the viability of the H-chondrite scenario within our framework is critically dependent on the precise topology of this thermodynamic crossover.

Quantitatively, our model simplifies the liquidus curves using a linear V-shaped parameterization (Figure \ref{fig:fefes}). 
Because the true thermodynamic liquidus curve is convex upward, characterized by a gentler slope near the end-members and a steeper drop approaching the eutectic cusp, the linear approximation could underestimate the liquidus temperature in the intermediate composition range around 9 wt.\% sulfur. 
If a more realistic, curved liquidus profile were implemented, the core would reach its higher melting temperature at an earlier stage of cooling. 
This would quantitatively shift the onset of the Fe-snow regime several hundred million years earlier, thereby altering the cumulative amount of generated buoyancy flux and the core solidification fraction at 4.6 Gyr. 
However, since the eutectic point remains a distinct thermodynamic cusp where the liquidus slope changes discontinuously in both linear and curved formulations, the physical transition between the Fe-snow and subsequent FeS-layer regimes remains robustly sharp. 
Despite these parameter and methodology sensitivities, this coupled framework demonstrates that combining geodetic constraints with the requirement to sustain an active magnetic dynamo can effectively rule out incompatible planetary bulk compositions.

It should be noted that the early transient warming of the core observed in our baseline models is a direct consequence of setting the initial core temperature precisely at its composition-dependent liquidus. 
Alternatively, if a superheated core resulting from rapid early differentiation were assumed \citep[e.g.][]{Solomon1979}, this initial warming phase would be absent, and the core would enter a strong cooling regime from the onset. 
While such a superheated initial state could advance the timing of dynamo initiation or shorten its early lifespan, it would not fundamentally alter the long-term thermochemical regime (Fe-snow or FeS-layer) required to sustain the magnetic field over 4.6 Gyr to the present day.

%
\subsection{Sensitivities to Hydrosphere Model Simplifications}
Although the thermal boundary condition and physical parameterization prescribed for the hydrosphere impose certain limitations, their overall impact on the deep interior evolution remains well-constrained. 
In our model, the temperature at the rocky mantle--hydrosphere boundary (RHB) is fixed at 300\,K as mentioned in Section 2.1. 
Crucially, because the heat transfer efficiency within the $\mathrm{H_2O}$ layer is orders of magnitude higher than that of the rocky mantle, the overall cooling rate of the deep interior is predominantly governed by the convective heat flux through the rocky mantle. Therefore, this simplified thermal boundary condition does not significantly alter the primary trajectory of the core-mantle thermal evolution.

Beyond these thermal effects, fixing the bulk density of the hydrosphere to either 1,200 or 1,300\,kg\,m$^{-3}$ introduces structural sensitivities regarding planetary mass distribution. 
In a fully thermodynamic volatile shell, the vertical density profile varies due to high-pressure ice phase transitions and the salinity of the subsurface ocean. 
Such density re-distributions shift the mass allocation required to satisfy the bulk moment of inertia (MoI) constraint, thereby subtly altering the permissible ranges of the rocky mantle thickness and the core radius. 
A thinner mantle reduces the total inventory of radiogenic heat sources, which subsequently accelerates core cooling and enhances the present-day surface magnetic field strength. 
As demonstrated in our results, this variance in the baseline hydrosphere density dictates the sharp bifurcation in bulk composition preference, favoring L and CI chondrites at 1,200\,kg\,m$^{-3}$ versus H chondrites at 1,300\,kg\,m$^{-3}$. 
Consequently, while the thermal simplification of the $\mathrm{H_2O}$ layer is physically robust, incorporating a comprehensive thermodynamic multi-layer hydrosphere model represents an important next step to further refine the exact structural boundaries of the primordial bulk composition.
%
\subsection{Effects of Early Differentiation vs. Protracted Core Formation}

The recent proposal by \citet{Trinh2026} regarding protracted core formation presents a compelling alternative to the classical "hot start" paradigm. 
In their model, Ganymede's current dynamo is powered by the ongoing segregation of iron from the rocky mantle, a process that avoids the need for a long-lived magnetic field maintained solely by secular cooling. 
However, while Ganymede's significantly lower moment of inertia factor \citep{GomezCasajus2022} strictly confirms its fully differentiated state today, the classical assumption of an early and efficient differentiation event is primarily motivated by the distinct structural dichotomy between Ganymede and Callisto \citep{Nagel2004}.

Our modeling approach, which assumes a fully differentiated state from 4.6\,Gyr ago, serves as a necessary end-member study to define the thermal and magnetic limits of the "hot start" scenario. While the ongoing differentiation model of \citet{Trinh2026} relies on the potential energy of falling iron diapirs, our model tests whether a static, cooling core can sustain the observed field strength under various mantle transport efficiencies and sulfur concentrations. These two scenarios represent the breadth of current uncertainties in icy satellite formation; thus, evaluating the long-term evolution of a fully formed core is essential for identifying observational signatures that might eventually distinguish between a cooling-driven and a growth-driven dynamo. By assuming a fully differentiated three-layer structure from the onset, we can isolate and evaluate the influence of core sulfur concentration and mantle heat transport on the evolution of magnetic field intensity, providing a necessary end-member comparison to delayed differentiation models.
%
\subsection{Sensitivities to Core Thermal Parameters}
\label{sec:core_thermal_parameters}
\begin{figure}
\centering
\includegraphics[width=0.75\textwidth]{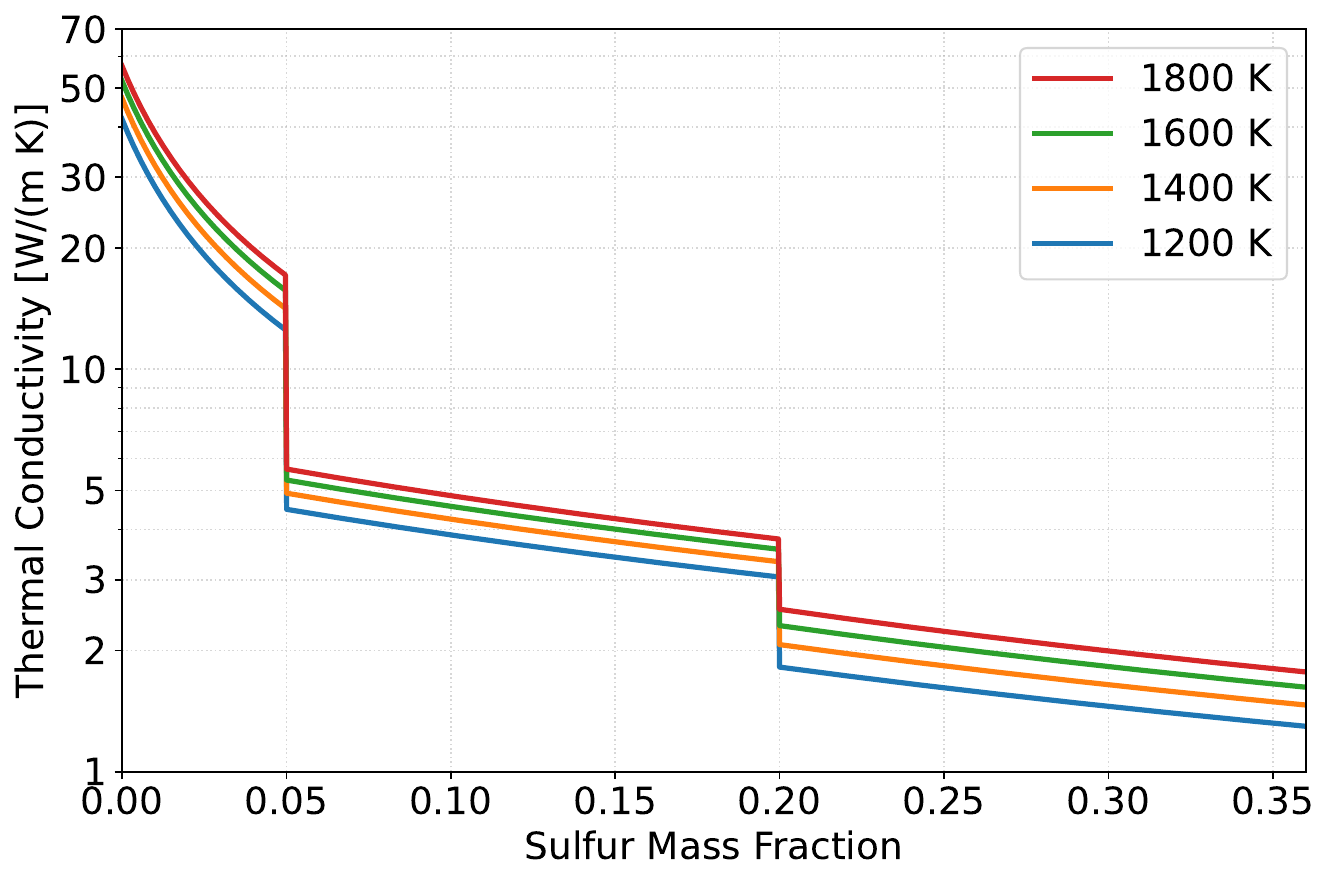}
\caption{
Core thermal conductivity ($k_{\mathrm{cond,c}}$) as a function of sulfur mass fraction ($x_{\mathrm{S}}$) at representative core temperatures ($T_c = 1200, 1400, 1600$, and $1800$ K). The values are dynamically computed using the Wiedemann-Franz law (Eq. \ref{eq:kcond}) with a constant Sommerfeld value ($L_0 = 2.445 \times 10^{-8}$ W\,$\Omega$\,K$^{-2}$) applied to the electrical resistivity model of \citet{Pommier2018b} (Eq. \ref{eq:rho_elc}). The profiles illustrate the strong compositional dependence of the thermal conductivity, spanning from over 60 W/(m K) near the pure-Fe endmember ($x_{\mathrm{S}} \le 0.05$) to approximately 1.5-3.5 W/(m K) in the sulfur-rich regime ($x_{\mathrm{S}} \ge 0.20$). The explicit mathematical discontinuities at $x_{\mathrm{S}}= 0.05$ and $x_{\mathrm{S}} = 0.20$ reflect the boundaries of the three piecewise linear compositional regimes parameterized from discrete experimental data.}
\label{fig:core_thermal_conductivity}
\end{figure}

The thermal boundary conditions and the internal buoyancy generation of the core dynamo are critically linked to the core thermal conductivity ($k_{\mathrm{cond,c}}$, Eq. \ref{eq:kcond}). 
This parameter determines the adiabatic heat flux ($F_{\mathrm{ad}}$) that core cooling must exceed to sustain purely thermal convection. 
In our model, rather than being held fixed, $k_{\mathrm{cond,c}}$ is dynamically computed at each grid point and time step. It is derived from the core electrical resistivity ($\rho_{\mathrm{el,c}}$) via the Wiedemann-Franz law, utilizing the theoretical Sommerfeld value $L_0=2.445 \times 10^{-8}~\text{W}\Omega\text{K}^{-2}$ as a constant proportionality factor. 
Crucially, the implemented $\rho_{\mathrm{el,c}}$ formulation is partitioned into three composition regimes with distinct temperature ($T_{\mathrm{c}}$) and sulfur concentration ($x_{\mathrm{S}}$) dependencies (Eq.\ref{eq:kcond}), based on the available discrete experimental data \citep{Pommier2018b}.
Note that this piecewise linear parameterization inevitably introduces mathematical discontinuities at the boundaries ($x_{\mathrm{S}}$ = 0.05 and 0.20), as explicitly visualized in Figure \ref{fig:core_thermal_conductivity}. 
Despite these local discontinuities, this formulation successfully captures the overarching macroscopic trend required for our core evolution models: a drastic reduction in thermal conductivity from over 60~Wm$^{-1}$K$^{-1}$) near the pure-Fe endmember to approximately 1.5-3.5~Wm$^{-1}$K$^{-1}$) in the sulfur-rich regime (0.20$\le x_{\mathrm{S}}\le$0.36). While the assumption of a constant Sommerfeld value under planetary core conditions, along with this empirical parameterization of $\rho_{\mathrm{el,c}}$, carries inherent thermodynamic uncertainties that propagate directly into the calculated core heat flux, we explicitly applied these standard relations to provide a self-consistent baseline for the current study. A more comprehensive parameter study addressing the full complexity of $k_{\mathrm{cond,c}}$ remains an important subject for future work.

For the H-chondritic composition, the initial sulfur content of approximately 9.4\,wt.\% sulfur places the core evolution within the intermediate regime ($0.05 \le x_{\mathrm{S}} < 0.20$). 
Within this specific window, the electrical resistivity exhibits high sensitivity to the linear coefficients of both $T_{\mathrm{c}}$ and $x_{\mathrm{S}}$. 
As the core cools and $x_{\mathrm{S}}$ varies due to top-down crystallization, the slopes defined by these regime-dependent coefficients dynamically govern the evolution of $\rho_{\mathrm{el,c}}$ and the subsequent variation in $k_{\mathrm{cond,c}}$. 
Any systematic errors within these empirical sub-regime formulations or physical departures from the ideal Lorenz number directly shift the dynamic trajectory of $F_{\mathrm{ad}}$. 
A higher assumed $k_{\mathrm{cond,c}}$ expands the fraction of heat carried by conduction alone, thereby suppressing early thermal dynamos and shifting the onset of the top-down Fe-snow regime to a later epoch. 
Conversely, a lower $k_{\mathrm{cond,c}}$ reduces the adiabatic threshold, accelerating the dynamic intersection between the core temperature and the liquidus profile.
These thermal conductivity uncertainties consequently introduce a quantitative uncertainty window of several hundred million years in the predicted dynamo lifetime and the remaining buoyancy flux at 4.6\,Gyr.

Beyond thermal conductivity, the thermodynamic parameters governing core buoyancy, specifically the thermal expansivity ($\alpha_c$) and the specific entropy of fusion ($\Delta S$), also carry inherent uncertainties due to the lack of complete experimental data across Ganymede's pressure (4 to 10 GPa) and wide sulfur concentration range. 
By adopting representative constants \citep[e.g.][]{Ruckriemen2018}, our model isolates the primary effects of bulk composition, though physical deviations in these parameters would introduce specific thermal feedbacks.
For instance, the thermal expansivity ($\alpha_c$) is generally expected to increase with higher sulfur concentrations. 
If $\alpha_c$ was significantly larger than our assumed constant, the adiabatic heat flux ($F_{ad}$) would proportionally increase, raising the threshold required to sustain purely thermal convection. This would delay or suppress early dynamo activity. 
Simultaneously, a higher $\alpha_c$ increases the efficiency with which secular cooling is converted into dynamo power, creating a physical trade-off between the instantaneous magnetic field strength and the duration of the dynamo.
Similarly, the specific entropy of fusion ($\Delta S$) directly determines the latent heat budget ($Q_{lat}$) during core solidification. 
If $\Delta S$ were larger than assumed, its impact would strictly depend on the active crystallization regime. 
In the Fe-snow regime, a larger $\Delta S$ intensifies the endothermic remelting of descending iron particles. 
This locally steepens the thermal gradient, temporarily enhancing convective vigour, but it simultaneously accelerates the global freezing of the core, thereby shortening the dynamo lifetime. Conversely, in the FeS-layer regime, a larger $\Delta S$ intensifies the exothermic release of latent heat at the top of the core. 
This acts as a powerful thermal buffer that slows secular cooling, potentially extending the dynamo's lifespan while reducing the instantaneous convective power. 
Despite these quantitative variations in dynamo intensity and lifetime, the fundamental structural bifurcation between the Fe-snow and FeS-layer regimes remains physically robust, as it is primarily governed by the bulk sulfur content and the topology of the liquidus curve.
%
\subsection{Observational Discrimination of Core Regimes by the JUICE Mission}

The numerical results and core dynamo regimes identified in this study offer critical benchmarks that can be tested by upcoming high-resolution magnetic field measurements from the JUICE spacecraft. 
Precise characterization of Ganymede's internal and external environments during its low-altitude orbital phase will provide tight constraints on the core's thermal and compositional states through four distinct observational windows.

First, the accurate separation of the intrinsic direct current (DC) field from the induction response of the conductive subsurface ocean is paramount. 
Resolving the time-variable induction field driven by Jupiter's magnetosphere to within a few nanoteslas will help better isolate the surface intensity of the core-generated field. 
Because this intensity correlates with the core surface heat flux and the cooling rate, narrowing the error margin of the intrinsic field strength will provide tighter bounds on the permissible combinations of core sulfur concentration and mantle heat transport efficiency within our model. 
However, it is important to note a key caveat: the dynamo scaling laws used to link the core heat flux to the magnetic field strength contain inherent uncertainties. 
Therefore, rather than expecting an exact numerical match, utilizing these high-precision observations to ensure the model outputs remain in the consistent order of magnitude will be a more robust approach to evaluating the viable interior regimes.

Second, determining the spherical harmonic spatial power spectrum up to higher-degree multipoles, specifically the quadrupole ($l=2$) and octupole ($l=3$), will provide a direct diagnostic of the internal convective geometry. 
A rapidly decaying power spectrum with degree $l$ would indicate that the dynamo is driven by a globally thick convective layer. 
While this supports a fully molten core regime without extensive stable stratification, it could also correspond to a scenario where a thin snow zone has formed but the dynamo is still operating in almost the entire core. 
Conversely, a flatter spectrum with relatively enhanced $l=2$ and $l=3$ components would imply that fluid motion is physically confined to a thin shell, arguing against convection throughout the whole core. 
Such a spectral signature could serve as observational evidence for a stratified core lower boundary, consistent with the Fe-snow or FeS-layer regimes \citep{Zhan2012,Christensen2015}. 
However, it is important to note that this signature could alternatively be produced by a thick thermally stable layer at the top of the core that similarly reduces the extent of the convecting region.

Third, the ratio of non-axisymmetric (m$\neq$0) to axisymmetric (m\,=\,0) components in these multipole fields can constrain the spatial uniformity of core cooling. A pronounced non-axisymmetric signature would reveal lateral heterogeneities in heat extraction at the core-mantle boundary, likely induced by lower mantle convection cells or plumes. 
Given Ganymede's known rotation rate ($\Omega=1.02\times10^{-5}$\,rad\,s$^{-1}$) and a typical core convective velocity ($U\sim10^{-4}$ m/s), the expected Rossby number ($Ro\,=\,U/\Omega\,R_c$) is on the order of $10^{-5}$. 
This low value indicates that Coriolis forces inherently dominate the bulk core fluid dynamics. 
Therefore, the degree of axisymmetry in the observed field will not merely serve to prove this rotational dominance, but rather test whether mantle-driven thermal anomalies are sufficiently powerful to break this strong baseline rotational constraint and imprint a non-axisymmetric signature.

Fourth, and most importantly, the comprehensive characterization of the hydrosphere via geodetic and geophysical payloads will establish an independent constraint on the planetary bulk composition, which we emphasize as the main strength and actual novelty of this coupled framework.
The JUICE mission will utilize its laser altimeter (GALA) and radio science experiment (3GM) to precisely measure Ganymede's tidal deformation and gravity field, effectively probing the thickness, stratification, and salinity-dependent density of the $\mathrm{H_2O}$ layer. 
Resolving the true bulk density of the hydrosphere will directly eliminate the principal degeneracy identified in our model. 
By determining whether the actual hydrosphere properties align closer to our 1,200 or 1,300 $\mathrm{kg\,m^{-3}}$ case, these geodetic constraints will select between the H chondrite and the L/CI chondrite models, shifting the mission's diagnostic power from localized core dynamics to the definitive identification of Ganymede's primordial origin.
%
\clearpage
%
\section{Conclusions}

In this study, we conducted long-term thermochemical evolution simulations of Ganymede's interior, explicitly including the core solidification process, across a wide range of interior structures permitted by various chondritic bulk compositions (H, L, LL, CI, CM, CO, CV, and CK) and geodetic constraints. 
By modeling the thermal history and magnetic field generation mechanisms, we utilized the observed surface magnetic field strength to constrain the permissible interior structures, specifically the sizes and densities of the metallic core, rocky mantle, and hydrosphere, and their corresponding bulk compositions.

The structural geometries permitted solely by the moment of inertia and bulk composition constraints vary significantly depending on the assumed $\mathrm{H_2O}$ layer density. 
In the low-density case ($1,200~\mathrm{kg\,m^{-3}}$), the initial structural filtering yields a core radius ranging from 550 to 850~km, while ruling out the H chondrite composition due to its insufficiently small moment of inertia. 
On the other hand, in the high-density case ($1,300~\mathrm{kg\,m^{-3}}$), the core radius ranges from 600 to 850~km, and the CI, CM, CV, and CK chondrite compositions are strictly excluded because they yield excessively large moments of inertia.

Our thermochemical evolution calculations demonstrated that a magnetic dynamo can be driven under all modeled core solidification regimes, including total melting, the Iron Snow regime, and the FeS-layer regime.
This result is in excellent agreement with previous dynamo models that investigated these individual regimes \citep{Zhan2012,Christensen2015,Ruckriemen2018}.
Crucially, applying the observed magnetic field strength as a constraint demonstrates the potential to narrow down the viable interior structures and bulk compositions. 
The models capable of sustaining the observed field strength preferentially require a relatively large core radius (approximately 750~km or larger) and a bulk core sulfur concentration distinct from the eutectic composition. 
These constraints are based on the 503\,--\,1222 nT compatibility interval for fields calculated with \(b_{\mathrm{dip}}=10\), corresponding to the range of 7\,--\,17 for \(b_{\mathrm{dip}}\).

This preference remains highly robust; although the inherent uncertainty in the dipole attenuation factor, $b_{\text{dip}}$, expands the acceptable window of magnetic field strengths, the resulting compatible ranges for the interior structure and bulk composition barely change. 
Ultimately, our numerical results demonstrate that while extremely low or near-eutectic sulfur concentrations generally fail to sustain a present-day dynamo, compositions with intermediate sulfur contents  maintain active dynamos across various structural conditions. 
When evaluating these viable models against a criterion to generate a field strength comparable to the observation, the compatible compositions become strongly dependent on the hydrosphere density. 
Specifically, L and CI chondrites yield parameter spaces capable of sustaining comparable field strengths under a low-density hydrosphere scenario (1,200 kg m-3), whereas the H chondrite composition can satisfy this criterion under a high-density scenario (1,300 kg m-3).
Therefore, characterizing the hydrosphere density provides the critical independent constraint required to break the degeneracy of Ganymede's primordial origin.

Future gravity and magnetic field observations by the JUICE mission, scheduled to arrive in the Jovian system in 2031, will provide a critical test for these model predictions. Finer constraints on Ganymede's moment of inertia, the detailed density profile of its $\mathrm{H_2O}$ layer, the tidal responses, and the rotational state will further refine the core solidification regime and evaluate the cosmochemical scenarios suggested in this work.

%



%
\printcredits

\section*{Acknowledgments}
This work was supported by JSPS KAKENHI Grant Numbers JP22K03700. 
We thank the reviewers for their constructive comments that improved the manuscript.

\bibliographystyle{cas-model2-names}

\bibliography{jkbibfile}



\end{document}